\documentclass[aps,prc,twocolumn,floatfix,showpacs,preprintnumbers,amsmath,amssymb,nofootinbib,superscriptaddress]{revtex4-1}
\usepackage[colorlinks=true,citecolor=blue,linkcolor=blue,anchorcolor=blue,filecolor=blue,
urlcolor=blue]{hyperref}
\usepackage{color}
\usepackage[dvipsnames,usenames]{xcolor}
\colorlet{orange}{orange!125!}
\usepackage{graphicx}
\usepackage{subfigure}
\usepackage{dcolumn}    
\usepackage{multirow}
\usepackage{bm}         
\usepackage{sidecap}
\usepackage{amsmath}
\usepackage{float}
\usepackage{adjustbox}
\usepackage{enumitem} 

\usepackage[utf8]{inputenc}

\usepackage{array}
\newcolumntype{d}[1]{D{.}{.}{#1}}

\usepackage{mathtools} 
\usepackage{mhchem} 

\usepackage{mathrsfs,natbib}
\usepackage{epsf,amssymb,amsbsy,amsfonts,amssymb,amsmath}
\usepackage{slashed}
\usepackage{comment}

\newcommand{\gmr}{\mathrm{ISGMR}}

\newcommand{\sat}{\mathrm{sat}}
\newcommand{\sym}{\mathrm{sym}}
\newcommand{\tov}{\mathrm{TOV}}

\newcommand{\nuc}{\mathrm{nuc}}
\newcommand{\tot}{\mathrm{tot}}

\newcommand{\cl}{\mathrm{cl}}

\begin{document}

\title{Low-energy nuclear physics and global neutron star properties}

\author{Brett V. Carlson}
\affiliation{Departamento de F\'isica, Instituto Tecnol\'ogico de Aeron\'autica, DCTA, 12228-900, S\~ao Jos\'e dos Campos, SP, Brazil}

\author{Mariana Dutra}
\affiliation{Departamento de F\'isica, Instituto Tecnol\'ogico de Aeron\'autica, DCTA, 12228-900, S\~ao Jos\'e dos Campos, SP, Brazil}

\author{Odilon Louren\c{c}o}
\affiliation{Departamento de F\'isica, Instituto Tecnol\'ogico de Aeron\'autica, DCTA, 12228-900, S\~ao Jos\'e dos Campos, SP, Brazil}

\author{J\'er\^ome Margueron}
\affiliation{Univ Lyon, Univ Claude Bernard Lyon 1, CNRS/IN2P3, IP2I Lyon, UMR 5822, F-69622, Villeurbanne, France}

\date{\today}

\begin{abstract}
We address the question of the role of low-energy nuclear physics data in constraining neutron star global properties, e.g., masses, radii, angular momentum, and tidal deformability, in the absence of a phase transition in dense matter. To do so, we assess the capacity of 415 relativistic mean field and non-relativistic Skyrme-type interactions to reproduce the ground state binding energies, the charge radii and the giant monopole resonances of a set of spherical nuclei. The interactions are classified according to their ability to describe these characteristics and we show that a tight correlation between the symmetry energy and its slope is obtained providing $N=Z$ and $N\ne Z$ nuclei are described with the same accuracy (mainly driven by the charge radius data). By additionally imposing the constraints from isobaric analog states and neutron skin radius in $^{208}$Pb, we obtain the following estimates: $E_{\sym,2}=31.8\pm0.7$~MeV and $L_{\sym,2}=58.1\pm 9.0$~MeV. We then analyze predictions of neutron star properties and we find that the 1.4$M_\odot$ neutron star (NS) radius lies between 12 and 14~km for the ``better'' nuclear interactions. We show that i) the better reproduction of low-energy nuclear physics data by the nuclear models only weakly impacts the global properties of canonical mass neutron stars and ii) the experimental constraint on the symmetry energy is the most effective one for reducing the uncertainties in NS matter. However, since the density region where constraints are required are well above densities in finite nuclei, the largest uncertainty originates from the density dependence of the EDF, which remains largely unknown.
\end{abstract}

\maketitle

\section{Introduction}

The modeling of neutron stars (NS) relies mostly on the present knowledge of nuclear physics, since nuclear properties determine the characteristics of its crust -- together with the electrons -- and the energetics of the  nucleon liquid in its outer core~\cite{Lattimer2016, Haensel2007, Steiner2005}. Nowadays however, most of the theoretical efforts necessary for the understanding of observational data, such as gravitational waves emitted from binary NS~\cite{ligo} or X-ray emissions from milli-second pulsars~\cite{nicer1a,nicer1b,nicer2a,nicer2b}, require for the most part the understanding of the NS inner core, where densities exceed by several units the saturation density of nuclear matter ($\rho_{\sat}\approx 2.7\times 10^{14}$~g~cm$^{-3}$). These new data question the impact of nuclear physics constraints, operating at or around saturation density, on the properties of supra-saturation density matter. To what extent do global properties of neutron stars, such as their masses, radii or tidal deformabilities, require accurate experimental nuclear data as complementary constraints? Is the extrapolation of nuclear physics models to higher densities predominantly controlled by nuclear physics data at saturation density? What is the impact of other uncertainties, such as for instance the isospin symmetry dependence of the equation of state (EoS), which is for the most part unknown, except close to saturation density and to isospin symmetry ($(N-Z)/A\lesssim0.25$).

Such questions were recently addressed by analyzing the correlations between a few nuclear empirical parameters (NEP), namely the symmetry energy $E_{\sym}$, its slope $L_{\sym}$ and the incompressibility modulus $K_\sat$, and NS global observables for a set of microscopic nuclear EoS derived within the Brueckner-Hartree-Fock (BHF) formalism~\cite{Wei2020}. No correlation were found, except the one between NS radius and tidal deformability for a 1.4~M$_\odot$ NS ($R_{1.4}$ and $\Lambda_{1.4}$) and the pressure of beta-stable matter at twice saturation density, as initially suggested in Ref.~\cite{Prakash2001}. In a different analysis based on the relativistic mean field (RMF) description of dense matter, a linear correlation between $\Lambda_{1.4}$ with $K_\sat$ and $L_{\sym}$ was however found, as well as an anti-correlation with $E_{\sym}$ and the effective mass~$m^*$~\cite{Souza2020}. 

Such a controversy implies that the question of the role of low-energy nuclear physics in the prediction of global properties of NS is not yet clarified. This motivates the new analysis presented in this paper. We perform a statistical analysis based on a large number of nuclear physics models (415 in total). We explore the question of the model dependence of the results by investigating various types of modeling: the Skyrme nuclear force~\cite{Bender2003} and two types of relativistic mean field (RMF) approaches, the RMF with non-linear couplings (\mbox{RMF-NL})~\cite{Bender2003,Reinhard1989} and the RMF with density dependence couplings (\mbox{RMF-DD})~\cite{Typel2018}). Note that RMF-DD models take into account, at least partially, the effect of BHF correlations at the mean field level. We have included such interactions in our analysis and we assume here that they serve as surrogates for more elaborate BHF models, such as these analysed in Ref.~\cite{Wei2020}. At variance with the analysis presented in Ref.~\cite{Wei2020}, we directly compare the nuclear models predictions in finite nuclei to experimental data.

Our approach is different from the one presented in Refs.~\cite{Stone2007,Dutra2012,Dutra2014}, where almost the same set of models were confronted to NS observables ignoring their adequacy in describing low energy nuclear physics properties. The two first papers compare non-relativistic Skyrme interactions~\cite{Stone2007,Dutra2012}, while the last one uses relativistic Hartree interactions as well as a smaller set of relativistic point interactions~\cite{Dutra2014}. In each study, only an extremely small number of the interactions were able to describe all of the nuclear matter properties considered. Surprisingly, none of the successful interactions are among those that provide the best fits to nuclear binding energies and charge radii. In the present paper we adopt a different strategy where we first select the models according to their ability to reproduce low-energy nuclear physics data. To do so, we perform a direct comparison in finite nuclei between model predictions and low energy nuclear data, namely we consider nuclear binding energies, charge radii, giant monopole energies and a constraint on the density dependence of the symmetry energy. We end up with different groups of models passing various constraints with different accuracies, as in the previous analyses. In the second step of the analysis, we propagate the model predictions to higher densities. This allows us to analyse the impact of low energy nuclear physics data on the predictions of global NS properties.

We show that the model dispersion at high density is weakly impacted by low-energy nuclear physics data, except for the data associated to the symmetry energy, while the largest source of uncertainties lies in the density dependence of the EoS, which is not constrained by low energy nuclear physics data. To be more precise, we find that the constraint to reproduce low-energy nuclear physics properties leads to the prediction that canonical mass neutron stars should have a radius between 12 and 14~km, if they are made of nucleons and leptons only. However, increasing the accuracy of the reproduction of the low-energy data is less effective than the missing information about the density dependence of the EoS at two to four times nuclear saturation density. So the confrontation of the nuclear equation of state (EoS) with low-energy nuclear physics data, while necessary, is not sufficient for an accurate prediction of the dense matter equation of state. While such a result may have been anticipated, at least qualitatively, our analysis provides quantitative estimates of the link between the goodness of nuclear models assessed in finite nuclei and their predictions for NS global properties.

In our analysis, we do not explore the impact of phase transitions on NS global properties while they are even more uncertain than the density dependence of dense nucleonic matter. For massive neutron stars, the dominant source of uncertainties comes indeed from the lack of a precise prediction for the new phase(s). The question whether present astrophysical data already indicate the existence of a phase transition is not yet a settled one, see for instance~\cite{Annala2020,Li2021,Somasundaram2022b} for a sample of recent papers on this subject. In the present work, we focus on the nuclear physics uncertainties, although our conclusion concerning the weak impact of nuclear physics data becomes even stronger in the case of phase transition(s) in dense matter.

The present paper is organised as follow: We first list the experimental data in Section~II, namely the nuclear binding energies, the nuclear charge radii, the isoscalar giant monopole resonance energy in $^{208}$Pb, and the density dependence of the symmetry energy, and discuss the uncertainty that we consider in the comparison between the models and the data. 
The binding energies and charge radii are considered only for a set of spherical nuclei in order to avoid the complication of including a pairing interaction and possible deformation effects. We then explain in Section~III how the EDF are classified and we show that our best selection defines a clear correlation between $E_{\sym}$ and $L_{\sym}$. We then calculate in Section IV masses and radii of NS based on the different groups. A further analysis of the density dependence of the symmetry energy is performed in Section V, considering the constraint of the NS mass. We next determine NS global properties from our best set of models and analyse the correlation between the radius, the mass and the central pressure at beta-equilibrium in Section VI. Other global properties such as tidal deformability and moment of inertia are studied in Section VII. We present our conclusions in Section VIII.

\section{Low energy nuclear experimental data and modeling}
\label{sec:data}

In this section we start with a quick discussion of the models and then present and discuss the nuclear experimental data employed in the model selection. 

\subsection{Modeling nuclear low energy properties}

In our analysis, we consider a set of Energy Density Functionals (EDFs), which have been found to be an effective tool for analysing the fundamental properties of finite nuclei and for connecting these to  nuclear matter properties~\cite{Bender2003}. EDFs can be employed over the full nuclide chart, except for very light nuclei with mass number $A\lesssim 10$. They have a number of free parameters (typically from 5 to 10) which are adjusted on low energy nuclear properties and are usually calibrated to reproduce the ground state energy of spherical nuclei or of the entire nuclear chart and charge radii. Some of them however are only adjusted to nuclear empirical parameters without being employed to describe finite nuclei. In our analysis, we consider a full set of existing EDFs (415 in total), independently of the way they have been adjusted. We consider both non-relativistic and relativistic mean field models. The former are employed in the Hartree-Fock or Hartree-Fock-Bogoliubov framework together with a zero-range Skyrme-type interaction or a finite range Gogny or M3Y type force~\cite{Bender2003,Stone2007,skyrmeligo}. The latter are typically used in a Hartree-Bogoliubov approach with a Lagrangian based on meson-exchange potentials~\cite{Bender2003,Reinhard1989,Typel2018}. In all cases the effective nucleon-nucleon interaction is the key to good agreement of the calculations with experimental data.

EDFs also predict nucleon densities, deformations and skin thicknesses,  as well as the nuclear EoS, which is a fundamental ingredient to determine the properties of neutron stars, see Ref.~\cite{Bender2003,Stone2007} for a complete review.

\subsection{Energies of doubly magic nuclei}

Doubly magic nuclei are often used to calibrate EDF models since they are spherical (no deformation) and have closed shells (no pairing). The many-body complexity is therefore reduced, which accelerates the search for the best set of parameters reproducing the experimental data. 
Introducing pairing and deformation would lead to an increase of the number of parameters in the model and  increase the subsequent uncertainties as well. There are about 13 doubly magic nuclei, see tables~\ref{tab:data:magic} and \ref{tab:rchsample}, which span the nuclear mass table from light to heavy nuclei, as well as from isospin symmetric to asymmetric nuclei. They allow an easy and tractable search for possible sources of uncertainties in the confrontation of mean field interactions with experimental data.

\begin{table}[tb]
\centering
\setlength{\tabcolsep}{1pt}
\renewcommand{\arraystretch}{1.3}
\caption{Binding energies $B$ for the 13 doubly magic nuclei which are considered in the present work. Here (-) stands for experimental error-bars smaller than the accuracy given in the table and $^\#$ identifies interpolated numbers. We also compare our reference values~\cite{AMDC2016} to the ones from Ref.~\cite{unedf}.}
\begin{ruledtabular}
\begin{tabular}{rrrd{9}d{9}}
$Z$ & $N$ & nucleus & \multicolumn{1}{c}{$B$ (MeV)} & \multicolumn{1}{c}{$B$ (MeV)}  \\
& & & \multicolumn{1}{c}{Ref.~\cite{AMDC2016}} & \multicolumn{1}{c}{Ref.~\cite{unedf}} \\
\hline
8  &  8  & $^{16}$O   & -127.6193(-) & -127.6172(-)\\ 
14 & 20  & $^{34}$Si  & -283.4289(140) & -283.4208(141) \\
20 & 20  & $^{40}$Ca  & -342.0521(-) & -342.0336(2) \\
20 & 28  & $^{48}$Ca  & -416.0009(1) & -415.9720(41)  \\
20 & 32  & $^{52}$Ca  & -438.3279(7) & -436.5522(6986) \\
20 & 34  & $^{54}$Ca  & -445.3642(500) & -.-  \\
28 & 20  & $^{48}$Ni$^\#$  & -348.7275(5000) & -.-  \\
28 & 28  & $^{56}$Ni  & -483.9956(4) & -483.9505(110) \\
28 & 50  & $^{78}$Ni$^\#$  & -641.5470(6000) & -.-  \\
40 & 50  & $^{90}$Zr  & -783.8972(1) & -783.7953(23)  \\
50 & 50  & $^{100}$Sn & -825.2944(3000) & -824.6295(7054)  \\
50 & 82  & $^{132}$Sn & -1102.8430(20) & -1102.6860(136) \\ 
82 & 126 & $^{208}$Pb & -1636.4301(11) & -1635.8927(12)  \\
\end{tabular}
\end{ruledtabular}
\label{tab:data:magic}
\end{table}

Let us first analyse the present situation in terms of the low-energy nuclear data. The experimental data we use in the present study are given in Table~\ref{tab:data:magic}. We have considered 13 doubly magic nuclei, including two for which the binding energy is not measured but extrapolated from neighbouring nuclei ($^{48}$Ni and $^{78}$Ni). Both of the latter are the first unmeasured-mass nucleus of their respective double beta-decay mass parabolas, which each contain five nuclei with measured masses and thus permit a fairly precise extrapolation of the unmeasured masses. The 13 nuclei are grouped into isospin symmetric ones (group S containing 4 nuclei) and the isospin asymmetric ones (group A with 9 nuclei). We also compare the binding energies we consider with the ones used by the UNEDF collaboration~\cite{unedf}, originating from averages of the AME2003~\cite{AME2003} mass table values with recent JYFLTRAP measurements~\cite{JYFLTRAP}. The latter values deviate from those we consider by less than about 0.2~MeV, except for $^{52}$Ca, $^{100}$Sn and $^{208}$Pb, where the differences are respectively 1.8, 0.7 and 0.5~MeV. These deviations are smaller than the criteria we will introduce in the following to assess the quality of the mean field interactions. These differences in the experimental values thus have little impact on the definitions of the groups of interactions we define in the following.

In the case of the Bruxelles-Montreal Skyrme interactions, a phenomenological Wigner correction $E_W$ is applied to the binding energy, which is given in term of the following expression,
\begin{eqnarray}
E_W &=& V_W \exp \left\{ -\lambda\left(\frac{N-Z}{A}\right)^2\right\} \nonumber \\
&&+V_W^\prime \vert N-Z\vert \exp \left\{ -\left(\frac{A}{A_0}\right)^2\right\} \, .
\end{eqnarray}

A spin-orbit interaction is added to the non-relativistic Skyrme force, see Appendix~\ref{ap:sof}, while the relativistic approach generates it naturally from the scalar and time components of the self-energies~\cite{Bender2003}.

In the following, we consider that the model accuracy in the prediction of binding energies $B$ is
\begin{equation}
\delta_{B}=2.0~\hbox{MeV}\, .
\end{equation}
This uncertainty is much larger than the experimental one, see Ref.~\cite{Dobaczewski2015} and references therein for more detailed discussions, and essentially reflects the limitation of the EDF approach. Note however that this uncertainty represents a per mil (0.1 \%) accuracy for $^{208}$Pb.

\subsection{Charge radii of doubly magic nuclei}

\begin{table*}[tb]
\centering
\setlength{\tabcolsep}{1pt}
\renewcommand{\arraystretch}{1.3}
\caption{Comparison of the experimental charge radii measured by different groups to those of a sample of effective nuclear interactions. Here we consider the values given in Ref.~\cite{Angeli2013}.
$^\dagger$: Using the 1995 PDG data (see the text), as in Ref.~\cite{Chabanat1997}.}
\begin{ruledtabular}
\begin{tabular}{rrrccccccccc}
$Z$ & $N$ & nucleus & $R_{ch}$~(fm) & $R_{ch}$~(fm)  & $R_{ch}$~(fm) & SLy5 & BSk18 & UNEDF0 & DD-ME2 & NL3* & NLRA1  \\
& & & \rm{Ref.}~\cite{Angeli2013} & \rm{Ref.}~\cite{unedf} & \rm{Ref.}~\cite{Fricke1995} & \rm{Ref.}~\cite{Chabanat1998} & \rm{Ref.}~\cite{BSK18} & \rm{Ref.}~\cite{unedf0} & Ref.~\cite{ddme2} & \rm{Ref.}~\cite{nl3s} & \rm{Ref.}~\cite{NLRA1} \\
\hline
8  &  8  & $^{16}$O        & 2.6991(52)  & 2.7010 & -.-                & 2.7975 & 2.8141 & 2.8138 & 2.7283  & 2.7346  & 2.7167\\ 
 & & & & & & & 2.7825$^\dagger$ & 2.7992$^\dagger$ & \\
20 & 20  & $^{40}$Ca    & 3.4776(19)  & 3.4780 & 3.4767(8)    & 3.5059 & 3.5200 & 3.4980 & 3.4651  & 3.4701  & 3.4664\\
 & & & & & & & 3.4939$^\dagger$ & 3.5081$^\dagger$ & \\
20 & 28  & $^{48}$Ca    &  3.4771(20)  & 3.4790 & 3.4736(8)    & 3.5262 & 3.5353 & 3.5204 & 3.4811  & 3.4701  & 3.4700 \\
 & & & & & & & 3.5137$^\dagger$ & 3.5228$^\dagger$ & \\
40 & 50  & $^{90}$Zr     &  4.2694(10)  & 4.2690 & 4.2692(10)  & 4.2859 & 4.2919 & 4.2716 & 4.2733  & 4.2631  & 4.2717\\
 & & & & & & & 4.2759$^\dagger$ & 4.2857$^\dagger$ & \\
50 & 82  & $^{132}$Sn  &  4.7093(76)  & -.-           & -.-             & 4.7198& 4.7410 & 4.7221 & 4.7172  & 4.7031  & 4.7141 \\ 
 & & & & & & & 4.7102$^\dagger$ & 4.7315$^\dagger$ & \\
82 & 126 & $^{208}$Pb &  5.5012(13)  & 5.4850 & 5.5013(7)   & 5.5001 & 5.5184 & 5.5021 & 5.5180  & 5.5085  & 5.5233\\
 & & & & & & & 5.4920$^\dagger$ & 5.5103$^\dagger$ & \\
\end{tabular}
\end{ruledtabular}
\label{tab:rchsample}
\end{table*}

We compare charge radii $R_{ch}$ from various compilations in Table~\ref{tab:rchsample}. They are in good agreement, with differences less than 0.01~fm, except for $^{208}$Pb, where the value taken in Ref.~\cite{unedf} is 0.016~fm smaller than those given in Refs.~\cite{Angeli2013,Fricke1995}. We also show in Table~\ref{tab:rchsample} the charge radii predicted by a set of Skyrme and relativistic approaches and explore the impact of changing the proton radius of the SLy5 and BSk18 interactions from the present adopted one to the one suggested in the 1995 PDG data. Some interactions have indeed been adjusted with different values for the proton radius, since its value has changed over time. The SLy5 interaction, for instance, was obtained with the 1995 PDG data~\cite{Chabanat1997}.

The nuclear charge radius $R_{ch}$ is related to the rms proton radius $\langle R_p^2\rangle$ as \cite{Bertozzi1972,Friar1975,Bender2003},
\begin{equation}
\langle R_{ch}\rangle^2 = \langle R_p^2\rangle + \langle r_p^2\rangle + \frac N Z \langle r_n^2\rangle + \langle R_{ch}^{so}\rangle^2 + \langle R_{ch}^{DF}\rangle^2 \, ,
\label{eq:rch}
\end{equation}
where the second term $\langle r_p^2\rangle= \frac 3 2 \sigma^2$ originates from the convolution of the point particle proton density with a proton Gaussian form factor (with width $\sigma$). The proton radius is further discussed below. The third term in Eq.~\eqref{eq:rch} is a correction induced by the negative electromagnetic contribution of the neutron charge density. It is defined as $\langle r_n^2\rangle=\frac 3 2 \hbar^2 /(m_N c)^2 \mu_n $, with $\mu_n$ the neutron magnetic moment. The spin-orbit charge distribution furnishes a magnetic dipole moment correction to the nuclear rms charge radius, $\langle R_{ch}^{so}\rangle^2$, the fourth term in Eq.~\eqref{eq:rch}, which reads
\begin{equation}
\langle R_{ch}^{so}\rangle^2=\frac{1}{Ze}\frac{\hbar}{m_{p}c}\sum_{nlj\tau}v_{nlj\tau}^{2}\mu_{\tau}^\prime\left(2j+1\right)\left\langle \vec{\sigma}\cdot\vec{l}\right\rangle _{lj}\,,
\label{eq:rso}
\end{equation}
where the $v_{nlj\tau}^2$ are the orbital occupation probabilities. The modified magnetic dipole moments $\mu_{\tau}^\prime$ are defined as $\mu_{n}^\prime=\mu_{n}$ and $\mu_{p}^\prime=\mu_{p}-1/2$~\cite{Friar1975}, and the $\mu_{\tau}$ are the intrinsic nucleon magnetic dipole moments, $\mu_{n}=-1.91304\mu_{N}$ and $\mu_{p}=2.79285\mu_{N}$, with $\mu_{N}= e\hbar/(2m_{p}c)$. Note that we have truncated the accuracy with which $\mu_n$ and $\mu_p$ are known since it does not impact the present analysis. Finally the spin matrix elements in Eq.~\eqref{eq:rso} are given in Appendix~\ref{ap:so}.

The last term $\langle R_{ch}^{DF}\rangle^2$ in Eq.~\eqref{eq:rch} is the Darwin-Foldy term, which is a relativistic correction considered only in non-relativistic approaches. We take it to have the value $\langle R_{ch}^{DF}\rangle^2=3/(4m_N^2)=0.03311$~fm$^2$~\cite{Friar1997,Jentschura2011}. We note that its value can be almost three times larger when the relativistic effective mass $M^{*}\approx 0.6 M $ is used in the non-relativistic reduction~\cite{Nishizaki1988}. However, in either case, the correction provided by the Darwin-Foldy term is small. Although a center-of-mass correction should also be considered in the comparison to experimental data, it is neglected in most calculations since the correction is usually small.

We take the proton and neutron charge radii from the 2020-2021 compilation of the Particle Data Group~\cite{pdg} (PDG), which provides for the proton $\sqrt{\langle r_p^2 \rangle} = 0.8409 \pm 0.0004$~fm and for the neutron $\langle r_n^2 \rangle = -0.1161\pm 0.0022$~fm$^2$.
Note that the value for the proton charge radius is still under debate, see Ref.~\cite{Karr2020} for a presentation of the actual situation. The PDG proton charge radius originates from $\mu p$ experiments, which however differs from $e p$ ones, suggesting $\sqrt{\langle r_p^2 \rangle} = 0.8751 \pm 0.0061$~fm. Considering the uncertainties in these values, they are incompatible and represent the largest uncertainty in the intrinsic nucleon properties. Interestingly, a global analysis of the proton and neutron elastic form factors in the light cone frame formulation have extracted $\sqrt{\langle r_p^2 \rangle} = 0.852 \pm 0.002_\mathrm{(stat.)}\pm0.009_\mathrm{(syst.)}$~fm and $\langle r_n^2 \rangle = -0.122\pm 0.004_\mathrm{(stat.)}\pm0.010_\mathrm{(syst.)}$~fm$^2$~\cite{Atac2021} in a good agreement with the PDG compilation~\cite{pdg}.

We have estimated the impact of the uncertainty in the proton charge radius on the nuclear charge radius as follows: Considering a typical uncertainty on the proton charge radius $\delta \sqrt{\langle r_p^2 \rangle}\approx 0.04$~fm and neglecting the smaller uncertainty from the neutron charge radius, the effect on the calculation of the nuclear charge radii is of the order of $\delta R_{ch} \approx \delta \langle r_p^2 \rangle / (2 R_{ch})\lesssim 2~10^{-4}$~fm (for a typical $R_{ch}\approx 5$~fm). This uncertainty is therefore much smaller than the experimental uncertainty across different groups~\cite{Angeli2013,unedf,Fricke1995} , see table~\ref{tab:rchsample} for a set of nuclei from $^{16}$O to $^{208}$Pb, as well as the model uncertainties for this observable. We considered SLy5~\cite{Chabanat1998}, BSK18~\cite{BSK18}, UNEDF0~\cite{unedf0}, DD-ME2~\cite{ddme2}, NL3*~\cite{nl3s}, NLRA1~\cite{NLRA1} in table~\ref{tab:rchsample}. The values of the nuclear charge radii obtained by elastic electron scattering from stable and exotic nuclei have been more systematically investigated for non-relativistic Skyrme and relativistic mean field interactions in Ref.~\cite{RocaMaza2008}. Note however that there is actually no estimate of the EDF uncertainty on the nuclear charge radius, to our knowledge, and we suggest below an empirical relation for it. We conclude that the present uncertainty in the proton charge radius has no impact on the following discussion.

In the past, the fits of nuclear EDFs have considered older estimates for the proton and neutron charge radii, which have varied more substantially. For instance in 1997, the Saclay-Lyon Skyrme interactions~\cite{Chabanat1997}, e.g. SLy5, employed $\langle r_p^2 \rangle = 0.634$~fm$^2$ (with $\sigma=0.65$~fm) and $\langle r_n^2 \rangle = -0.12674558$~fm$^2$, originating from the 1995 PDG compilation. For SLy5 and BSK18, we compute the charge radii obtained by taking the values for the proton and neutron charge radii from the 1995 PDG compilation. Note also that in 2003 the values $\langle r_p^2 \rangle = 0.74$~fm$^2$ and $\langle r_n^2 \rangle = -0.117$~fm$^2$ were considered in Ref.~\cite{Bender2003}. Although larger, these variations of nucleon charge radii impact the nuclear charge radius by about $0.01$~fm (for a typical $R_{ch}\approx 5$~fm), which is still smaller than the uncertainty we associate in the following to the model predictions. The fluctuation of the proton charge radius reported in the past will thus not impact the present analysis.

Finally, the following empirical expression for the charge radius,
\begin{equation}
\langle R_{ch}^\mathrm{emp}\rangle^2 \approx \langle R_p^2\rangle + 0.64~\mathrm{fm}^2\, ,
\label{eq:rchemp}
\end{equation}
has sometimes been considered instead of Eq.~\eqref{eq:rch}, see for instance the discussion in Ref.~\cite{Bender2003}. The difference between Eqs.~\eqref{eq:rch} and \eqref{eq:rchemp} is of the order of 0.02~fm for the lightest nuclei, e.g. $^{16}$O, and decreases to about 0.0001~fm for $^{132}$Sn and $^{208}$Pb. This is the largest source of theoretical uncertainty in the estimate of the nuclear charge radius. 

In summary, by considering both experimental and theoretical uncertainties and by including the uncertainties in using the empirical formula~\eqref{eq:rchemp} instead of \eqref{eq:rch}, we come to the following estimate of the nuclear charge radius uncertainties which can be used in the confrontation of EDF modeling of nuclear data,
\begin{equation}
\delta_{R_{ch}}\approx 0.1 A^{-1/3}~\mathrm{fm}. 
\end{equation}
We will see in the following that such a loose uncertainty in the nuclear charge radius is still able to filter out many nuclear EDFs.

\subsection{Isoscalar giant monopole resonance (ISGMR) collective mode}

\begin{table*}[tb]
\centering
\setlength{\tabcolsep}{1pt}
\renewcommand{\arraystretch}{1.3}
\caption{Experimental value for the ISGMR centroid energy $E_\mathrm{GMR}$ in $^{208}$Pb compared to predictions from various nuclear EDFs. For consistency with the theoretical calculations, we report in this table the ISGMR experimental centroid energy defined as $\sqrt{m_1/m_{-1}}$ and provided in Ref.~\cite{Garg2018}. The incompressibility modulus $K_\mathrm{sat}$, the skewness parameter $Q_\mathrm{sat}$ and the parameters $p_\mathrm{c}$, $K_\mathrm{c}$, and $M_\mathrm{c}$ are also given for the EDFs.}
\begin{ruledtabular}
\begin{tabular}{rrrlccccccccc}
$Z$ & $N$ & nucleus & $E_\mathrm{GMR}^\mathrm{exp.}$ (MeV) & SLy5 & BSk18 & UNEDF0 & RATP & SGII & SIII & DD-ME2 & NL3* & NLRA1\\
 & & &  $\sqrt{m_1/m_{-1}}$ & Ref.~\cite{Chabanat1998} & Ref.~\cite{BSK18} & Ref.~\cite{unedf0} & Ref.~\cite{ratp} & Ref.~\cite{sgii} & Ref.~\cite{siii} & Ref.~\cite{ddme2} & Ref.~\cite{nl3s} & Ref.~\cite{NLRA1}\\
\hline
82 & 126  & $^{208}$Pb   &  13.50(10) \cite{Garg2018} & 13.77(1) &  14.02(0) & 13.65(1) &  14.12(1)&  13.44(1) & 16.79(1) & 14.08(1) & 14.77(1) & 15.50(1)\\
\hline
 & & & $K_\mathrm{sat}$ (MeV) & 230 & 242 & 230 & 240 & 215 & 355 & 251 & 258 & 285 \\
 & & & $Q_\mathrm{sat}$ (MeV) & -364 & -364 & -404 & -350 & -381 & 101 & 479 & 122 & 279 \\
 & & & $p_\mathrm{c}$ (MeV.fm$^{-3}$) & -0.653 & -0.675 & -0.659 & -0.673 & -0.608 & -0.822 & -0.589 & -0.650 & -0.678 \\
 & & & $K_\mathrm{c}$ (MeV) & 35.3 & 36.0 & 36.7 & 35.4 & 34.8 & 27.4 & 23.4 & 35.7 & 31.9 \\
 & & & $M_\mathrm{c}$ (MeV) & 1141 & 1202 & 1147 & 1188 & 1066 & 1717 & 992 & 1160 & 1271  \\
\end{tabular}
\end{ruledtabular}
\label{tab:data:gmr}
\end{table*}

The isoscalar giant monopole resonance energy is also used in the estimation of the adequacy of a nuclear EDF for NS properties, since it is correlated with the incompressibility modulus~\cite{Blaizot1980,Blaizot1995}. The latter determines the variation of the energy density as the nucleon density departs from the saturation density in symmetric nuclear matter (SM). It thus provides important information about the density dependence of the EoS, fundamental for the determination of NS properties. For recent reviews of the incompressibility in finite nuclei and nuclear matter, see for instance Refs.~\cite{Stone2014,Garg2018}.

The energy of the ISGMR can be calculated using the sum rule approach, which provides a fast and consistent way to get the centroid of the ISGMR energy in deeply bound nuclei. It is defined as~\cite{Bohigas1979}
\begin{equation}
E_\gmr = \sqrt{\frac{m_1}{m_{-1}}} \, ,
\end{equation}
where the $k$th energy-weighted sum rule is
\begin{equation}
m_k = \sum_l (E_l)^k \vert\langle l\vert\hat{Q}\vert 0\rangle\vert^2 \, ,
\end{equation}
with $E_l$ the collective excitation energy and $\hat{Q}=\sum_{i=1}^A r_i^2$ the isoscalar monopole transition operator. The moment $m_1$ is evaluated in terms of a double commutator using the Thouless theorem~\cite{Thouless1961},
\begin{equation}
m_1 = 2 A \frac{\hbar^2}{m_N} \langle r^2 \rangle \, ,
\end{equation}
where $A$ is the nucleon number, $m_N$ the nucleon mass, and $\langle r^2 \rangle$ the rms radius. In the constrained Hartree-Fock (CHF) approach~\cite{Bohigas1979,Colo2004} the moment $m_{-1}$ is obtained from the derivative of the expectation value of the monopole operator,
\begin{equation}
m_{-1} = -\frac 1 2 \left[\frac{\partial}{\partial \lambda} \langle\lambda\vert\hat{Q}\vert\lambda\rangle\right]_{\lambda=0}\, ,
\end{equation}
where $\vert\lambda\rangle$ is the ground-state energy of the constrained Hamiltonian,
\begin{equation}
\hat{H}_\mathrm{constr.} = \hat{H} + \lambda \hat{Q} \, .
\end{equation}

In Table~\ref{tab:data:gmr}, the experimental value and theoretical predictions for the ISGMR centroid are given for $^{208}$Pb. It has been estimated that an uncertainty of about 0.2-0.4~MeV in the centroid can be translated into an uncertainty of about 15~MeV in the incompressibility modulus~\cite{Avogadro2013}. Precision of the experimental results and of the theoretical calculations for the centroid energy are thus essential. 
Considering that the present uncertainty in $K_\sat$ is of the order of 20~MeV~\cite{Garg2018}, we have fixed the uncertainty in the model prediction for the ISGMR centroid energy to be
\begin{equation}
\delta_{\gmr}=0.7~\hbox{MeV}\, .
\label{eq:deltagmr}
\end{equation}

We also report in Table~\ref{tab:data:gmr} a set of parameters defined in uniform matter. 
The incompressibility modulus $K_\sat$ and the skewness parameter $Q_\sat$ are nuclear empirical parameters (NEP) encoding the density dependence of the energy per particle in SM as,
\begin{equation}
e_\mathrm{SM}(n) = E_\sat + \frac 1 2 K_\sat x^2 + \frac 1 6 Q_\sat x^3 + \dots
\end{equation}
with $x=(n-n_\sat)/3n_\sat$. We can check that the models predicting $K_\sat=230\pm 20$~MeV~\cite{Garg2018} also predict in $^{208}$Pb $E_\gmr=13.50\pm 0.7$~MeV, confirming \textsl{a posteriori} the relation~\eqref{eq:deltagmr}. Note also the large differences predicted by these EDFs for the parameter $Q_\sat$ for the models with good incompressibilities: between -400 and -350~MeV for the non-relativistic EDFs and an opposite sign for the relativistic ones. It has been suggested that these systematic differences are at the origin of the model dependence in the $E_\gmr$-$K_\sat$ correlation~\cite{Khan2012, Margueron2018a}.

The other parameters $p_c$, $K_c$ and $M_\mathrm{c}$ reported in Table~\ref{tab:data:gmr} are the pressure, incompressiblity and $M_c$-parameter, 
\begin{eqnarray}
p(n) &=& n^2 \frac{\partial e}{\partial n} \, , \\
K(n) &=& \frac{18}{n} p(n) + 9 n^2 \frac{\partial^2 e}{\partial n^2} \, , \\
M(n) &=& 3 n \frac{\partial K(n)}{\partial n} \, , 
\label{eq:mc}
\end{eqnarray}
defined at the crossing density $n_\mathrm{c}=0.71(1)n_\sat$~\cite{Khan2012}.

Note that the values of the pressure and incompressibilities, $p_\mathrm{c}$ and $K_\mathrm{c}$ are quite constant for the models with good incompressibilities. The value of $M_\mathrm{c}$ is in agreement with the one suggested in Ref.~\cite{Khan2012}, namely $M_\mathrm{c}=1100\pm70$~MeV.

\subsection{The symmetry energy}
\label{sec:esym}

The symmetry energy is a crucial quantity guiding the exploration of asymmetric nuclear matter, such as beta-equilibrium matter in NS's. We shall, however, distinguish between the global symmetry energy $e_\mathrm{sym}$ defined as
\begin{equation}
e_\mathrm{sym}(n) = e_\mathrm{NM}(n) - e_\mathrm{SM}(n) \, ,
\end{equation}
and its quadratic contribution $e_\mathrm{sym,2}$,
\begin{equation}
e_\mathrm{sym,2}(n) = \frac 1 2 \frac{\partial^2 e(n,\delta)}{\partial \delta^2}\bigg\vert_{\delta=0} \, ,
\end{equation}
where $e(n,\delta)$ is the energy per particle in asymmetric matter, $\delta=(n_n-n_p)/n$ the isospin asymmetry, and $e_\mathrm{NM}$ the energy per particle in neutron matter (NM). The quadratic contribution to the symmetry energy $e_\mathrm{sym,2}$ is the quantity which is probed by nuclear physics experiments, since the isospin parameter remains small ($\delta\lesssim 0.25$), while the properties of neutron matter, with large asymmetries, are better described by $e_\mathrm{sym}$. The difference between the symmetry energy and its quadratic contribution, $e_\sym-e_{\sym,2}$, represents the non-quadraticities, which are often found to be small (2-3\% of the symmetry energy), see Ref.~\cite{Somasundaram2021} and references therein for a recent study. In the literature, these two quantities are usually not distinguished, although formally, they are different. The two representations of the symmetry energy $e_\sym$ and $e_{\sym,2}$ can be expanded in terms of the density parameter  $x=(n-n_\sat)/(3n_\sat)$ as,
\begin{align}
e_\mathrm{sym}(n) &= E_{\sym} + L_{\sym} x + \frac{1}{2}K_{\sym} x^2
+ \frac{1}{6}Q_{\sym} x^3 + \dots, \label{eq:esym}\\
e_\mathrm{sym,2}(n) &= E_{\sym,2} + L_{\sym,2}\,x + \frac{1}{2}K_{\sym,2}\, x^2 \nonumber\\
&+ \frac{1}{6}Q_{\sym,2}\, x^3 + \dots,
\label{eq:esym2}
\end{align}
where $E_{\sym}$, $L_{\sym}$, $K_{\sym}$, and $Q_{\sym}$ are nuclear empirical parameters (NEP) and $E_{\sym,2}$, $L_{\sym,2}$, $K_{\sym,2}$, and $Q_{\sym,2}$ are quadratic nuclear empirical parameters (QNEP).

\begin{figure}[tb] 
\centering
\includegraphics[width=0.50\textwidth]{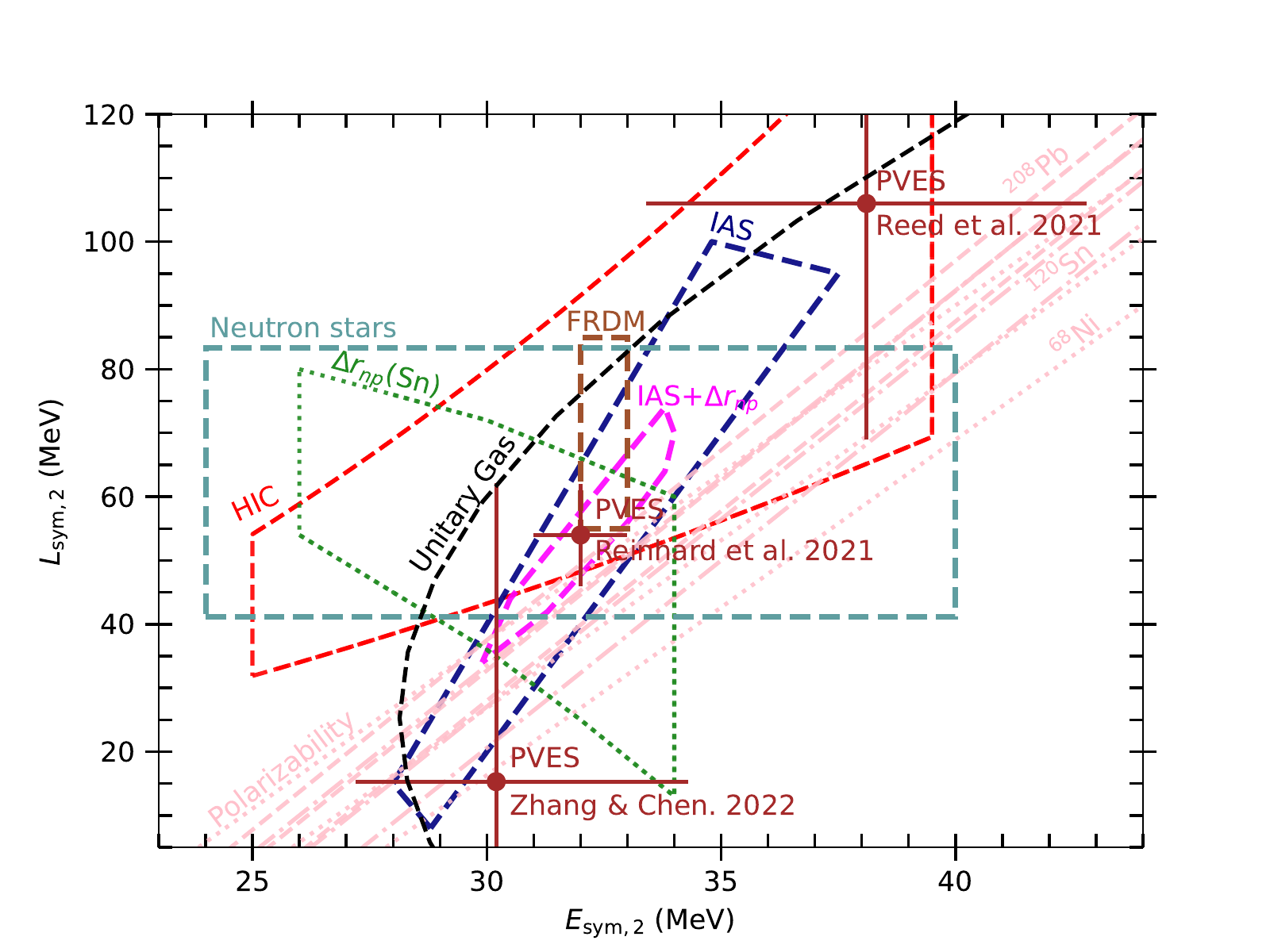}
\caption{Correlation between the symmetry energy $E_{\sym,2}$ and its slope $L_{\sym,2}$ at saturation density. See text for more details on the various constraints.} 
\label{fig:ELsymExp}
\end{figure}

There are several experimental constraints for the symmetry energy in finite nuclei, see Refs.~\cite{Lattimer2013,Wei2020} for a detailed presentation of these. Adopting the $E_{\sym,2}$-$L_{\sym,2}$ representation, we show a few of them in Fig.~\ref{fig:ELsymExp}, including the recent ones from the analyses of the PREX-II and CREX parity-violating electron scattering (PVES) experiments. 

Before discussing these recent results, let us first present the others: ``HIC'': constraints inferred from isospin diffusion in heavy ion collisions (HICs)~\cite{Tsang2009}; ``Polarizability'': constraints on the electric dipole polarizability of $^{208}$Pb, $^{120}$Sn and $^{68}$Ni~\cite{RocaMaza2015}; ``$\Delta r_{np}$(Sn)'': constraints deduced from the analysis of neutron skin thickness in Sn isotopes~\cite{Chen2010}; ``FRDM'': constraint from the finite-range droplet mass model calculations~\cite{FRDM}; ``IAS'': constraint deduced from the analysis of the excitation energy of the isobaric analog state (IAS) based on Skyrme-Hartree-Fock calculations~\cite{Danielewicz2013}; ``IAS+$\Delta r_{np}$'': combination of the IAS constraint and neutron skin in $^{208}$Pb~\cite{Danielewicz2013}.
Two additional constraints are also represented, while they formally refer to the global symmetry energy NEP ($E_{\sym,2}$ and $L_{\sym,2}$: ``Neutron Stars'': horizontal constraint obtained from a Bayesian analysis of mass and radius observations of NSs by considering the 95\% confidence values for $L_{\sym}$~\cite{Steiner2013}; ``Unitary Gas'': the analysis of the unitary gas predictions for the symmetry energy parameters~\cite{Tews2017} permits the values to the right of the curve.

Also shown in Fig.~\ref{fig:ELsymExp} are analyses of the PREX-II~\cite{Adhikari2021} and CREX~\cite{Adhikari2022} PVES experiments: There are indeed big differences between the analysis by Reed et al~\cite{Reed2021} ($E_{\sym,2}=38.1\pm 4.7$~MeV, $L_{\sym,2}=106\pm 37$~MeV) and the one by Reinhard et al.~\cite{Reinhard2021} ($E_{\sym,2}=32\pm 1$~MeV, $L_{\sym,2}=54\pm 8$~MeV), which also includes the electric dipole polarizability. Another analysis by Zhang and Chen~\cite{Zhang2022} combining PREX-II and CREX using  a Bayesian inference find a very low centroid for $L_{\sym,2}$ ($E_{\sym,2}=30.2^{+3.0}_{-4.1}$~MeV, $L_{\sym,2}=15.3^{+41.5}_{-46.8}$~MeV). It has indeed been pointed our that the results of PREX-II and CREX are in disagreement~\cite{Reinhard2022,Yuksel2022}. There are large differences among the various PVES analyses. One of the tightest constraints in the $E_{\sym,2}$-$L_{\sym,2}$ diagram shown in Fig.~\ref{fig:ELsymExp} is the one referred to as ``IAS+$\Delta r_{np}$''~\cite{Danielewicz2013}. We will investigate the role of this constraint in the following analysis.

\section{Combined analysis of the modeling reproducing low energy nuclear data}

Since we have different types of low-energy nuclear physics data, we face the difficulty of assembling them together in a meaningful way. We suggest two ways of performing the assessment, each of them providing interesting results about the interactions.

\subsection{The groups $G_i$ and $D_i$}

The first method is a global assessment, in which all nuclei contribute equally to the variance of each type of observable, where the variances $\sigma_i$ for the binding energies ($i=B$), the charge radii ($i=R_{ch}$) and the ISGMR energy ($i$=ISGMR) are defined as,
\begin{eqnarray}
\sigma_{B}^2 &=& \frac{1}{N_{B}} \sum_{i} \left[\frac{B_i(\mathrm{exp})-B_i(\mathrm{model})}{\delta_{B}} \right]^2 \, , \\
\sigma_{R_{ch}}^2 &=& \frac{1}{N_{R_{ch}}} \sum_{i} \left[\frac{R_{ch,i}(\mathrm{exp})-R_{ch,i}(\mathrm{model})}{\delta_{R_{ch}}(A_i)} \right]^2 \, , \\
\sigma_{\gmr}^2 &=& \frac{1}{N_{\gmr}} \times
\nonumber\\
&& \sum_{i} \left[\frac{E_{\gmr,i}(\mathrm{exp})-E_{\gmr,i}(\mathrm{model})}{\delta_{\gmr}}\right]^2,\qquad
\end{eqnarray}
with $N_{B}=13$ (see Table~\ref{tab:data:magic}), $N_{R_{ch}}=6$ (see Table~\ref{tab:rchsample}), and $N_{\gmr}=1$ (see Table~\ref{tab:data:gmr}). The uncertainties $\delta_B$, $\delta_{R_{ch}}(A_i)$ and $\delta_{\gmr}$ have been introduced in Section~\ref{sec:data}. The groups built on this global assessment will be called G$_i$.

In the second method, the variances of the binding energy and the charge radius of the symmetric $N=Z$ and asymmetric $N\ne Z$ nuclei are accumulated separately.
We evaluate the following rms deviations for the symmetric nuclei,
\begin{eqnarray}
\sigma_{B,S}^2 &=& \frac{1}{N_{B,S}} \sum_{i\in S} \left[\frac{B_i(\mathrm{exp})-B_i(\mathrm{model})}{\delta_{B}} \right]^2 \, , \\
\sigma_{R_{ch},S}^2 &=& \frac{1}{N_{R_{ch},S}} \sum_{i\in S} \left[\frac{R_{ch,i}(\mathrm{exp})-R_{ch,i}(\mathrm{model})}{\delta_{R_{ch}}(A_i)} \right]^2 \, , 
\end{eqnarray}
for asymmetric nuclei,
\begin{eqnarray}
\sigma_{B,A}^2 &=& \frac{1}{N_{B,A}} \sum_{i\in A} \left[\frac{B_i(\mathrm{exp})-B_i(\mathrm{model})}{\delta_{B}} \right]^2 \, , \\
\sigma_{R_{ch},A}^2 &=& \frac{1}{N_{R_{ch},A}} \sum_{i\in A} \left[\frac{R_{ch,i}(\mathrm{exp})-R_{ch,i}(\mathrm{model})}{\delta_{R_{ch}}(A_i)}\right]^2 \, ,
\end{eqnarray}
and finally for the ISGMR energy, which remains the same as in the previous case. We include calculations for the following nuclei in the groups described above,
\begin{itemize}
\item $(B,S)$: $^{16}$O, $^{40}$Ca, $^{56}$Ni, $^{100}$Sn.
\item $(B,A)$: $^{34}$Si, $^{48}$Ca, $^{52}$Ca, $^{54}$Ca, $^{48}$Ni, $^{78}$Ni, $^{90}$Zr, $^{132}$Sn, $^{208}$Pb.
\item $(R_{ch},S)$: $^{16}$O, $^{40}$Ca.
\item $(R_{ch},A)$: $^{48}$Ca, $^{90}$Zr, $^{132}$Sn, $^{208}$Pb.
\item $(\gmr)$: $^{208}$Pb. 
\end{itemize}
We thus have $N_{B,S}=4$ and $N_{B,A}=9$, $N_{R_{ch},S}=2$ and $N_{R_{ch},A}=4$, and $N_{\gmr}=1$. We note that the rms deviations of the global approach are simply the renormalized sums of the deviations of this second approach. In the following, the groups built upon this more detailed approach are called D$_i$.

\begin{figure}[tb] 
\centering
\includegraphics[scale=0.47,trim=10 10 0 0, clip=true]{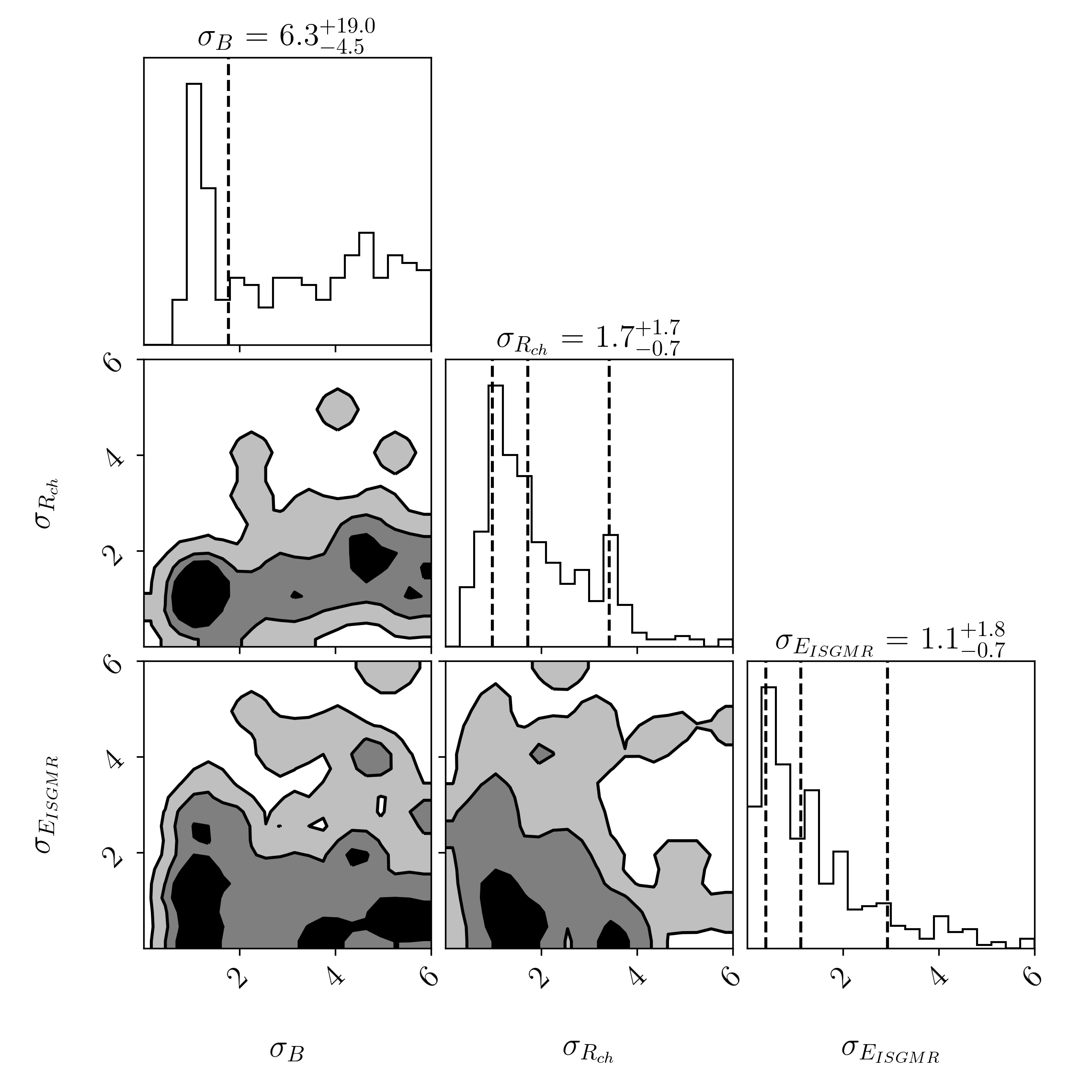}
\caption{Representation of the rms deviations for the observables $B$, $R_{ch}$ and $E_{\gmr}$.}
\label{fig:corner}
\end{figure}

We show in Fig.~\ref{fig:corner} the distribution of the rms deviations $\sigma_i$ associated with the observables $i=E$, $i=R_{ch}$ and $i=E_{\gmr}$, for all the interactions considered (415 in total). Note that for these three observables, the main peak is systematically located at about $\sigma_i<2$, which supports our choices for the associated uncertainties presented in Section~\ref{sec:data}.

In the following, we sort the modeling according to the rms deviation and attribute to them a set of letters, namely the three letters $L_{B}L_{R_{ch}}L_{E_\gmr}$ for the groups G$_i$ and the five letters $L_{B,S}L_{B,A}:L_{R_{ch},S}L_{R_{ch},A}:L_{E_\gmr}$ for the groups D$_i$, where the letters $L$ are:
\begin{itemize}
\item $L=A$, if $\sigma<1$; 
\item $L=B$, if $1<\sigma<2$; 
\item $L=C$, if $2<\sigma<3$; 
\item $L=D$, if $\sigma>3$.
\end{itemize}

The complete list of the scores for each parametrization analyzed in this work is given in the supplemental material. As examples, we obtain the following scores for the two approaches (global versus detailed) in the cases of the relativistic NLSV1, and the non-relativistic RATP and SLy4 Skyrme forces: 
\begin{itemize}
\item NLSV1: ABC, BA:AB:C,
\item RATP: BBA, BC:BB:A,
\item SLy4: BBA, BB:BB:A.
\end{itemize}
The relativistic NLSV1 interaction reproduces the binding energies better than the charge radii, which are better reproduced than the ISGMR energy. In detail, the binding energies (charge radius) of the $N\ne Z$ nuclei are better (worse) reproduced than the $N=Z$ ones. For the non-relativistic models, we observed that they are scored identically (BBA) in the general analysis, but a more detailed analysis shows that the SLy4 is better than the RATP at reproducing the binding energies in $N\ne Z$ nuclei. This illustrates the differences in the global and detailed approach, which will be further analysed in the following.

\begin{table}[tb]
\tabcolsep=0.05cm
\def\arraystretch{1.5}
\centering
\caption{Number of EDFs passing the filters imposed by the groups G$_i$ and D$_i$, and D$_{4\sym}$. The number of EDFs in each groups for which M$_{\tov}\geq 1.6$M$_\odot$ and M$_{\tov}\geq 2.0$M$_\odot$ are also counted. See the text for more details.}
\begin{tabular}{lcccccccccc}
\hline\noalign{\smallskip}
\rm          & D$_0$/G$_0$ & D$_1$ & G$_1$ & D$_2$ & G$_2$ & D$_3$ & G$_3$ & D$_4$ & G$_4$ & D$_{4\sym}$ \\
\noalign{\smallskip}\hline\noalign{\smallskip}
Total        & 374 & 81 & 90 & 66 & 74  & 61 & 74 & 45 & 54 &  22 \\
M$_{\tov}\geq 1.6$M$_\odot$ & 312 & 77 & 85 & 65 & 72  & 61 & 72  & 45 & 52 & 22 \\
M$_{\tov}\geq 2.0$M$_\odot$ & 198 & 49 & 53  & 44 & 49 & 41 & 49 & 25 & 29 & 12 \\
\noalign{\smallskip}\hline
\label{tabgroups}
\end{tabular}
\end{table}

Based on the criteria described above, we separate the interactions submitted to the finite nucleus constraints into five different groups, as follows, 
\begin{itemize}
\item D$_0$ and G$_0$: groups containing all the interactions considered,
\item D$_1$ and G$_1$: groups containing interactions with a letter rank from $A$ to $C$ over all types of data,
\item D$_2$ and G$_2$: groups containing interactions with a letter rank of $A$ or $B$ for the binding energies, 
\item D$_3$ and G$_3$: groups containing interactions with a letter rank of $A$ or $B$ for the binding energies and charge radii,
\item D$_4$ and G$_4$: groups containing interactions with a letter rank of $A$ or $B$ for the binding energies, charge radii and GMR energies.
\item D$_{4\sym}$: This group imposes on top of D4 the constraint ``IAS+$\Delta r_{np}$'', as detailed in the following.
\end{itemize}

\begin{figure}[tb] 
\centering
\includegraphics[width=0.50\textwidth]{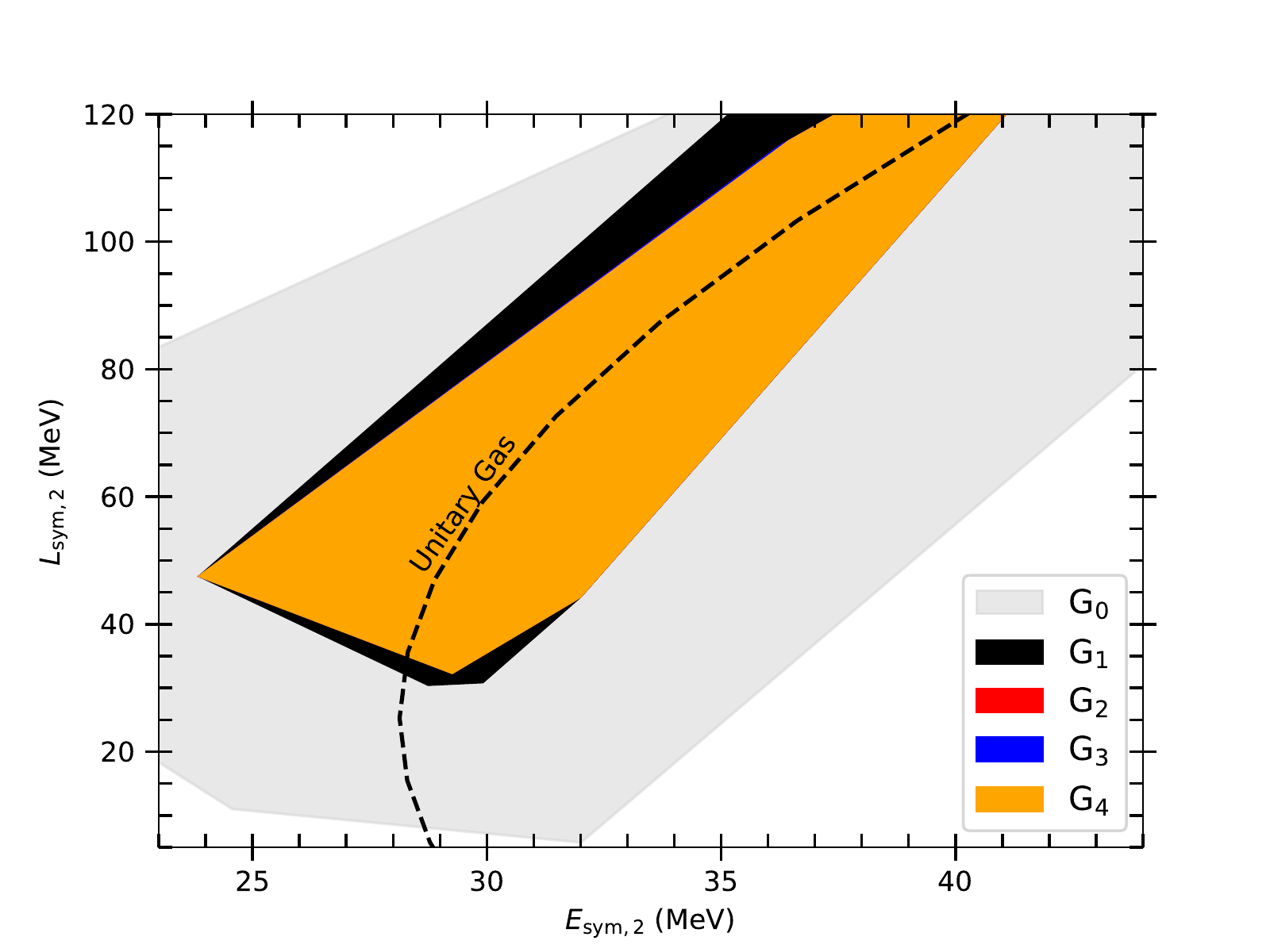}
\includegraphics[width=0.50\textwidth]{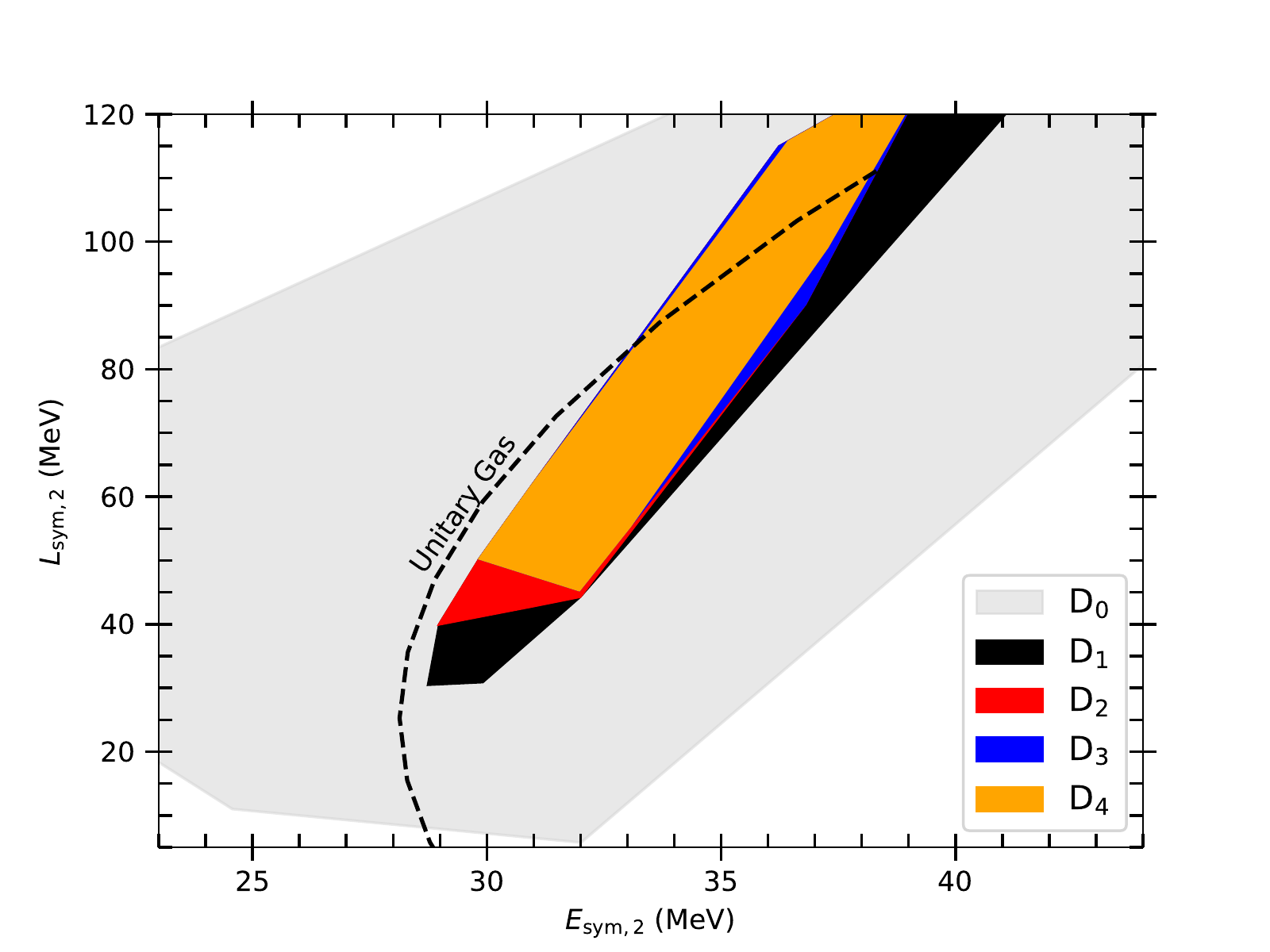}
\caption{Correlation between the symmetry energy and its slope for the groups G$_i$ (top panel) and D$_i$ (bottom panel). The unitary gas boundary is shown for reference.} 
\label{fig:ELsym_orders}
\end{figure}

The number of interactions surviving the conditions imposed on the different groups D$_i$ and G$_i$ are shown in Table~\ref{tabgroups} by the line denoted by ``total''. We also count the number of interactions that permit a neutron star of mass M$_{\tov}\geq 1.6$M$_\odot$ and M$_{\tov}\geq 2.0$M$_\odot$ in the case of TOV hydrostatic equilibrium, as is detailed in the following Section~\ref{sec:MR}. We remark that the D$_0$ (or G$_0$) group is composed of $374$ parametrizations rather than $415$, the total number of interactions. This is due to the fact that a number of problematic interactions have been discarded, due to one of the following conditions: (i) spinodal instability (negative values of the sound speed) above $n_\sat$ or (ii) negative value of the pressure in stellar matter. 

\begin{figure}[tb] 
\centering
\includegraphics[scale=0.35]{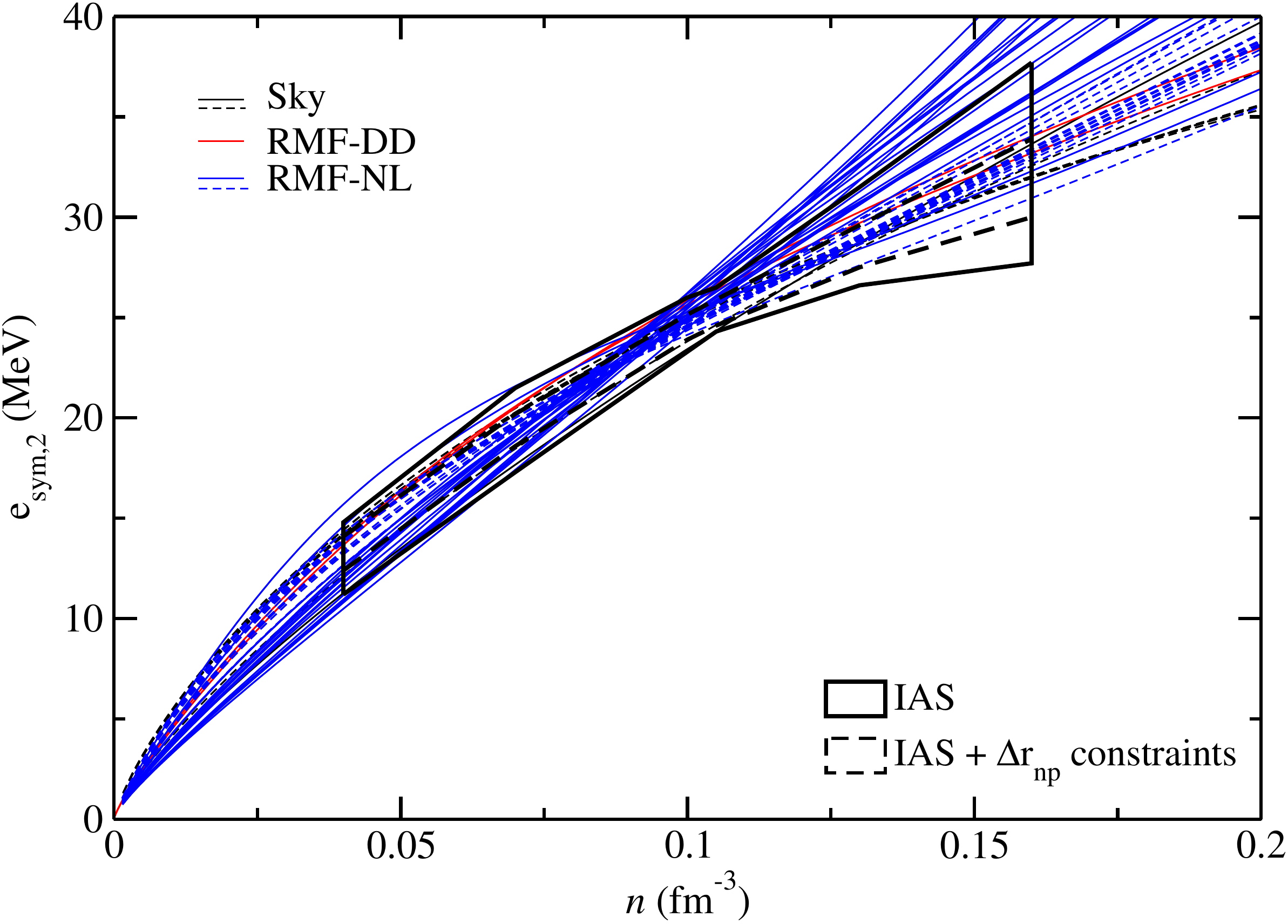}
\caption{Symmetry energy as a function of the density $n$ for the interactions of the D$_4$ group confronted with the IAS contour and the IAS+$\Delta r_{np}$ one. Dashed curves: interactions satisfying the IAS + $\Delta r_{np}$ constraint. Full curves: interactions not compatible with this constraint. See text for more details.} 
\label{fig:esym}
\end{figure}

\subsection{Impact of the groups $G_i$ and $D_i$ on the $E_{\sym,2}$-$L_{\sym,2}$ correlation}
\label{sec:ELcor}

We compare in Fig.~\ref{fig:ELsym_orders} the impact of the different groups G$_i$ and D$_i$ on the $E_{\sym,2}$-$L_{\sym,2}$ correlation. It is clear that the groups D$_i$ are better correlated than the groups G$_i$, reflecting the constraint that $N=Z$ and $N\ne Z$ nuclei are reproduced with same accuracy. Already the group D$_2$ is better correlated than the group G$_2$, showing that the goodness of the models to reproduce data (the difference between G$_1$ and G$_2$ or D$_1$ and D$_2$) is less effective than the condition imposed on the D$_2$ models (difference between D$_2$ and G$_2$). The additional condition also appears through the charge radii, i.e. D$_3$ removes the lower values of $L_{\sym,2}$, while the additional constraint on the ISGMR in $^{208}$Pb plays a small but non-negligible role (difference between D$_3$ and D$_4$).

By comparing the contours G$_4$ with D$_4$, we see clearly the impact of imposing that $N=Z$ and $N\ne Z$ nuclei are reproduced with the same accuracy. In G$_4$ for instance, a poor reproduction of $N\ne Z$ nuclei could be compensated by a better description of $N=Z$ ones, e.g., as occurs for the SKa interaction. This is not true in the group D$_4$, which creates a tighter correlation in the $E_{\sym,2}$-$L_{\sym,2}$ diagram, see bottom panel in Fig.~\ref{fig:ELsym_orders}. 

In conclusion, Fig.~\ref{fig:ELsym_orders} shows the effectiveness of the condition that $N=Z$ and $N\ne Z$ nuclei are reproduced with same accuracy on the $E_{\sym,2}$-$L_{\sym,2}$ correlation. The constraints on the charge radii and ISGMR centroid energy also play an additional role for the D$_i$ groups, but have almost no impact on the G$_i$ groups.

\subsection{The group D$_{4\sym}$ with an additional symmetry energy constraint}\label{d4sym-esym}

\begin{figure}[tb] 
\centering
\includegraphics[width=0.50\textwidth]{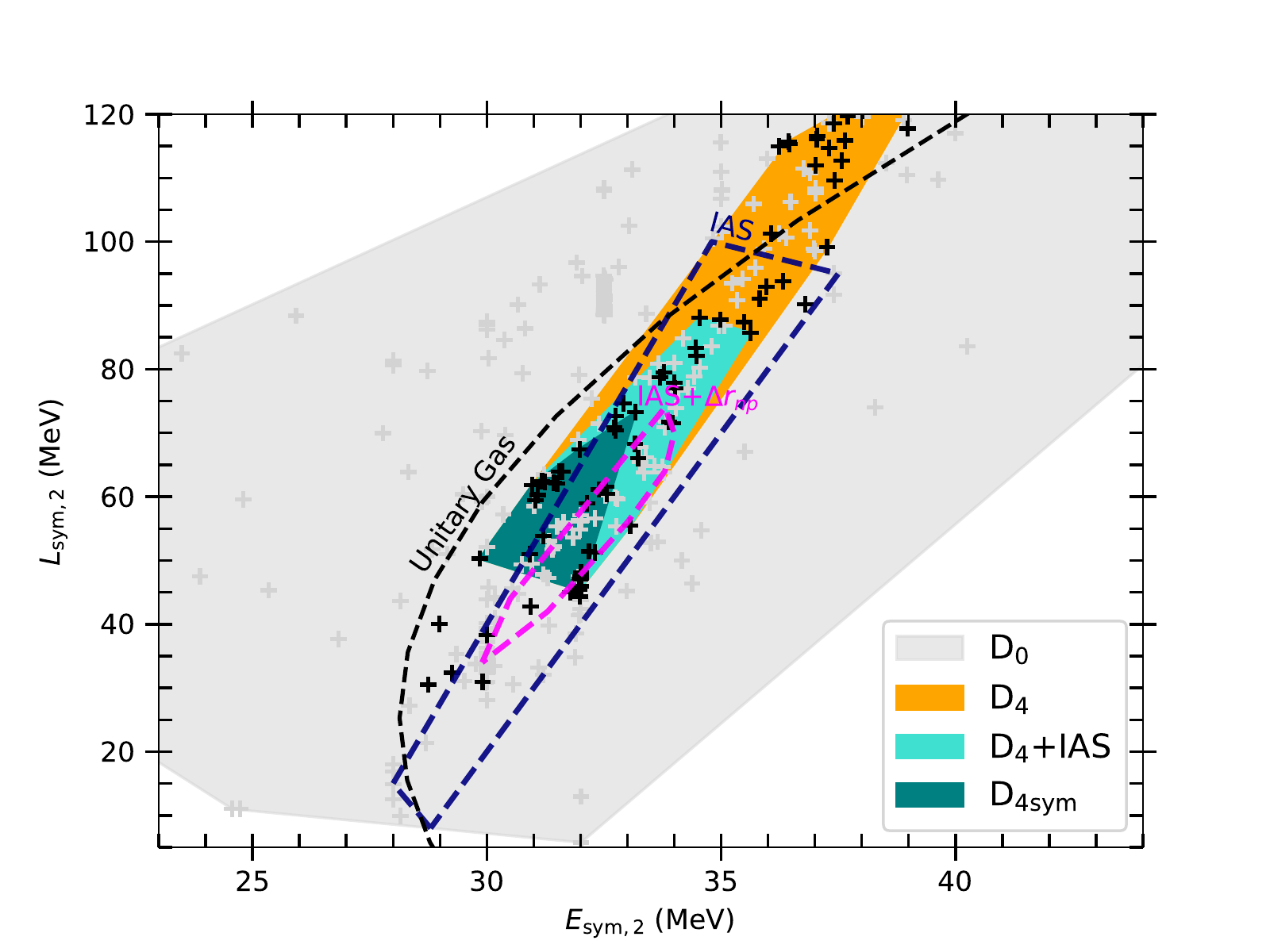}
\caption{Correlation between the symmetry energy and its slope for the groups D$_i$. The contour of the group D$_0$=G$_0$ is shown in light grey, while the contour in dark grey includes D$_4$ group. The blue (green) contour includes the D$_4$+IAS (D$_{4\sym}$) group, defined by the constraints shown in Fig.~\ref{fig:esym}. A few other experimental constraints from Fig.~\ref{fig:ELsymExp} are shown for reference.} 
\label{fig:ELsym}
\end{figure}

We now detail how the group D$_{4\sym}$, see Table~\ref{tabgroups}, is obtained by adding symmetry energy constraints on top of the D$_4$ group, which has been shown in Section~\ref{sec:ELcor} to naturally constrain $E_{\sym,2}$ and $L_{\sym,2}$. Among the set of experimental constraints for the symmetry energy shown in Section~\ref{sec:esym}, we decided to investigate the impact of IAS+$\Delta r_{np}$~\cite{Danielewicz2013} since it is obtained from low-energy nuclear data. Another reason for investigating these constraints is that they are provided as contours in the density dependence of the symmetry energy, see Fig.~\ref{fig:esym}. We can then filter our interactions in the D$_4$ group according to their ability to fit inside these contours, as illustrated in  Fig.~\ref{fig:esym}.

Two contours are shown in Fig.~\ref{fig:esym}: the one representing the IAS constraint alone as well as  one representing the IAS+$\Delta r_{np}$ constraints together. The different colors reflect the different kinds of interactions analyzed, namely, black for Skyrme, blue for RMF-NL, and red for RMF-DD. 
For each model, we compute a loss function defined as $\chi^2=1/N\sum_i (e_{\sym,2}^{av}(i)-e_{\sym,2}^{model})^2/(\Delta e_{\sym,2}(i))^2$, where the index $i=1$ to $N$ scans over the data. Models with $\chi^2<1$ are accepted. The interactions compatible with the IAS + $\Delta r_{np}$ constraint are represented by dashed curves and define the D$_{4\sym}$ group.

We now show in Fig.~\ref{fig:ELsym} the region of the $E_{\sym,2}$-$L_{\sym,2}$ which is populated by the D$_{4\sym}$ group. For a better understanding, we also show the role of the groups D$_0$, D$_4$, D$_4$+IAS and finally D$_{4\sym}$. The contour D$_4$ isolates a subgroup allowed by HIC and is partially excluded by the unitary gas constraint. We show how the constraint from IAS (D$_4$+IAS) and IAS + $\Delta r_{np}$ (D$_{4\sym}$)~\cite{Danielewicz2013} reduces the viable models shown in the $E_{\sym,2}$-$L_{\sym,2}$ diagram. Including the IAS constraint (D$_4$+IAS) reduces the contour to a smaller group which makes it compatible with neutron stars, and finally the contour D$_{4\sym}$ reduces it even further inside the $\Delta r_{np}$(Sn) experimental constraint. All the new contours (D$_4$, D$_4$+IAS and D$_{4\sym}$) are given in the supplemental material.

\begin{figure}[tb] 
\centering
\includegraphics[width=0.50\textwidth]{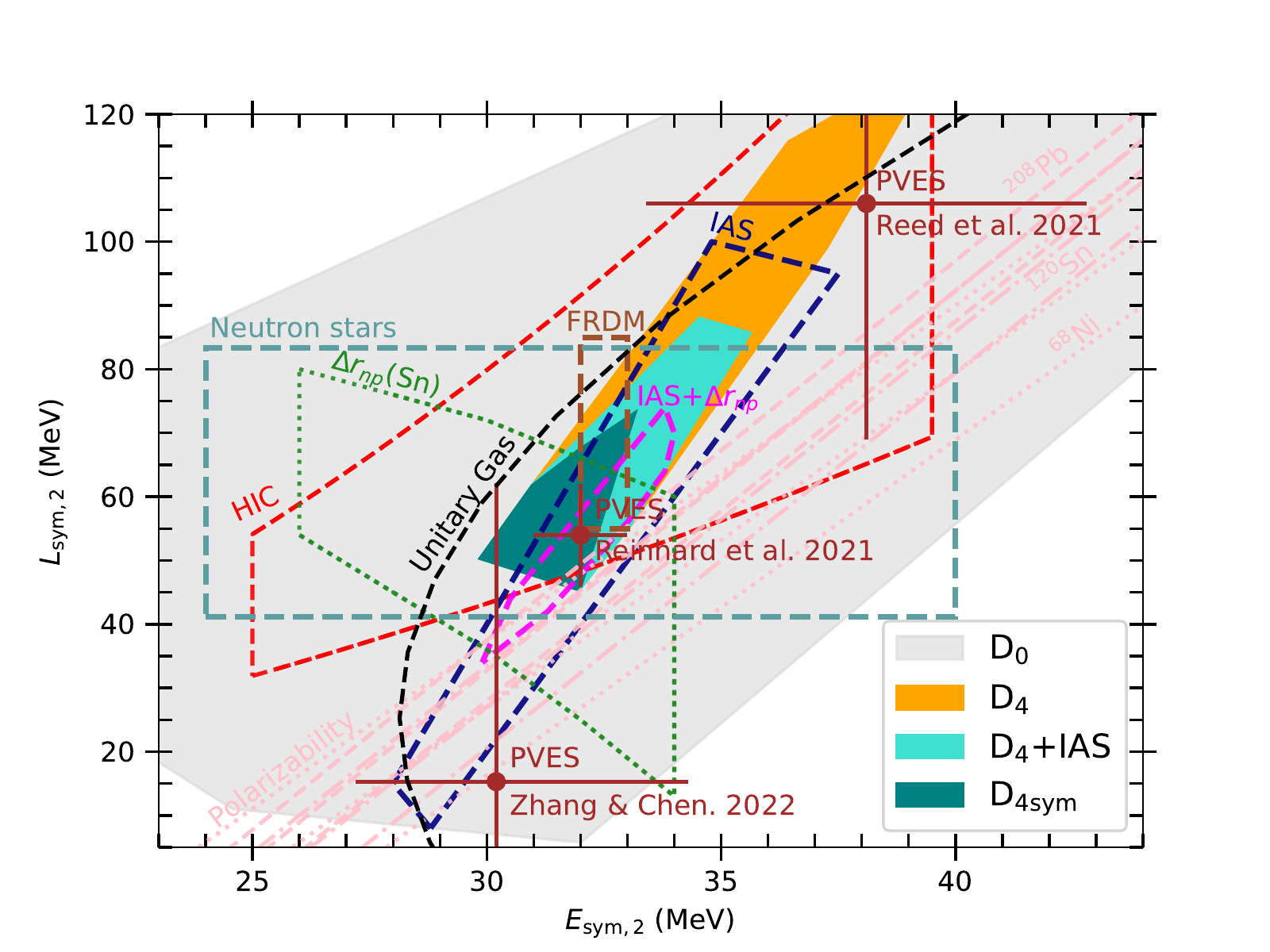}
\caption{Same as Fig.~\ref{fig:ELsymExp} including the new contours from the present analysis: D$_4$, D$_4$+IAS, D$_{4\sym}$. The light-grey band represent the contour of all explored interactions (group D$_0$=G$_0$).}
\label{fig:ELsymNew}
\end{figure}

One could also remark that, despite the good overlap between the IAS experimental constraint and our D$_4$+IAS contour, some of our interactions are outside the IAS experimental constraint. This is because we have considered a larger number of interactions, including relativistic approaches, which haven't been considered in Ref.~\cite{Danielewicz2013} where the contours are based on results solely from Skyrme models. The same remark could be made about the comparison of the group D$_{4\sym}$ and the IAS+$\Delta r_{np}$ experimental constraint.

\begin{table}[tb]
\tabcolsep=0.5cm
\def\arraystretch{1.5}
\centering
\caption{$E_{\sym,2}$ and $L_{\sym,2}$ centroid and standard deviation evaluated for the interactions in the groups: D$_4$, D$_4$+IAS and D$_{4\sym}$.}
\begin{tabular}{lcc}
\hline\noalign{\smallskip}
group     & $E_{\sym,2}$ & $L_{\sym,2}$ \\
          & (MeV) & (MeV) \\
\noalign{\smallskip}\hline\noalign{\smallskip}
D$_4$        & 33.5$\pm$2.4 & 73.4$\pm$23.3 \\
D$_4$+IAS    & 32.3$\pm$1.2 & 62.9$\pm$12.3 \\
D$_{4\sym}$ & 31.8$\pm$0.7 & 58.1$\pm$9.0 \\
\noalign{\smallskip}\hline
\label{tab:centroids}
\end{tabular}
\end{table}

Finally, the centroid and the standard deviation evaluated among the interactions forming the groups D$_4$, D$_4$+IAS and D$_{4\sym}$ are given in Table~\ref{tab:centroids}. Note the small dispersion obtained for the D$_{4\sym}$ group.

\section{Masses and radii of neutron star}
\label{sec:MR}

We now reach the second stage of our analysis where the EDFs that have been successfully selected by their goodness in reproducing finite nuclei properties are confronted with their predictions for NS properties. The EoS for dense matter and NS are detailed in Appendix~\ref{sec:densematter}. In this section, we discuss their predictions.

\begin{figure}[tb] 
\centering
\includegraphics[width=0.55\textwidth,trim=120 60 0 350, clip=true]{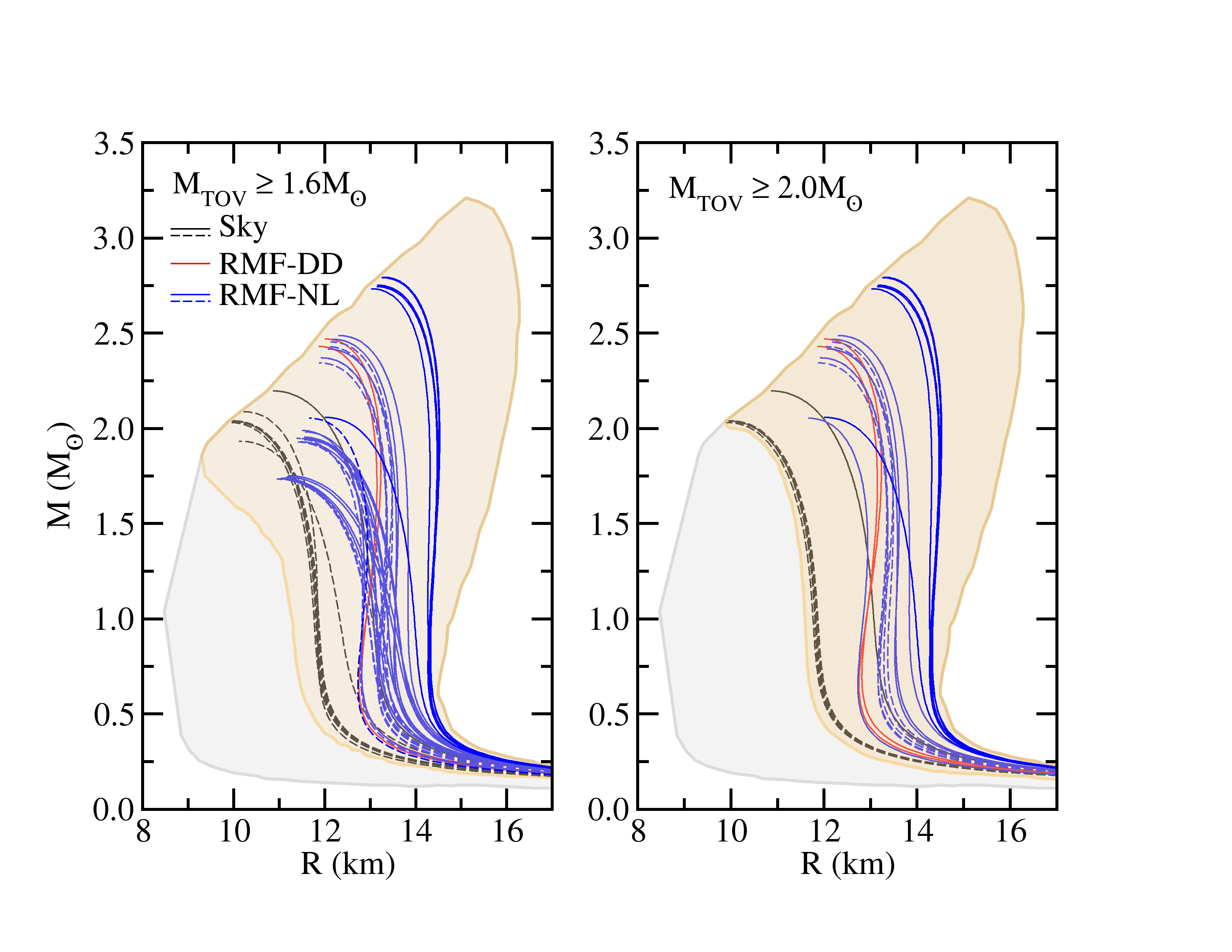} 
\includegraphics[width=0.55\textwidth,trim=120 100 0 350, clip=true]{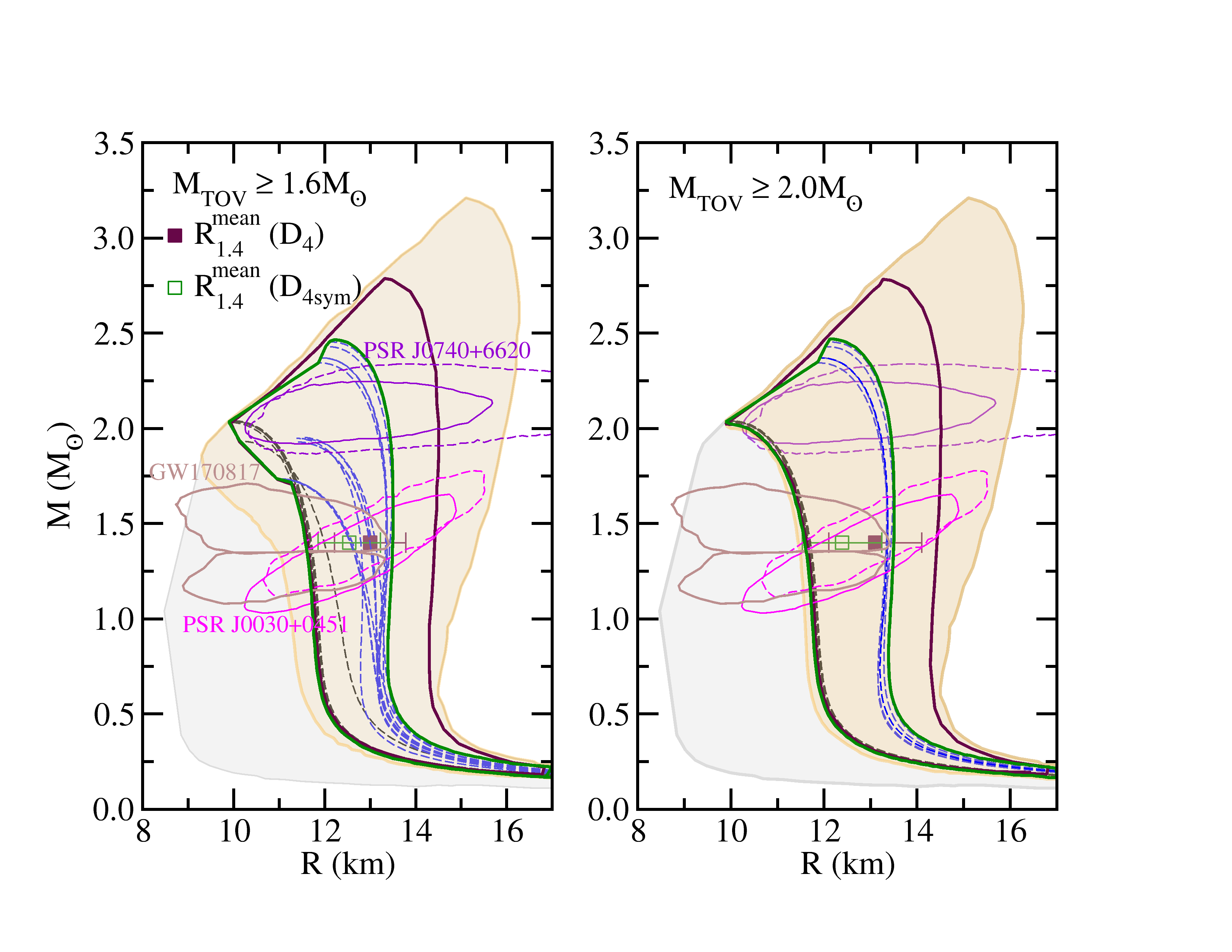}
\caption{Mass-radius diagram for various groups: The $D_0$ group is shown as a grey band, the $D_0$ group with the restriction on M$_\tov$, namely, $1.6$M$_\odot$ (left panels) and $2.0$M$_\odot$ (right panels). Curves in top panels: interactions of the D$_4$ group following the same notation as in Fig.~\ref{fig:esym}. In the bottom panels are shown the contours (dark brown ones) constructed from the interactions of the D$_4$ group in which  M$_\tov\geq1.6$M$_\odot$ (left) and M$_\tov\geq2.0$M$_\odot$ (right). Also in the bottom panels, we display the contours (green ones) defined by the interactions of the D$_{4\sym}$ subgroup and the curves that define it (dotted lines on top panel). Finally, violet and magenta contours represent  the mass-radius constraints of the NICER mission for PSR J0030+0451~\cite{nicer1a,nicer1b} and PSR J0740+6620~\cite{nicer2a,nicer2b} at the $90\%$ confidence level. Dashed (solid) lines indicates the data from  Miller et al.~\cite{nicer1a,nicer2a} (Riley et al. \cite{nicer1b,nicer2a}). The constraint determined from LIGO and Virgo Collaboration on the GW170817 event~\cite{ligo} is represented by brown contours.} 
\label{fig:mr2}
\end{figure}

The properties of non-rotating NS are obtained from the solution of the the Tolman-Oppenheimer-Volkoff (TOV) equations~\cite{tov39,tov39a,glen} written as ($G=c=1$) 
\begin{align}
\frac{dp_\tot(r)}{dr}&=-\frac{[\rho_\tot(r) + p_\tot(r)][m(r) + 4\pi r^3p_\tot(r)]}{r^2[1-2m(r)/r]},
\label{eq:tov1}
\\
\frac{dm(r)}{dr}&=4\pi r^2\rho_\tot(r),
\label{eq:tov2}
\end{align}
whose solution is determined by the initial conditions $p_\tot(0)=p_c$ (central pressure) and $m(0) = 0$.  In Eqs.~\eqref{eq:tov1} and \eqref{eq:tov2}, the energy density $\rho_\tot$ and pressure $p_\tot$ are given from Eqs.~\eqref{eq:totaled}-\eqref{eq:totalp}. The maximum value of $M$ for a given EoS is called M$_\tov$. The radius corresponding to a given mass, e.g., $1.4$M$_\odot$, is called $R_{1.4}$.

The break-down density above which the nucleonic EoS is replaced by an EoS with new degrees of freedom, e.g. hyperons or quarks, is not known. Turning the discussion of the break-down density into NS masses is easier in terms of observational data. For instance, a NS with mass $1.2/1.6/2.0$M$_\odot$ corresponds to central densities of $\approx$1.7-3/2-4.5/2.3-6$n_\sat$, where the larger central densities are obtained for the softer EoSs. From these numbers, it is reasonable to extrapolate the nucleonic EoS up to about $1.6$M$_\odot$, while the NS with $2.0$M$_\odot$ are considered as an extreme nucleonic scenario. In the following, we explore two cases where M$_\tov\geq1.6$M$_\odot$ and M$_\tov\geq2.0$M$_\odot$.

At very low mass (below M$_\odot$) the core EoS is connected to a crust EoS. The necessity of having a unified approach for both the crust and the core~\cite{DH2001,fortin} has been pointed out. Nevertheless, a piece-wise approach, in which the EoS in the core is connected to another EoS in the crust, is also widely used when a precision in the NS radius of about 100~m is sufficient or when detailed information of the crust-core transition is not required. In this work we adopt the procedure used in Refs.~\cite{Margueron2018b} in which the SLY interaction of Ref.~\cite{DH2001}, based on the SLy4 Skyrme parametrization~\cite{Chabanat1998}, is used for the crust region up to $n=0.1n_\sat$. As in Refs.~\cite{Margueron2018b}, a uniform matter EoS starts at $n_\sat$ and a log scale cubic spline takes care of smoothly connecting the lower limit of the crust to the upper limit of the core. Such a prescription allows a good description of the crust and the core, provided they can be smoothly connected. Exceptions exist however. For instance if the core EoS is described by an interaction with a value for $L_{\sym,2}$ much larger than the one used to describe the crust, difficulties in connecting the pressure in the crust and the core regions appear. This, however, is not the case for the interactions selected in the D$_4$ group.

Mass-radius profiles are shown in Fig.~\ref{fig:mr2} for various sets of EoSs.  The condition M$_\tov\geq2.0$M$_\odot$ removes a small region in the MR diagram corresponding to the softer EoSs for which M$_\tov\geq1.6$M$_\odot$. We represent the individual contributions of the interactions belonging to the D$_4$ group with the same convention as detailed in Fig.~\ref{fig:esym}. Dashed curves represent the interactions satisfying this constraint, namely, black for Skyrme, and blue for RMF-NL interactions with constant coupling constants. In the bottom panels of Fig.~\ref{fig:mr2}, the envelopes of the groups D$_4$ (dark brown) and D$_{4\sym}$ (green) are compared. The stiffest EoSs from the D$_4$ group are excluded in the D$_{4\sym}$ group, since they require a symmetry energy out of the boundaries shown in Fig.~\ref{fig:esym}. We obtain $R^{\rm{mean}}_{1.4}=13.00\pm0.78$ (12.53$\pm0.69$)~km for the D$_4$ (D$_{4\sym}$) group for M$_\tov\geq 1.6$M$_\odot$ and $R^{\rm{mean}}_{1.4}=13.10\pm1.00$ (12.38$\pm0.87$)~km for M$_\tov\geq 2.0$M$_\odot$. 

The contours related to the observational constraints from NICER~\cite{nicer1a,nicer1b,nicer2a,nicer2b} and for the GW170817 event detected by LIGO and Virgo Collaboration~\cite{ligo} are indicated in the bottom panels. A good overlap between the NICER contours and the groups D$_4$ and D$_{4\sym}$ is obtained, illustrating the agreement between the present constraints from nuclear physics and the ones from observations of neutron stars. A similar conclusion was obtained in Ref.~\cite{Tews2018} for the gravitational waves constraint extracted from GW170817.

\begin{figure}[tb] 
\centering
\includegraphics[scale=0.37]{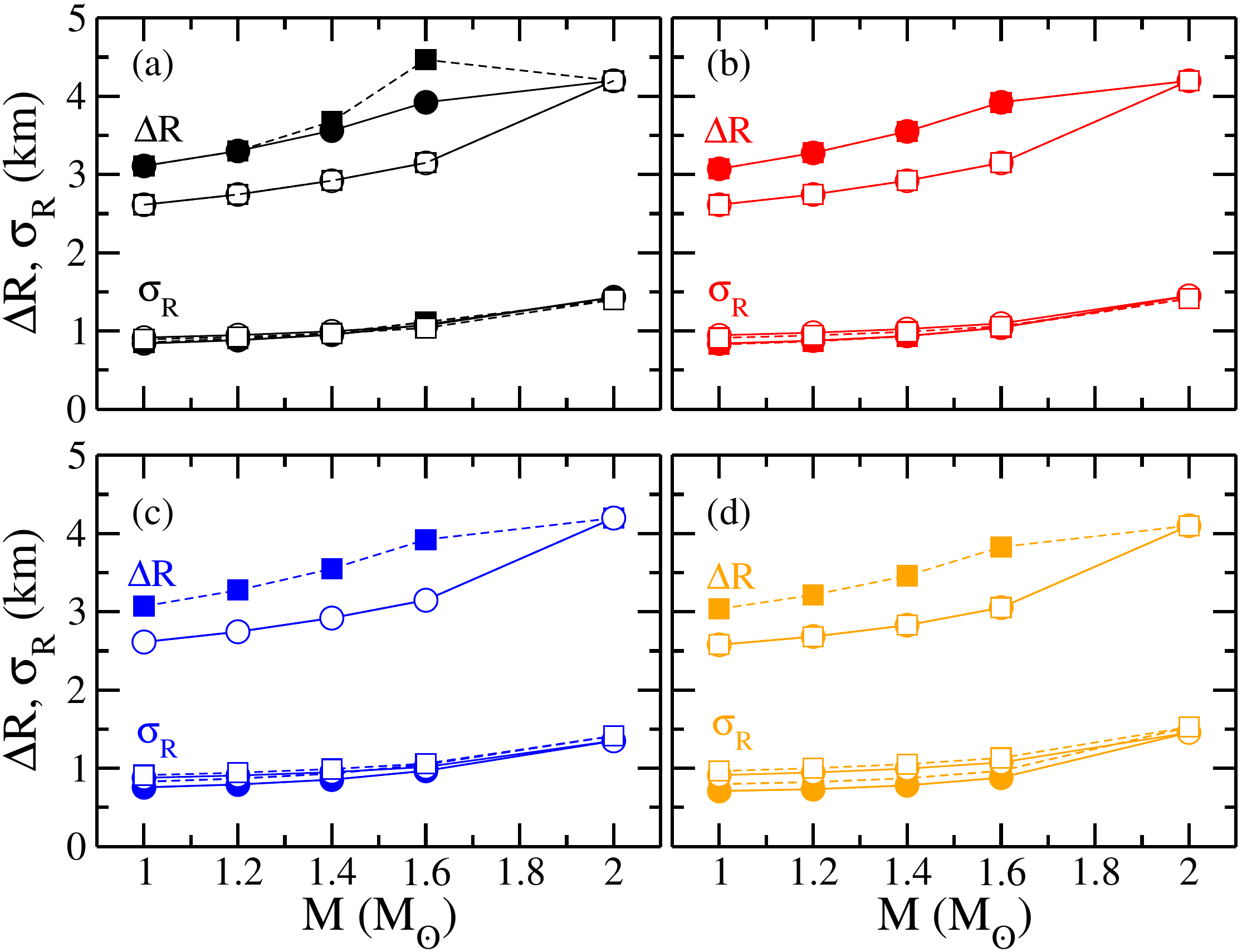}
\caption{Radius dispersion represented by the largest radius uncertainty $\Delta R$ and its standard deviation $\sigma_R$ as a function of the mass M, for the groups D$_i$ (solid lines) and G$_i$ (dashed lines) with $i=1$ (black), 2 (red), 3 (blue), and 4 (orange). The results correspond to those interactions satisfying M$_\tov\geq 1.6$M$_\odot$ (closed symbols) and M$_\tov\geq 2.0$M$_\odot$ (open symbols). See text for more details.}
\label{fig:statr}
\end{figure}

We now perform a more detailed study of the different groups D$_i$ and G$_i$ ($i=1$ to 4) to understand the impact of the different low energy nuclear constraints we have considered. In Fig.~\ref{fig:statr}, we compare the largest radius uncertainty $\Delta R$, namely, the difference between the maximum and the minimum radius, with the standard deviation $\sigma_R$ defined as
\begin{eqnarray}
\sigma^2_R=\frac{1}{n}\sum_{i=1}^{n}\big(R_i - \left<R\right>\big)^2\, ,
\end{eqnarray}
where $i$ runs over the nuclear interactions belonging to the groups D$_i$ (solid lines) or G$_i$ (dashed lines). As expected, we have $\sigma_R< \Delta R$ and both quantities increase as functions of the mass. For a canonical mass NS and the D$_4$ group, we obtain $\Delta R\approx 2.8$~km and $\sigma_R\approx 0.8$~km (assuming only that M$_\tov\geq 1.6$M$_\odot$). The number of interactions belonging to each group is given in Table~\ref{tabgroups}. Comparing D$_1$/G$_1$ and D$_2$/G$_2$ one measures the impact of better accuracy in the reproduction of the masses: the impact is very small in general. Note however a small reduction of $\Delta R$ between G$_1$ and G$_2$ at the mass 1.6M$_\odot$. Then, comparing D$_2$/G$_2$ and D$_3$/G$_3$ as well as D$_3$/G$_3$ and D$_4$/G$_4$, one can see the successive impact of an improved reproduction of the charge radii and ISGMR. Note the reduction of $\Delta R$ induced by the condition on the charge radius in the groups $D_i$, which is not visible for the groups $G_i$. This shows that the requirement to reproduce the charge radius in $N=Z$ and $N\ne Z$ with the same accuracy is the main condition which breaks the degeneracy between the groups $G_i$ and $D_i$.

The striking result from Fig.~\ref{fig:statr} is however the very weak dependence of the radius uncertainty, represented here by $\sigma_R$ and $\Delta R$, across the increasing index $i$ of the groups D$_i$ and G$_i$. This feature indicates that a more accurate reproduction of experimental masses, charge radii and the GMR energy in $^{208}$Pb does not have a major impact on the modelling of global NS properties, here M and R. There is however still an impact from the requirement that $N=Z$ and $N\ne Z$ nuclei are described with same accuracy, which is mainly given by the charge radius data.

\begin{figure*}[tb]
\begin{center}
\includegraphics[width=0.475\textwidth]{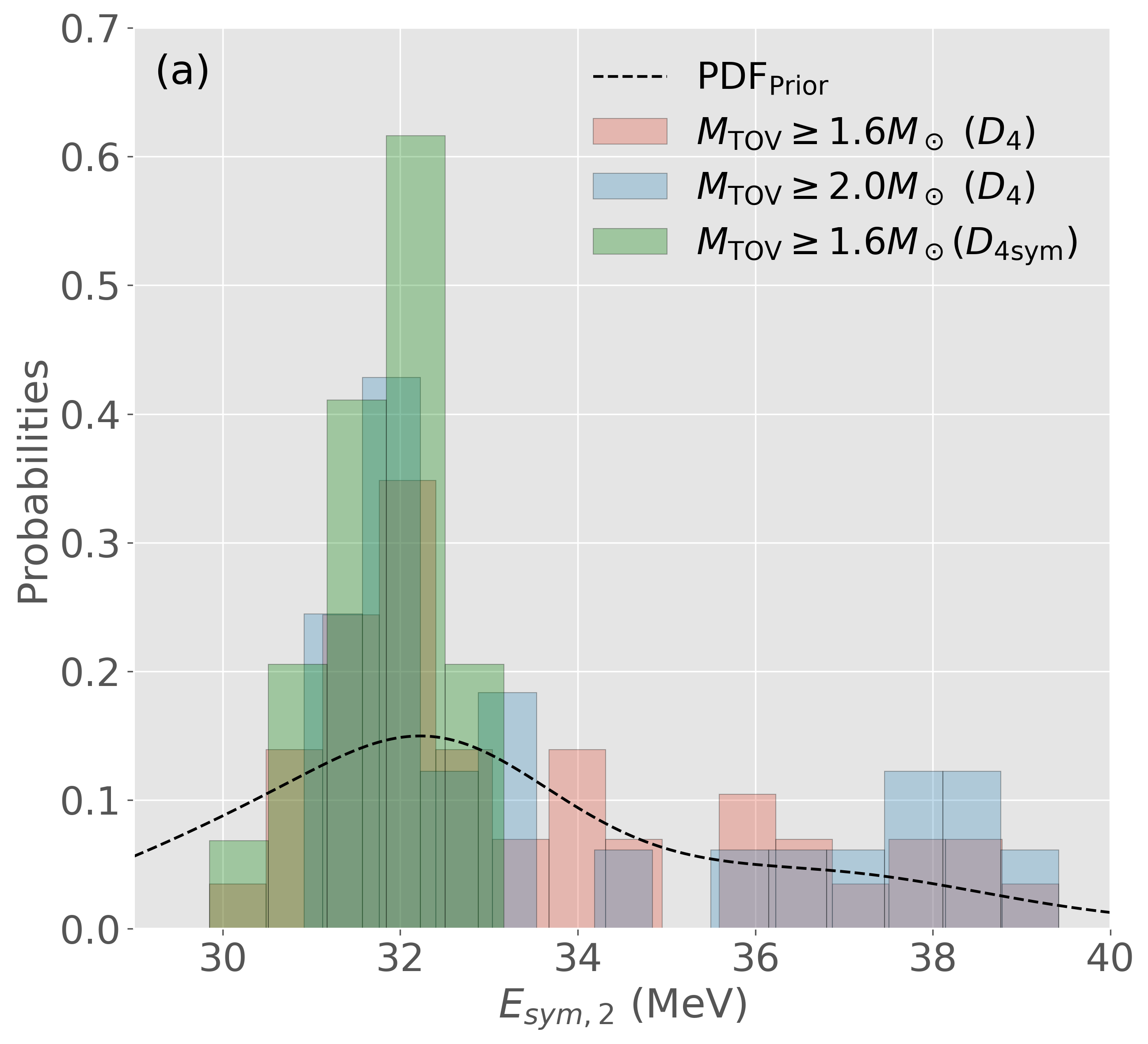} 
\includegraphics[width=0.475\textwidth]{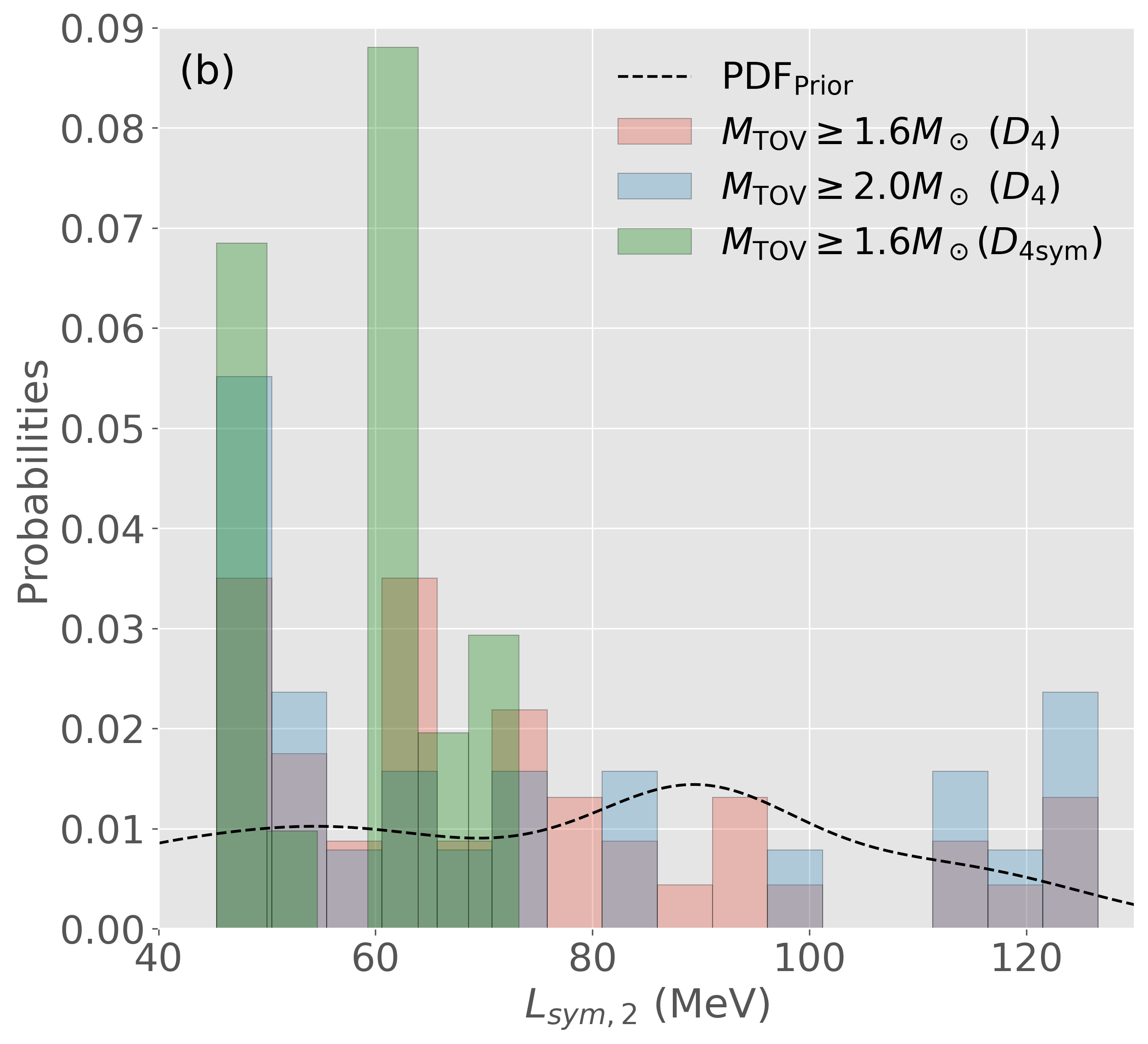} 
\includegraphics[width=0.495\textwidth]{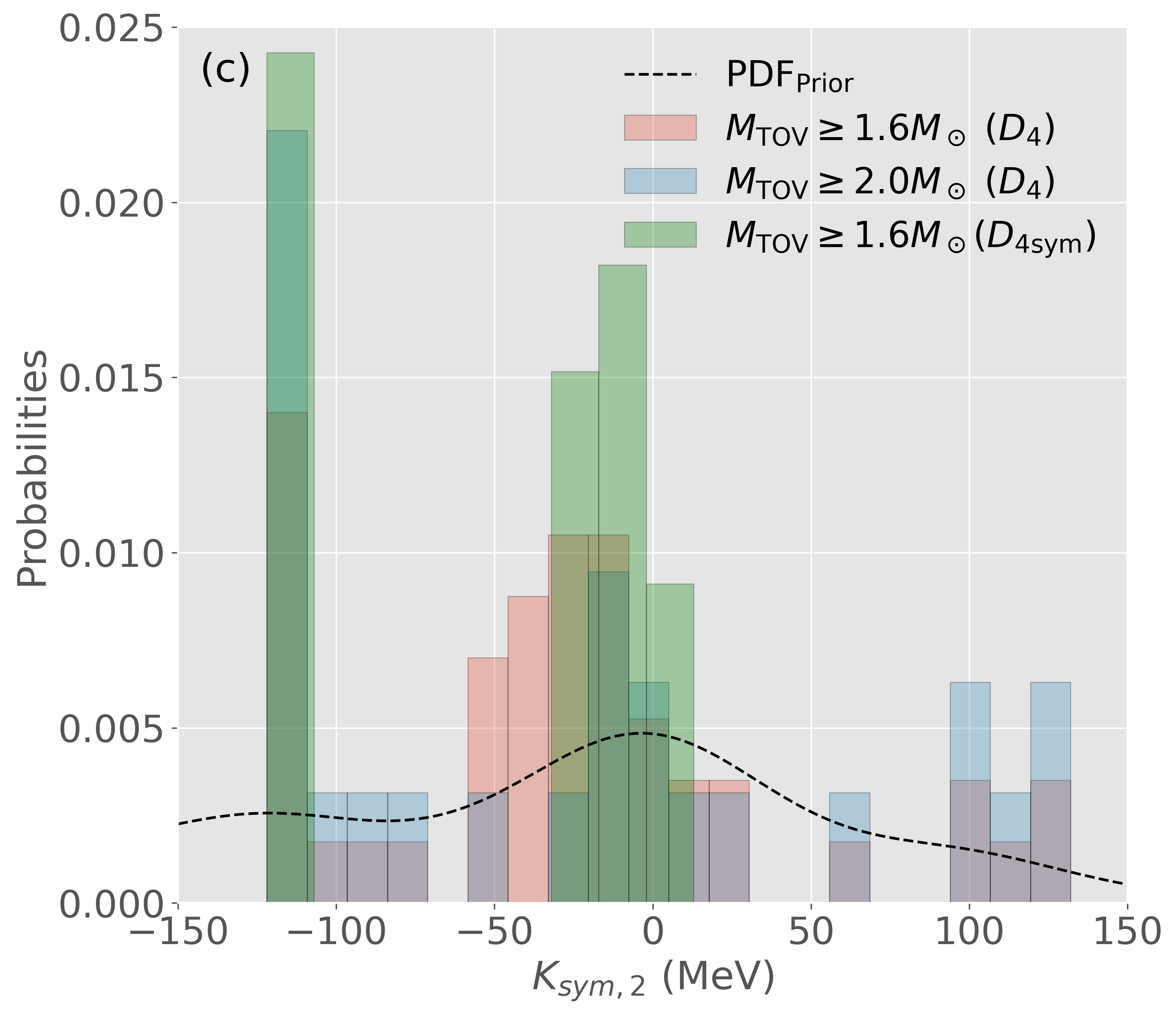}
\includegraphics[width=0.495\textwidth]{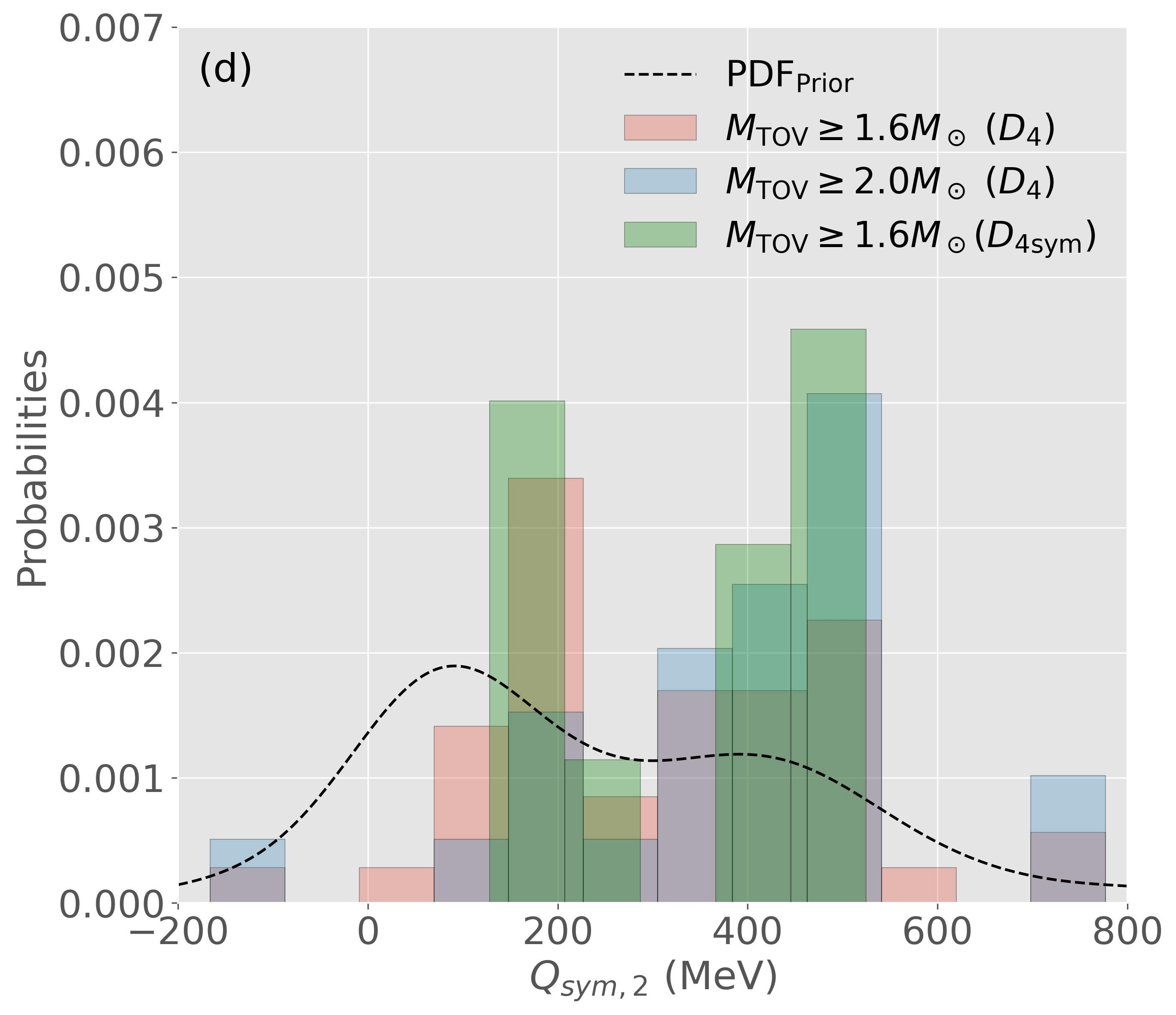}
\vspace{-0.5cm}
\caption{Normalized distributions of several isovector QNEPs, in particular, the symmetry energy ($E_{\sym,2}$) and its derivatives ($L_{\sym,2}$, $K_{\sym,2}$, $Q_{\sym,2}$). The probability density functions (PDF) for the priors from interactions in the D$_0$ group (black dashed curves) are also presented in each panel.}
\label{fig:hist}
\end{center}
\end{figure*}

Our finding suggests that a rough reproduction of low-energy nuclear physics properties (experimental masses, charge radii and ISGMR energy), as given in model D$_4$ for instance, is sufficient, provided that $N=Z$ and $N\ne Z$ nuclei are described with same accuracy. The gain in improving the modeling reproducing these experimental data is not effective for the prediction of NS global properties. The reason is that the extrapolation from finite nuclei located at around $n_{\sat}$ and close to isospin symmetry $\delta\lesssim 0.25$ to canonical mass NS with densities above $2n_\sat$, where matter is neutron rich $\delta\sim 1$,  requires the knowledge of the density and isospin dependence of the nuclear EoS, which represents a large and effective source of uncertainties. However, low-energy nuclear physics properties should be more important in determining the properties of the crust, such as the mass and the charge of nuclear clusters ($A_\cl$, $Z_\cl$), see for instance Refs.~\cite{Wolf2013,Fortin2016,Antic2019,Grams2022a,Grams2022b}. 

\section{Further analysis of the density dependence of the symmetry energy}

In the previous section, we illustrated the needs for a better understanding of the density dependence of the nuclear EoS for the prediction of NS global properties. Since the symmetry energy is the most important term in the EoS driving the density dependence of the EoS and since the quadratic nuclear empirical parameters (QNEPs) allow for a simple representation of this density dependence, we now analyse directly the QNEPs.

We present the normalized distributions (ND) of the QNEPs $E_{\sym,2}$, $L_{\sym,2}$, $K_{\sym,2}$, and $Q_{\sym,2}$ in Fig.~\ref{fig:hist} for various scenarios. We also present the prior distribution represented by the black dashed lines. This prior is obtained from the D$_0$ group.

\begin{figure*}[tb] 
\centering
\includegraphics[scale=0.55]{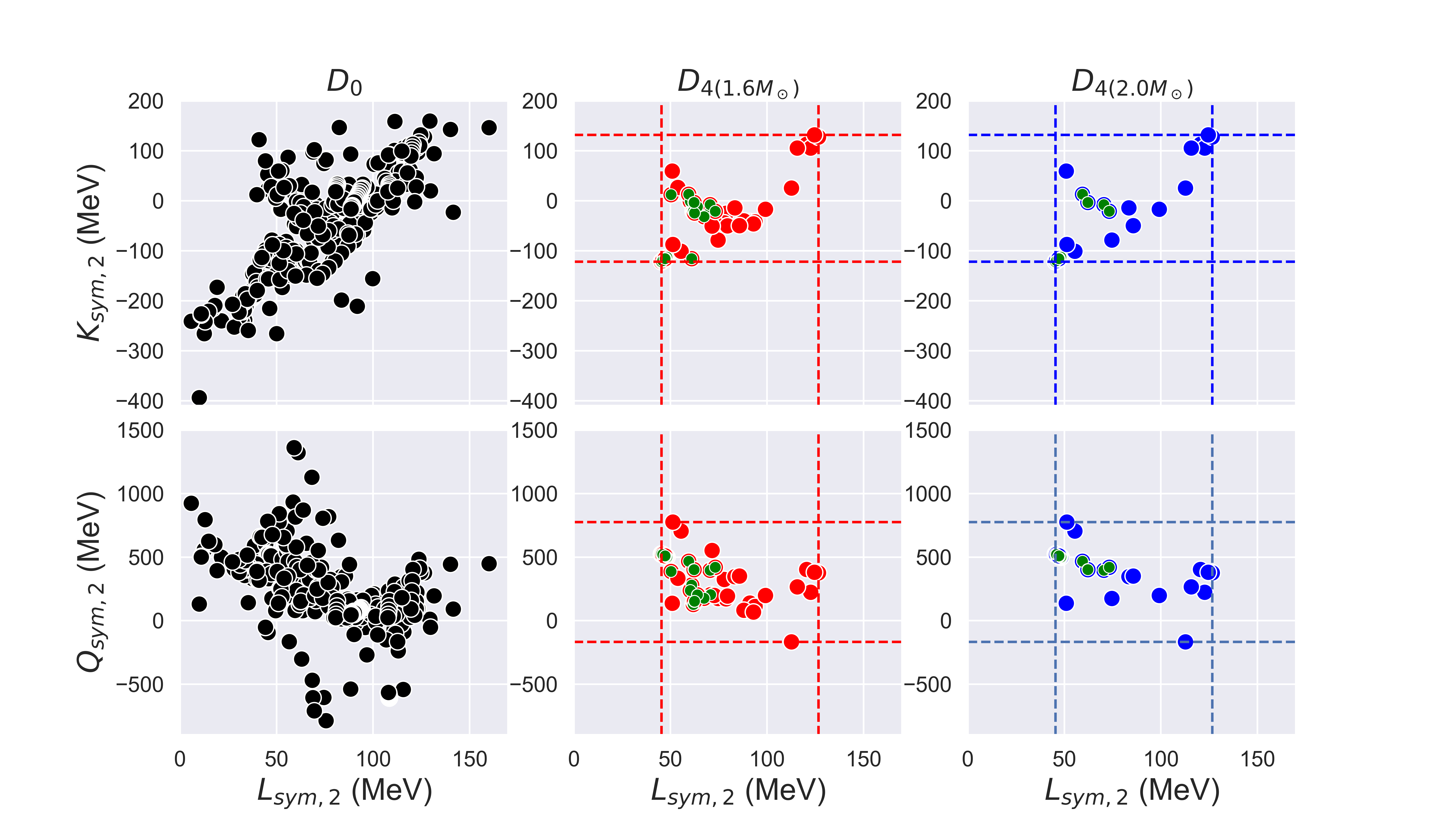}
\caption{Representation of correlations between $K_{\sym,2}$ ($Q_{\sym,2}$) and $L_{\sym,2}$ for the interactions of the D$_0$ group (left panels) and D$_4$ group with the restriction of reaching at least $1.6M_\odot$ (middle panels) and $2.0M_\odot$ (right panels). The green symbols represent the interactions of the D$_{4\sym}$ subgroup. The boundaries values (dashed lines) are given in Table~\ref{tab:boundaries}.}
\label{fig:bivariate}
\end{figure*}

The positions of the ND peaks for $E_{\sym,2}$ are almost identical to that of the prior, showing no effect of the interaction selection or the condition on M$_\tov$ for this QNEP. The ND for the group D$_{4\sym}$ is more peaked than the others. For $L_{\sym,2}$, the prior is almost flat while the posterior distributions indicate a preference for lower values of $L_{\sym,2}$ (about 50-70~MeV). In the case of group D$_{4\sym}$, the $L_{\sym,2}$ normalized distribution is double peaked, as is the distribution for $K_{\sym,2}$. For the latter, one maximum is located in the region $-50$ to $20$~MeV and another around $-100$~MeV. With regard to the D$_4$ group, notice that the relative size of the peaks in $K_{\sym,2}$ changes with the mass constraint: the peak at -100~MeV is preferred if M$_\tov\geq 2.0$M$_\tov$, while the peak around 0~MeV is preferred if M$_\tov\geq 1.6$M$_\tov$. For $Q_{\sym,2}$, the PDF is very broad, with most of the models lying between 100 an 600~MeV. The centroid of the normalized distribution conditioned by M$_\tov\geq 2.0$M$_\odot$ is, however, larger than the one conditioned by M$_\tov\geq 1.6$M$_\odot$. This shows that both $K_{\sym,2}$ and $Q_{\sym,2}$ are impacted by the condition on M$_\tov$.

\begin{table}[tb]
\tabcolsep=0.1cm
\def\arraystretch{1.5}
\centering
\caption{Boundaries for the quantities presented in Fig~\ref{fig:bivariate}. All quantities are in MeV.}
\begin{tabular}{lrrr}
\hline\noalign{\smallskip}
\rm          & D$_0$ & D$_{4\,(1.6)} = $D$_{4\,(2.0)}$ & D$_{4\sym\,(1.6)}$ \\
\noalign{\smallskip}\hline\noalign{\smallskip}
$L_{\sym,2}$ & 5.75 -- 160.06 & 45.36 -- 126.60 & 45.36 -- 73.23\\  
$K_{\sym,2}$ & -393.73 -- 159.57 & -121.90 -- 132.12 & -121.90 -- 12.96\\
$Q_{\sym,2}$ & -786.14 -- 1361.45 & -166.17 -- 776.91 &  128.26 -- 524.75\\
\noalign{\smallskip}\hline
\label{tab:boundaries}
\end{tabular}
\end{table}

Another view of these data is shown in Fig.~\ref{fig:bivariate}, where we plot the correlations between the QNEPs ($K_{\sym,2}$ versus $L_{\sym,2}$ in the top panels and $Q_{\sym,2}$ versus $L_{\sym,2}$ in the bottom panels) for interactions in the D$_0$ group (left panels) and D$_4$ groups central and right panels. These two last cases also show the impact of M$_\tov$. We observe that there are no strong correlations among the  QNEPs visible in this figure. The reason is that a low value of $L_{\sym,2}$ could be compensated by larger values of $K_{\sym,2}$ and/or $Q_{\sym,2}$. There is however a set of limits that we can extract from Fig.~\ref{fig:bivariate}. This set is given in Table~\ref{tab:boundaries}. 

The lower boundary in $L_{\sym,2}$ is due to two constraints. The first one is the reproduction of the low-energy nuclear physics properties, which produces a lower limit of the order of 30~MeV, as shown in Fig.~\ref{fig:hist}. The requirement to reproduce large masses, such as $1.6M_\odot$ or $2.0M_\odot$ pushes this lower limit up to about 45~MeV. It is however important to keep in mind that the boundaries we obtain are strongly impacted by the hypothesis that we have made concerning the absence of phase transition above saturation density. The case of a phase transition has been studied in Ref.~\cite{Xie2021}. 

The range of values for the isovector QNEPs reported in Table~\ref{tab:boundaries} could however be compared to other boundaries suggested in the literature. In the following we assume that the NEPs and QNEPs in the isospin channel are similar. For instance, by analysing non-relativistic Skyrme, relativistic mean field and relativistic Hartree-Fock models, it has been suggested that $K_{\sym,2}=(-100\pm 100)$~MeV~\cite{Margueron2018a}. From observational constraints based on X-ray emission from seven NSs in globular clusters, a value $K_{\sym}=-85^{+82}_{-70}$~MeV was preferred~\cite{Baillot2019}. These values are consistent with the recent analysis based on GMR energies in $^{90}$Zr, $^{116}$Sn and $^{208}$Pb using the Skyrme model, leading to $K_{\sym,2}=(-120\pm 40)$~MeV~\cite{Sagawa2019}. From the analysis of GW170817, it was suggested that $-259\mbox{ MeV} \leq K_{\sym} \leq 32$~MeV~\cite{Carson2019} and, in a similar analysis, the following values were suggested~\cite{Guven2020} $K_{\sym}=(440\pm 210)$~MeV, $K_{\sym}=(560\pm 150)$~MeV, and $K_{\sym}=(260\pm 240)$~MeV, depending on the observational PDF for $\tilde{\Lambda}$ extracted from GW170817~\cite{Abbott2019,De2018,Coughlin2019}. The value taken for $K_{\sym}$, however, was shown to be well anti-correlated with $L_{\sym}$, mainly driven by the condition to reproduce 2M$_\odot$. Since the PDFs for $\tilde{\Lambda}$ from GW170817 prefer low values for $L_{\sym}$, this explains why large values for $K_{\sym}$ were obtained in Ref.~\cite{Guven2020}. More recently, a compilation of 16 results from independent analyses of neutron star observational data since GW170817 lead to the following expectation, $K_{\sym}=(-107\pm 88)$~MeV~\cite{Li2021}. We also mention a recent analysis based on the latest results from the NICER observatory, where it was found that the lower radius limit of J0740 by Riley et al.~\cite{nicer2b} only requires $K_{\sym}$ to be higher than about $-150$~MeV, depending somewhat on the value of the skewness of the symmetry energy (unconstrained by the data).

All these results have to be taken with caution however, since they have been obtained with different nuclear models and corresponding systematical uncertainties that are difficult to estimate. Different priors have also been considered for the results based on a Bayesian statistical approach, which impact the results. We note, however, that our present findings for $K_{\sym,2}$ are in agreement with the predictions from these analyses. 

In conclusion, we have extracted constraints on the QNEPs determining the density dependence of the symmetry energy. These constraints are given in Table~\ref{tab:boundaries} for the different groups: D$_0$, D$_4$ and D$_{4\sym}$ conditioned by M$_\tov$. One could remark that these QNEPs are still largely unknown, although they are crucial for precise predictions of NS properties. However, the ranges for these QNEPs given in Table~\ref{tab:boundaries} represent the best evaluation of these QNEPs based on low energy nuclear experiments and conditioned by M$_\tov$.

\section{Neutron star global properties}

In this section we further elaborate on the role of the symmetry energy in the determination of NS global properties. We then introduce a refined classification of the EDFs based on the properties of the symmetry energy and we analyse the impact of this classification on NS global properties.

\subsection{Mass-radius relation of neutron stars}

\begin{figure}[tb] 
\centering
\includegraphics[scale=0.35]{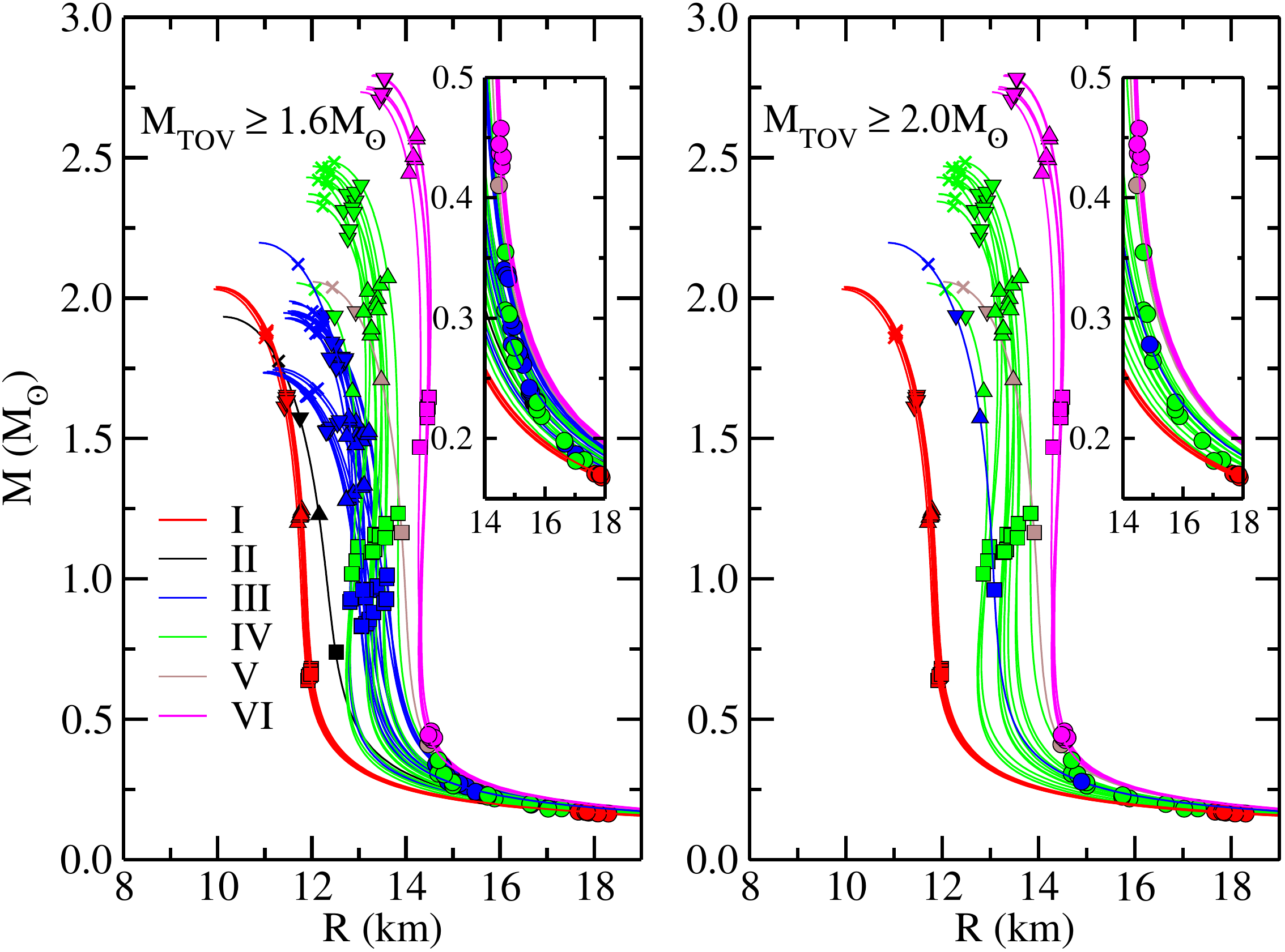}
\caption{Mass-radius profiles obtained from the interactions of the D$_4$ group, restricted to those producing a neutron star of at least $1.6M_\odot$ (left panel) and $2.0M_\odot$ (right panel). The points show the values at which the central density corresponds to saturation density $n_\sat$ (circles), $2n_\sat$ (squares), $3n_\sat$ (triangles up), $4n_\sat$ (triangles down) and $5n_\sat$ (x symbol).} 
\label{fig:mr1}
\end{figure}

The mass-radius (MR) relations for the interactions of the group D$_4$ are shown in Fig.~\ref{fig:mr1}, in a similar manner as in the top panels of Fig.~\ref{fig:mr2}. In addition, we have grouped the interactions in D$_4$ into six different sets according to their MR relations. These sets are shown in color as indicated in the legend.

\begin{table*}[tb]
\centering
\caption{Properties of the interactions belonging to the D$_4$ group. All entries are in MeV, except for the saturation density $n_\sat$ given in fm$^{-3}$, and the dimensionless effective mass $m^* = M^*(n_\sat)/M_{\mbox{\tiny nuc}}$. $E_\sat$ is the binding energy, $K_{\rm NM}=K_\sat+K_{\sym,2}$, $Q_{\rm NM}=Q_\sat+Q_{\sym,2}$, and $K_{\rm \tau,v}=K_{\sym,2}-6L_{\sym,2}-Q_\sat L_{\sym,2}/K_\sat$. The symbol~$\checkmark$ refers to those interactions that also produce neutron stars with M$_\tov \geq 2.0$M$_\odot$ and/or and belong to the D$_{4\sym}$ group. See the text for the definition of the subgroups I to VI.}
\setlength{\tabcolsep}{1pt}
\renewcommand{\arraystretch}{1.3}
\begin{ruledtabular}
\begin{tabular}{l|cc|l|c|c|c|c|c|c|c|c|c|c|c|c|c}
& $2.0M_\odot$ & D$_{4\sym}$ & interaction & Ref. & $n_\sat$ & $E_\sat$ & $K_\sat$ & $Q_\sat$ 
& $E_{\sym,2}$ & $L_{\sym,2}$ & $K_{\sym,2}$ & $Q_{\sym,2}$ & $K_{\rm NM}$ & $Q_{\rm NM}$ & $K_{\rm \tau,v}$ & $m^*$ \\ \hline
\multirow{7}{*}{I} & $\checkmark$ &  $\checkmark$ & SLy3 & \cite{Chabanat-thesis}& 0.160 & -15.94 & 229.51 & -362.56 & 31.97 & 45.36 & -121.90 & 524.75 & 107.61 & 162.18 & -322.39 & 0.70 \\ 
& $\checkmark$ & $\checkmark$  & SLy4 & \cite{Chabanat1998} & 0.160 & -15.97 & 229.91 & -363.11 & 32.00 & 45.94 & -119.73 & 521.53 & 110.18 & 158.43 & -322.83 & 0.69 \\ 
& $\checkmark$ & $\checkmark$  & SLy230b & \cite{Chabanat1997} & 0.160 & -15.97 & 229.91 & -363.10 & 32.01 & 45.97 & -119.72 & 521.50 & 110.19 & 158.40 & -322.92 & 0.69 \\
& $\checkmark$ & $\checkmark$  & SLy0 & \cite{Chabanat-thesis}& 0.160 & -15.97 & 229.66 & -364.01 & 31.98 & 47.11 & -116.23 & 508.68 & 113.43 & 144.67 & -324.23 & 0.70 \\ 
& $\checkmark$ & $\checkmark$  & SLy8 &\cite{Chabanat-thesis} & 0.160 & -15.97 & 229.89 & -363.27 & 32.00 & 47.18 & -115.59 & 509.88 & 114.31 & 146.61 & -324.09 & 0.70 \\ 
& $\checkmark$ & $\checkmark$  & SLy2 & \cite{Chabanat-thesis}& 0.161 & -15.99 & 229.92 & -364.21 & 32.00 & 47.46 & -115.13 & 506.52 & 114.79 & 142.31 & -324.69 & 0.70 \\ 
& $\checkmark$ & $\checkmark$ & SLy5 & \cite{Chabanat1998} & 0.161 & -15.99 & 229.92 & -364.16 & 32.01 & 48.15 & -112.76 & 500.67 & 117.16 & 136.51 & -325.38 & 0.70 \\ \hline
\multirow{1}{*}{II}  & & $\checkmark$ & SD1 & \cite{sd1} & 0.156 & -15.70 & 231.91 & -376.39 & 32.00 & 60.94 & -115.87 & 281.29 & 116.04 & -95.10 & -382.59 & 1.00 \\ %
\hline
\multirow{19}{*}{III} & & $\checkmark$ & IUFSU* & \cite{iufsus} & 0.150 & -16.02 & 235.67 & -259.49 & 29.85 & 50.30 & 12.20 & 388.15 & 247.87 & 128.67 & -234.20 & 0.61 \\ 
& &  & SINPA & \cite{sinpab} & 0.151 & -16.00 & 202.55 & -58.73 & 31.20 & 53.85 & 26.58 & 334.85 & 229.13 & 276.12 & -280.92 & 0.58 \\
& & $\checkmark$ & BSR8 & \cite{BSR} & 0.147 & -16.04 & 230.95 & -290.85 & 31.08 & 60.25 & -0.74 & 238.23 & 230.22 & -52.62 & -286.36 & 0.61 \\ 
& & $\checkmark$ & BSR15 & \cite{BSR} & 0.146 & -16.03 & 226.82 & -512.29 & 30.97 & 61.79 & -21.36 & 128.26 & 205.47 & -384.03 & -252.54 & 0.61 \\
& & $\checkmark$ & FSUGZ06 & \cite{fsugz}& 0.146 & -16.05 & 225.06 & -503.17 & 31.18 & 62.42 & -24.49 & 153.31 & 200.57 & -349.86 & -259.47 & 0.61 \\ 
& & $\checkmark$ & BSR16 & \cite{BSR} & 0.146 & -16.05 & 224.98 & -503.17 & 31.24 & 62.33 & -24.17 & 152.29 & 200.82 & -350.88 & -258.75 & 0.61 \\ 
&  & $\checkmark$ & BSR9 & \cite{BSR} & 0.147 & -16.07 & 232.50 & -297.11 & 31.61 & 63.89 & -11.32 & 202.86 & 221.18 & -94.25 & -313.03 & 0.60 \\ 
& & $\checkmark$ & FSUGZ03 & \cite{fsugz} & 0.147 & -16.07 & 232.48 & -297.13 & 31.54 & 63.98 & -11.66 & 203.43 & 220.82 & -93.69 & -313.79 & 0.60 \\ 
& & $\checkmark$ & BSR17 & \cite{BSR} & 0.146 & -16.05 & 221.67 & -489.45 & 31.98 & 67.44 & -31.58 & 176.65 & 190.09 & -312.80 & -287.31 & 0.61 \\ 
& & $\checkmark$ & BSR10 & \cite{BSR} & 0.147 & -16.06 & 227.41 & -255.13 & 32.72 & 70.83 & -16.51 & 205.04 & 210.89 & -50.09 & -362.04 & 0.60 \\ 
& &  & SINPB & \cite{sinpab} & 0.150 & -16.05 & 206.40 & -449.21 & 33.96 & 71.55 & -50.60 & 552.47 & 155.80 & 103.26 & -449.21 & 0.59 \\ 
& &  & BSR18 & \cite{BSR} & 0.146 & -16.05 & 221.13 & -485.73 & 32.74 & 72.65 & -42.24 & 199.39 & 178.89 & -286.35 & -318.55 & 0.61 \\
& $\checkmark$ &  & SKa & \cite{ska} & 0.155 & -15.99 & 263.16 & 300.13 & 32.91 & 74.62 & -78.46 & 174.54 & 184.70 & 474.66 & -441.08 & 0.61 \\
& &  & BSR12 & \cite{BSR} & 0.147 & -16.10 & 232.35 & -290.31 & 34.00 & 77.90 & -44.23 & 324.15 & 188.12 & 33.85 & -414.30 & 0.61 \\ 
& &  & BSR11 & \cite{BSR} & 0.147 & -16.08 & 226.75 & -312.37 & 33.69 & 78.78 & -24.72 & 172.54 & 202.03 & -139.83 & -388.86 & 0.61 \\ 
& &  & BSR19 & \cite{BSR} & 0.147 & -16.08 & 220.83 & -484.25 & 33.78 & 79.47 & -50.13 & 194.70 & 170.70 & -289.55 & -352.70 & 0.61 \\ 
& &  & BSR20 & \cite{BSR} & 0.146 & -16.09 & 223.25 & -507.75 & 34.54 & 88.03 & -39.90 & 82.74 & 183.35 & -425.02 & -367.86 & 0.61 \\ 
& & & BSR13 & \cite{BSR} & 0.147 & -16.13 & 228.64 & -294.46 & 35.82 & 91.07 & -41.68 & 138.98 & 186.96 & -155.48 & -470.82 & 0.60 \\ 
& & & BSR21 & \cite{BSR} & 0.145 & -16.12 & 220.32 & -468.20 & 35.96 & 92.94 & -46.01 & 67.45 & 174.30 & -400.75 & -406.16 & 0.60 \\ 
& & & BSR14 & \cite{BSR} & 0.147 & -16.18 & 235.47 & -317.10 & 36.32 & 93.85 & -41.95 & 112.53 & 193.51 & -204.57 & -478.66 & 0.61 \\ \hline
\multirow{11}{*}{IV} &  & & FSUGarnet & \cite{fsugarnet} & 0.153 & -16.23 & 229.61 & -13.12 & 30.92 & 50.96 & 59.44 & 138.08 & 289.06 & 124.96 & -249.21 & 0.58 \\ 
& $\checkmark$ &  & DD-ME2 & \cite{ddme2} & 0.152 & -16.14 & 250.92 & 478.75 & 32.30 & 51.25 & -87.19 & 776.91 & 163.73 & 1255.67 & -492.45 & 0.57 \\ 
& $\checkmark$ &  & DD-ME1 & \cite{ddme1} & 0.152 & -16.20 & 244.72 & 316.66 & 33.06 & 55.45 & -101.05 & 705.59 & 143.66 & 1022.25 & -505.50 & 0.58 \\ 
& $\checkmark$ & $\checkmark$ & BSR1 & \cite{BSR} & 0.148 & -16.02 & 239.89 & -35.68 & 31.04 & 59.41 & 12.96 & 468.10 & 252.85 & 432.42 & -334.65 & 0.61 \\ 
& $\checkmark$ & $\checkmark$ & BSR2 & \cite{BSR} &  0.149 & -16.03 & 239.93 & -48.06 & 31.50 & 62.02 & -3.14 & 403.21 & 236.79 & 355.15 & -362.81 & 0.61 \\ 
& $\checkmark$ & $\checkmark$ & FSUGZ00 & \cite{fsugz} & 0.149 & -16.03 & 240.00 & -47.74 & 31.43 & 62.16 & -3.46 & 402.48 & 236.54 & 354.74 & -364.05 & 0.61 \\ 
& $\checkmark$ & $\checkmark$ & BSR3 & \cite{BSR} & 0.150 & -16.09 & 230.55 & -114.72 & 32.74 & 70.45 & -7.76 & 397.59 & 222.78 & 282.86 & -395.42 & 0.60 \\ 
& $\checkmark$ & $\checkmark$  & BSR4 & \cite{BSR} & 0.150 & -16.08 & 238.57 & 4.00 & 33.17 & 73.23 & -20.71 & 420.06 & 217.85 & 424.06 & -461.34 & 0.61 \\ 
& $\checkmark$ & & BSR5 & \cite{BSR} & 0.151 & -16.12 & 235.71 & -10.96 & 34.46 & 83.37 & -14.16 & 346.84 & 221.55 & 335.88 & -510.53 & 0.61 \\ 
& $\checkmark$ & & BSR6 & \cite{BSR} & 0.149 & -16.13 & 235.75 & -7.59 & 35.62 & 85.68 & -49.55 & 352.00 & 186.20 & 344.41 & -560.86 & 0.60 \\ 
& $\checkmark$ & & BSR7 & \cite{BSR} & 0.149 & -16.18 & 231.80 & -19.80 & 37.26 & 99.14 & -16.97 & 198.47 & 214.83 & 178.66 & -603.32 & 0.60 \\ \hline
\multirow{1}{*}{V} & $\checkmark$ & & FSUGold2 & \cite{fsugold2} & 0.150 & -16.26 & 237.69 & -149.95 & 37.57 & 112.68 & 25.38 & -166.17 & 263.07 & -316.13 & -579.61 & 0.59 \\ \hline
\multirow{4}{*}{VI} & $\checkmark$ & & Q1 & \cite{q1} & 0.148 & -16.10 & 241.86 & 8.70 & 36.44 & 115.71 & 105.65 & 266.72 & 347.51 & 275.43 & -592.77 & 0.60 \\ 
& $\checkmark$ & & FAMA1 & \cite{fama1} & 0.148 & -16.00 & 200.05 & -303.20 & 38.01 & 120.53 & 113.22 & 403.17 & 313.27 & 99.97 & -427.27 & 0.60 \\
& $\checkmark$ & & NL3* & \cite{nl3s} & 0.150 & -16.31 & 258.25 & -122.04 & 38.68 & 122.63 & 105.56 & 223.95 & 363.81 & 101.92 & -688.19 & 0.59 \\
& $\checkmark$ & & E & \cite{eer} & 0.150 & -16.13 & 221.43 & 20.87 & 38.58 & 124.57 & 132.12 & 381.38 & 353.54 & 402.25 & -627.06 & 0.58 \\
& $\checkmark$ & & ER & \cite{eer} &  0.149 & -16.16 & 220.49 & -24.93 & 39.42 & 126.60 & 127.62 & 377.17 & 348.11 & 352.24 & -617.67 & 0.58 \\ 
\end{tabular}
\end{ruledtabular}
\label{tab:bulk}
\end{table*}

These sets could also be sorted by the values of the NEPs $L_{\sym,2}$, $K_{\sym,2}$ and the condition that M$_\tov\geq 2$M$_\odot$ as shown in table~\ref{tab:bulk}: set I consists of interactions from the group D$_4$ for which $L_{\sym,2}\leq 50$~MeV; set II, III and IV have $50~\mbox{MeV}<L_{\sym,2}\leq 100~\mbox{MeV}$, and additionally set II contains the single EoS for which $K_{\sym,2}\leq -110$~MeV, while sets III and IV have larger values for $K_{\sym,2} > -110$~MeV. The difference between set III and IV is that all EoSs from set IV satisfy the condition M$_\tov\geq 2$M$_\odot$, while models in set III do not, except for the SKa parametrization.
Finally, sets V and VI are EoSs with large values for $L_{\sym,2}\geq 100$~MeV. Set V has a value of $K_{\sym,2}$ of about $25$~MeV, while set VI has $K_{\sym,2}\geq 100$~MeV.

The analysis of Fig.~\ref{fig:mr1} and Table~\ref{tab:bulk} leads to the following conclusions: The main difference among the different sets is coming from the value of $L_{\sym,2}$: low values (set I), intermediate values (sets II, III and IV) and large values (set V and VI). The value of $L_{\sym,2}$ determines the stiffness of the EoS as shown in Fig.~\ref{fig:mr1}: the softer the EoS, the lower the radius and the larger the central density for a given mass M. To a certain degree, the stiffness of the EoS is determined by the value of $K_{\sym,2}$. In order to make this more clear in Fig.~\ref{fig:mr1}, we have separated set II from sets III and IV according to the value of $K_{\sym,2}$.

We also indicate in Table~\ref{tab:bulk} the EoSs which belong to the subgroup D$_{4\sym}$. They are the EoSs for which $L_{\sym,2}< 90$~MeV. Note that the precise upper value for $L_{\sym,2}$ actually depends  on the value of $K_{\sym,2}$.

In conclusion, we have analysed the dominant role of $L_{\sym,2}$ in globally controlling the MR diagram, with some additional contribution from $K_{\sym,2}$. In Fig.~\ref{fig:mr1} however, one observes some correlations between $L_{\sym,2}$ and the masses/radii at fixed central densities. In the following, we analyse these correlations in more detail.

\subsection{Radius and mass (individual analysis)}

\begin{figure}[tb] 
\centering
\includegraphics[scale=0.35]{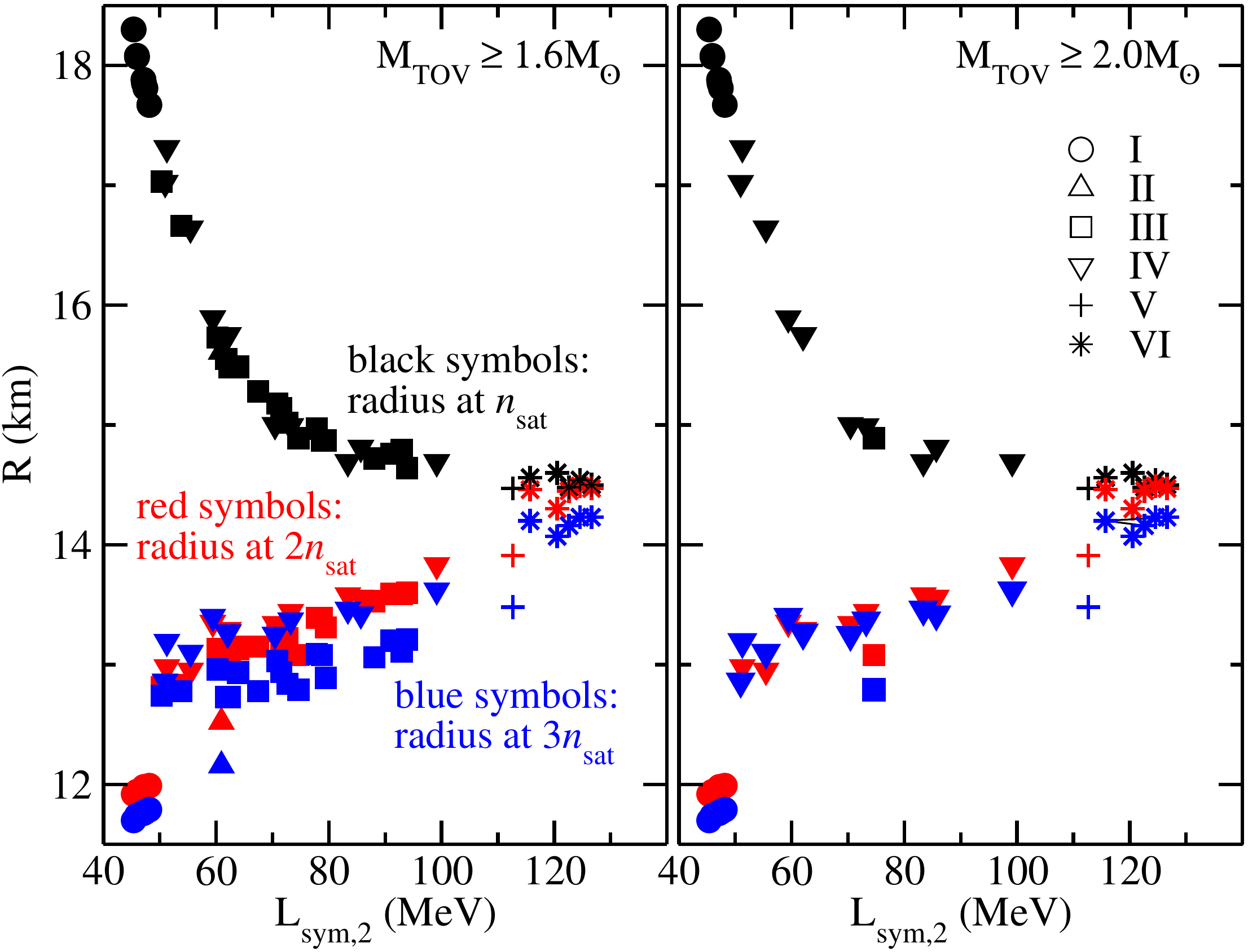}
\caption{Neutron star radius at $n_\sat$ (black symbols), $2n_\sat$ (red symbols), and $3n_\sat$ (blue symbols) as a function of $L_{\sym,2}$, obtained from the interactions of the D$_4$ group and restricted to those 
satisfying M$_\tov\geq 1.6$M$_\odot$ (left panel) and M$_\tov\geq 2.0$M$_\odot$ (right panel). The symbols correspond to the sets I to VI as indicated in the legend.} 
\label{fig:rl}
\end{figure}

We represent in Fig.~\ref{fig:rl} the correlation between the NS radius R -- extracted at different densities ($n_\sat$, $2n_\sat$ and $3n_\sat$) -- and $L_{\sym,2}$. Note that the correlation is opposite at $n_\sat$ to that at higher densities: At $n_\sat$ the radius decreases as $L_{\sym,2}$ increases, while above, the radius increases as a function of $L_{\sym,2}$. The reason is simple: having a larger value of $L_{\sym,2}$ implies a softer EoS below $n_\sat$. So the anti-correlation at $n_\sat$ reflects that the EOS is softer at low densities for larger values of $L_{\sym,2}$. At higher densities, the situation is different since the larger the value of $L_{\sym,2}$, the stiffer the EoS is above $n_\sat$. The EoSs are so stiff that they change the MR relation: above saturation density, the radius is weakly impacted by the mass. Since stiffer EoSs above $n_\sat$ imply larger values for $L_{\sym,2}$, the radius is correlated with $L_{\sym,2}$ in this region.

\begin{figure}[tb] 
\centering
\includegraphics[scale=0.35]{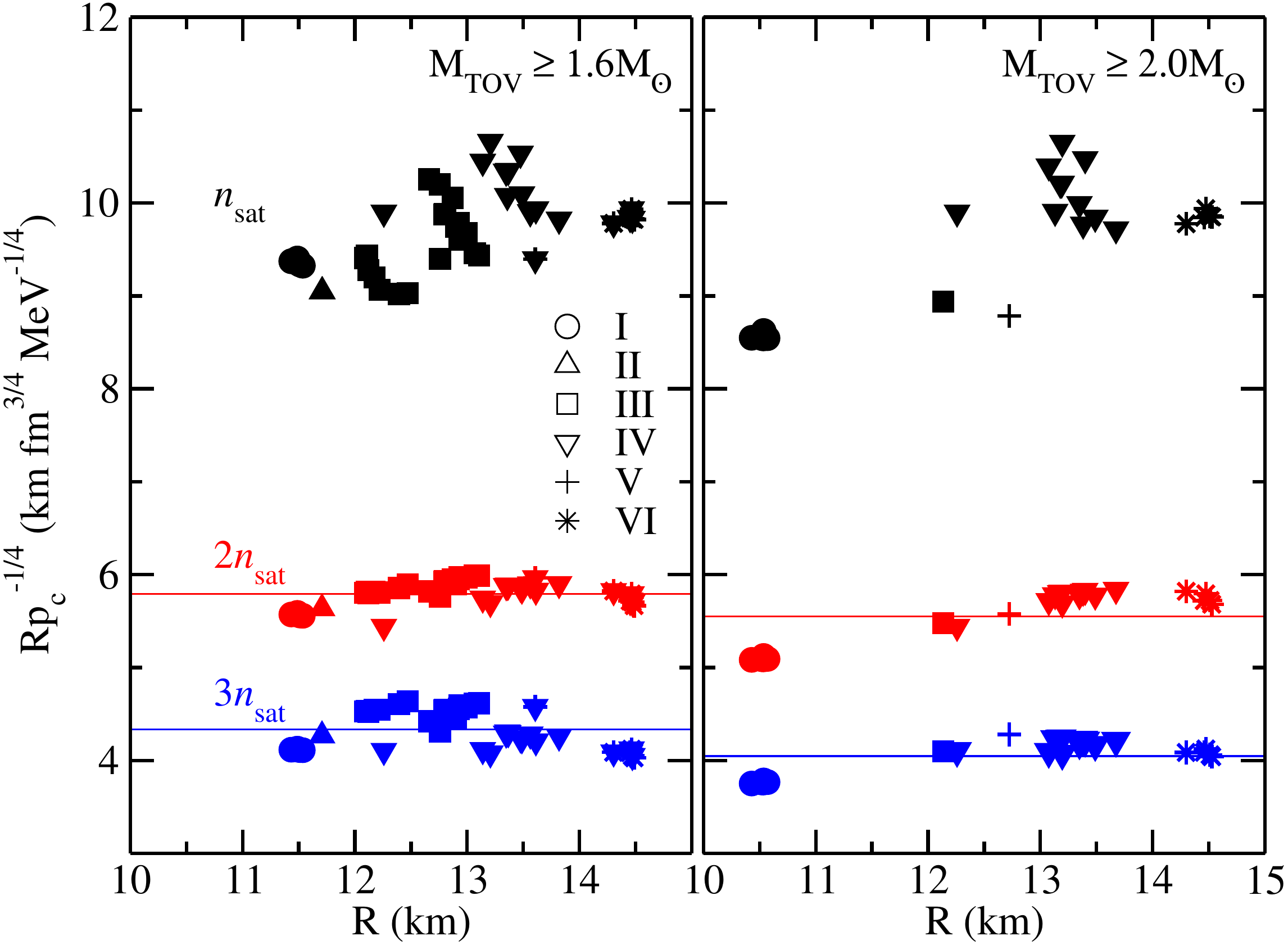}
\caption{Empirical relation between pressure (in units of MeV.fm$^{-3}$) and the radius (in km) obtained from the interactions of the D$_4$ group and restricted to those satisfying M$_\tov\geq 1.6$M$_\odot$ (left panel) and M$_\tov\geq 2.0$M$_\odot$ (right panel). The symbols correspond to the sets I to VI as indicated in the legend. The solid lines represent the best values for the product $Rp_c^{-1/4}$, namely, left (right) panel: $5.79$ ($5.55$) and $4.33$ ($4.05$), for the red and blue lines, respectively. All numbers in units of \mbox{km fm$^{3/4}$ MeV$^{-1/4}$}.} 
\label{fig:rp}
\end{figure}

We now test the empirical relation between R and $p_c$, suggested in Ref.~\cite{Prakash2001}. The pressure $p_c$ is the central pressure of the NS (at $\beta$-equilibrium). We show the quantity $Rp_c^{-1/4}$ as a function of $R$ in Fig.~\ref{fig:rp}. The correlation between R and $p_c$ is better at $2n_\sat$ and $3n_\sat$ compared to $n_\sat$, as already noted in Ref.~\cite{Prakash2001}. The product $Rp_c^{-1/4}$ is weakly correlated with the radius, if the pressure is taken at $2n_\sat$ and $3n_\sat$. Since the radius is well correlated with $L_{\sym,2}$ at these densities, see Fig.~\ref{fig:rl}, the product $Rp_c^{-1/4}$ is also weakly correlated with $L_{\sym,2}$ at $2n_\sat$ and $3n_\sat$. At $n_\sat$ however, the points are much less aligned than at $2n_\sat$ and $3n_\sat$. The dispersion between the points reflects the (dominant) effect of $L_{\sym,2}$  as well as that of $K_{\sym,2}$. 
In NS, the central pressure near saturation density is dominantly given by $L_{\sym,2}$~\cite{Lattimer2016}, while other NEPs contribute more as the density increases~\cite{Margueron2018a,Margueron2018b}. 

The correlation suggested in Ref.~\cite{Prakash2001} at $2n_\sat$ and $3n_\sat$ reflects at least two interesting features: first it shows the weak influence of the radial distribution of the pressure in NSs, since it is mostly the central value which fixes the NS radius, and second, it hides the contribution of the different NEPs to the density dependence of the pressure, since the correlation furnishes the central pressure $p_c$ directly.

\begin{figure}[tb] 
\centering
\includegraphics[scale=0.36]{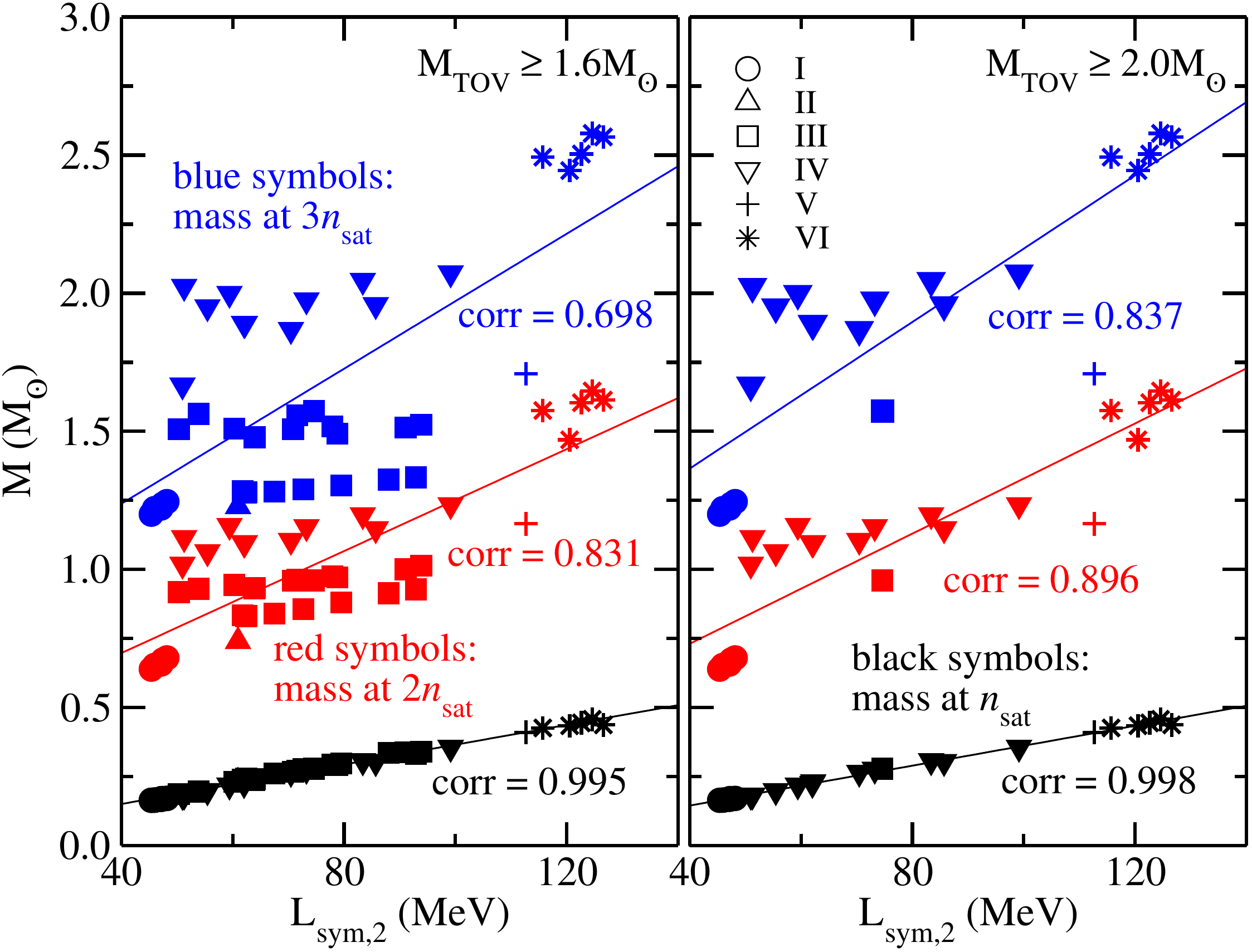}
\caption{Neutron star masses corresponding to central densities of $n_\sat$ (black symbols), $2n_\sat$ (red symbols), and $3n_\sat$ (blue symbols) as a function of $L_{\sym,2}$, obtained from the interactions of the D$_4$ group and restricted to those satisfying M$_\tov\geq 1.6$M$_\odot$ (left panel) and M$_\tov\geq 2.0$M$_\odot$ (right panel). Full lines: fitting curves with the respective correlation coefficients.}
\label{fig:ml}
\end{figure}

With regard to the correlation between NS mass and $L_{\sym,2}$, notice that in Fig.~\ref{fig:mr1} one observes the relation between the mass M at fixed central density and the radius, which reflects the influence of $L_{\sym,2}$. To make this clearer, we explicitly represent in Fig.~\ref{fig:ml} the correlation between the mass M and $L_{\sym,2}$ at different central densities: $n_\sat$, $2n_\sat$ and $3n_\sat$. The correlation is almost perfect at $n_\sat$ with the correlation coefficient being 0.995 (0.998) for M$_\tov\geq 1.6$M$_\odot$ (M$_\tov\geq 2.0$M$_\odot$). However, it becomes broader at higher densities. This reflects the role of other empirical parameters governing the density dependence of the EoS, as for instance $K_\sym$ or $Q_\sat$. It is also interesting to observe that the correlations are very close for $n_\sat$ and 2$n_\sat$ when conditioned by M$_\tov\geq 2.0$M$_\odot$ in comparison to the M$_\tov\geq 1.6$M$_\odot$ case, reflecting the weak impact of M$_\tov$ on this correlation. The same is not true for 3$n_\sat$, that presents a better correlation when the 2M$_\tov\geq 2.0$M$_\odot$ condition is applied.

\section{Other global properties of neutron stars}

In this last section of the paper, we analyse global properties of NS that have not yet been analysed, namely the moment of inertia and the tidal deformability.

\subsection{Moment of inertia}

In the low spin regime, as suggested by Hartle and Sharp~\cite{inertia1}, the rotation of a NS is much smaller than the Kepler frequency, allowing us to assume that the NS remains spherical. The moment of inertia is therefore expressed as~\cite{inertia1,inertia2}
\begin{eqnarray}
I = \frac{8\pi}{3}\int_0^Rdrr^4\epsilon\left(1 + \frac{p}{\epsilon}\right)\frac{\bar{\omega}}{\Omega}e^{\lambda-\Phi},
\end{eqnarray}
where $\bar{\omega}$ is the local spin frequency, which represents the correction from general relativity to the asymptotic angular momentum $\Omega$. The local angular momentum is $\omega=\Omega-\bar{\omega}$. Furthermore, $e^\lambda=[1-m(r)/r]^{-1/2}$ and $\Phi$ is the gravitational potential solution of the equation
\begin{eqnarray}
\frac{d\Phi(r)}{dr} = \frac{m(r) + 4\pi r^3p(r)}{r^2[1-2m(r)/r]} \, ,
\label{eq:phi}
\end{eqnarray}
with the boundary condition $\Phi(R)=\frac 1 2 \ln (1-2M/R)$.

\begin{figure}[tb] 
\includegraphics[width=0.55\textwidth,trim=170 45 0 350, clip=true]{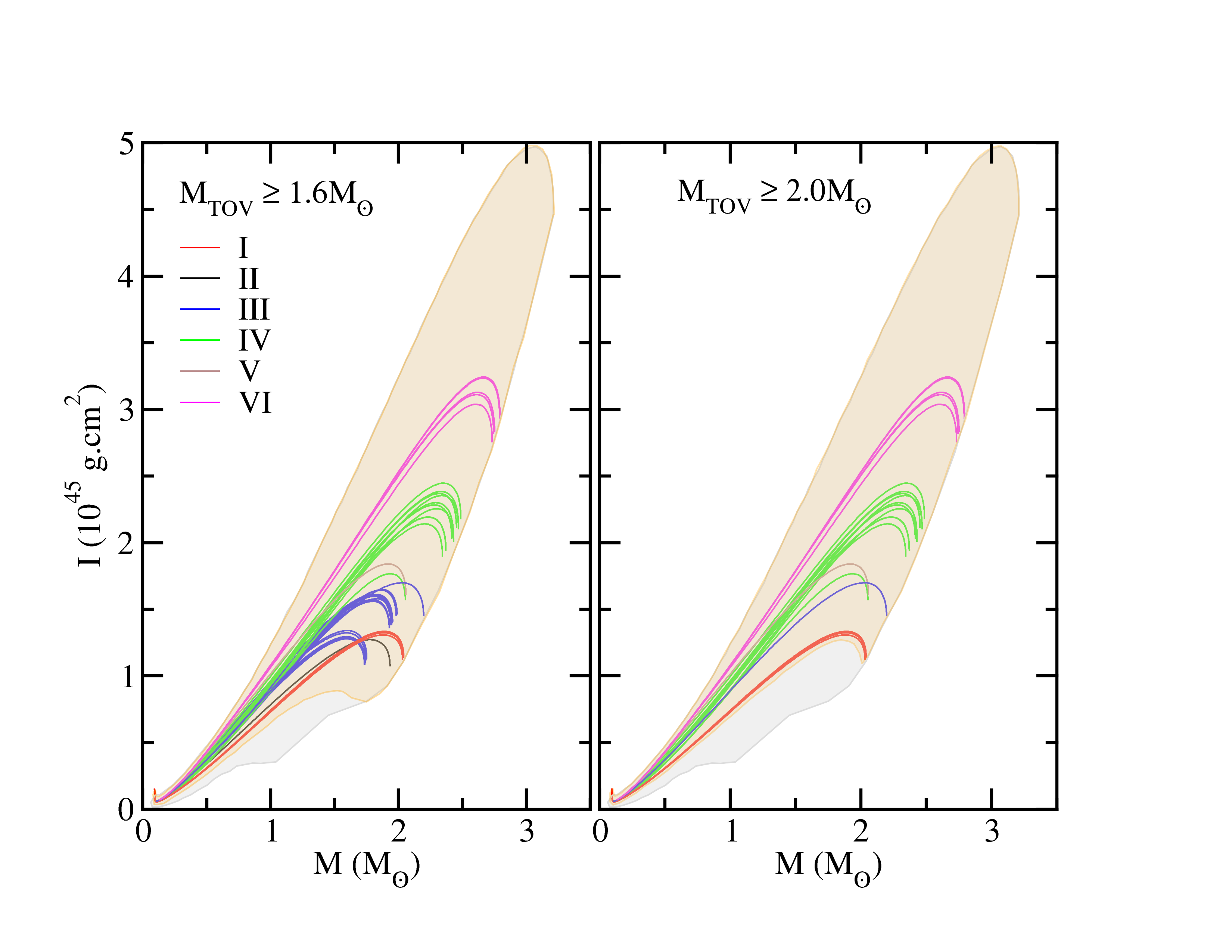}
\includegraphics[width=0.55\textwidth,trim=170 100 0 350, clip=true]{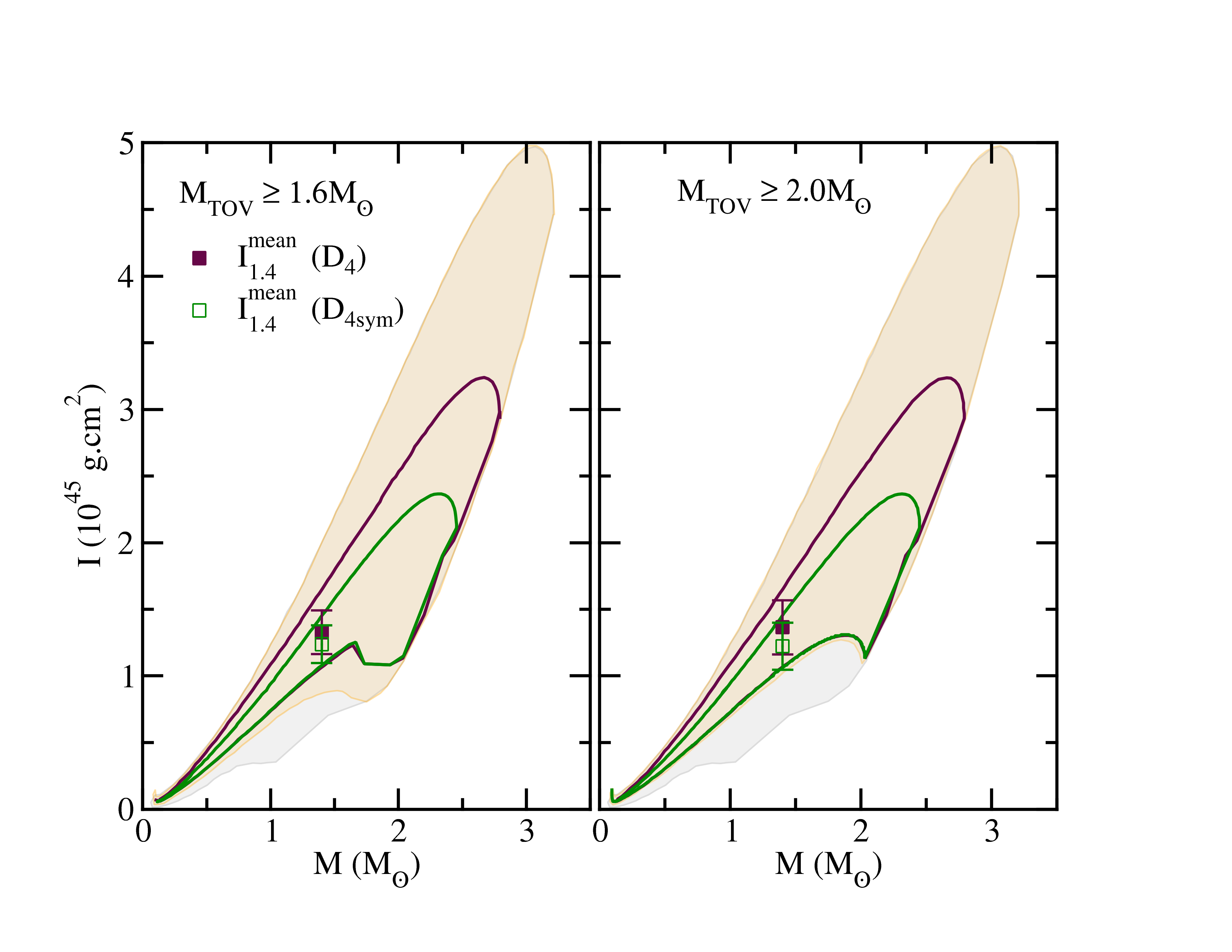}
\caption{$I$-$M$ correlation obtained from the D$_0$ group (grey band in all panels). Orange band: regions from the D$_0$ group conditioned by M$_\tov$. Top panels: the curves are from the subgroups of D$_4$ as given in Table~\ref{tab:bulk} and the legend is the same as in Fig.~\ref{fig:mr1}. Bottom panels: the dark brown contour is constructed from the interactions of the D$_4$ group conditioned by M$_\tov$ (as in the top panels) and the green contour delimits the predictions based on the D$_{4\sym}$ group.}
\label{fig:inertia}
\end{figure}

We then investigate how the moment of inertia is influenced by the low energy nuclear experimental constraints. We show in the top panels of Fig.~\ref{fig:inertia} the moment of inertia $I$ as a function of the mass M for the six sets, as indicated in the legend. There is a reasonable ordering of the moment of inertia as a function of the sets: the moment of inertia increases for increasing values of $L_{\sym,2}$. We obtain $I^{\rm{mean}}_{1.4}=1.33\pm0.16$ (1.24$\pm0.14$)~$10^{45}$g.cm$^2$ for the D$_4$ (D$_{4\sym}$) group for M$_\tov\geq 1.6$M$_\odot$ and $I^{\rm{mean}}_{1.4}=1.36\pm0.20$ (1.22$\pm0.18$)~$10^{45}$g.cm$^2$ for M$_\tov\geq 2.0$M$_\odot$. 

\begin{figure}[tb] 
\centering
\includegraphics[scale=0.35]{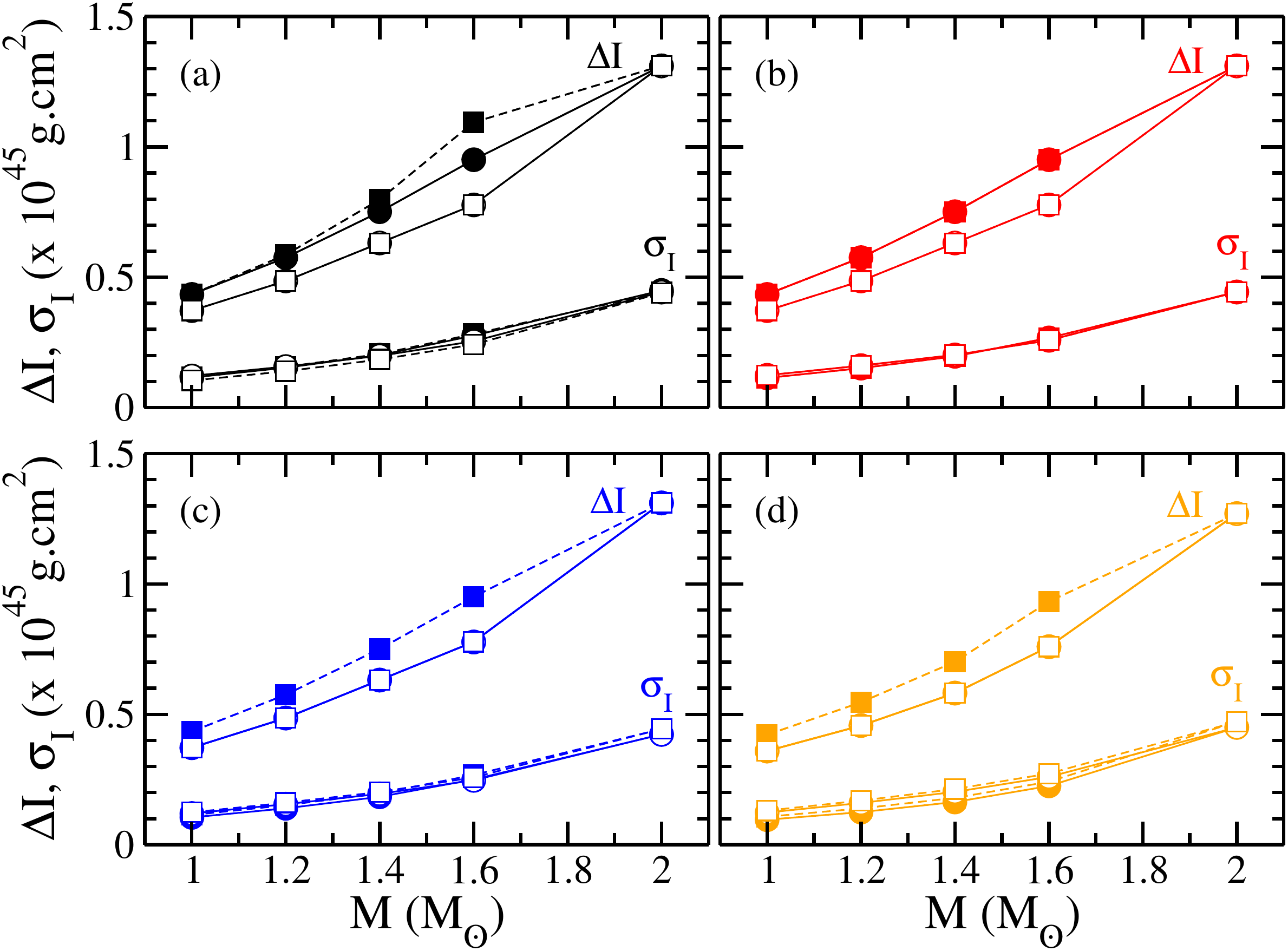}
\caption{$\Delta I$ (range) and $\sigma_I$ (standard deviation) as a function of the mass M, for the groups D$_i$ (solid lines) and G$_i$ (dashed lines) with $i=1$ (black), 2 (red), 3 (blue), and 4 (orange). The results correspond to those interactions satisfying M$_\tov\geq 1.6$M$_\odot$ (closed symbols) and M$_\tov\geq 2.0$M$_\odot$ (open symbols). See text for more details.}
\label{fig:statI}
\end{figure}

A more systematical investigation of the moment of inertia is shown in Fig.~\ref{fig:statI}, where the effects of the groups D$_1$-D$_4$ and G$_1$-G$_4$ are given in the four panels. As in Fig.~\ref{fig:statr}, we represent the largest uncertainties $\Delta I=I_{\mbox{\tiny max}}-I_{\mbox{\tiny min}}$ and the standard deviation $\sigma_I=(1/n)\sum_i (I_i-\left<I\right>)^2$. The uncertainty measured by these two quantities increases as a function of the mass M. We also confirm our previous conclusions from Fig.~\ref{fig:statr}: there is only a very limited impact due to a better description of the low-energy nuclear data. An improvement is seen when using the D$_i$ groups rather than the G$_i$ ones, showing that the condition to describe equally well the $N=Z$ and $N\neq Z$ nuclei also plays a role here. However, the uncertainty in the moment of inertia is generated by the unknown density dependence of the EoS, as we have already discussed in the case of the radius.

\subsection{Tidal deformability}

Finally, we address the question of the tidal deformability, which is probed by coalescing neutron stars and carried away by gravitational waves emitted during the last orbits before merger. We analyse its correlation with the experimental nuclear data. 

\begin{figure}[tb] 
\centering
\includegraphics[width=0.55\textwidth,trim=10 170 0 350, clip=true]{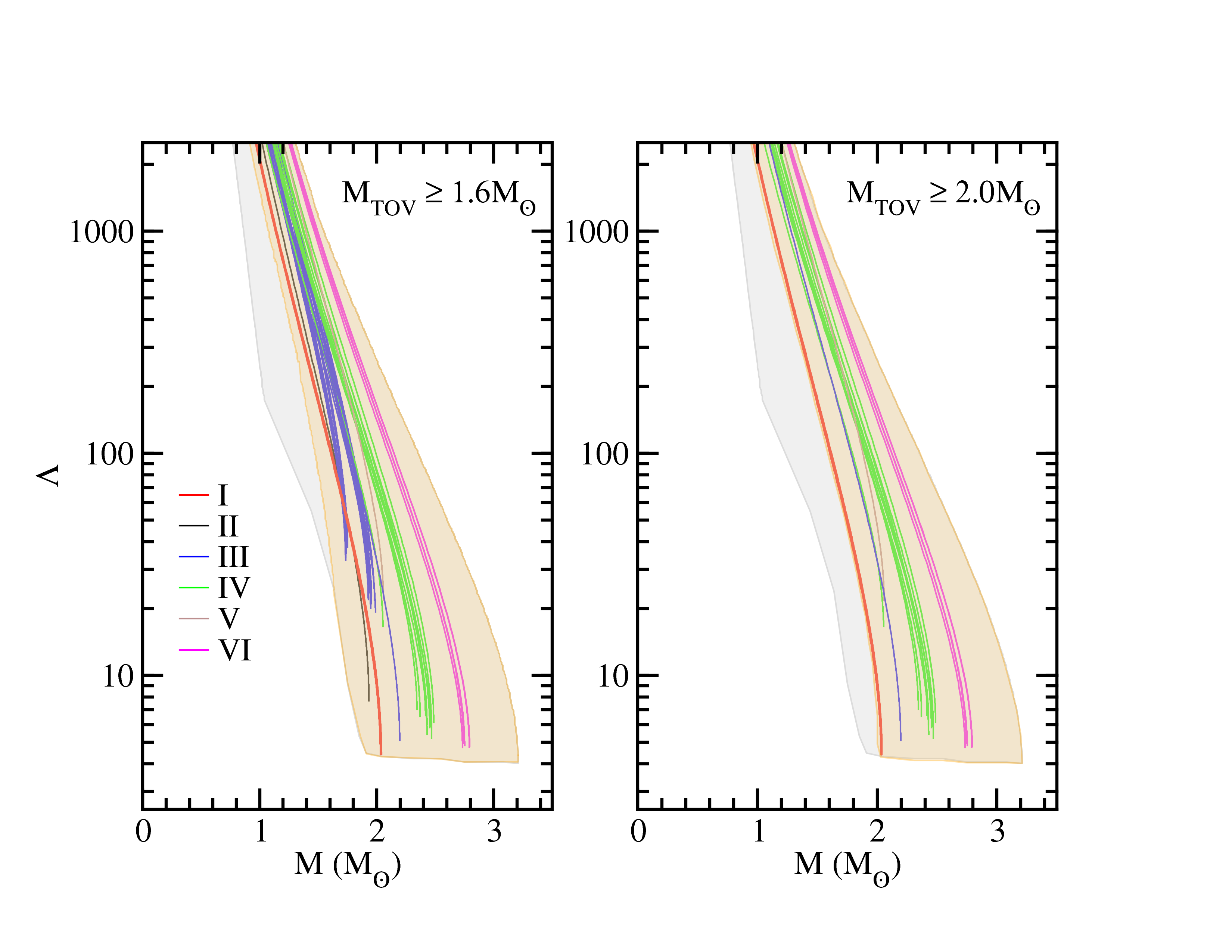}
\includegraphics[width=0.55\textwidth,trim=10 170 0 350, clip=true]{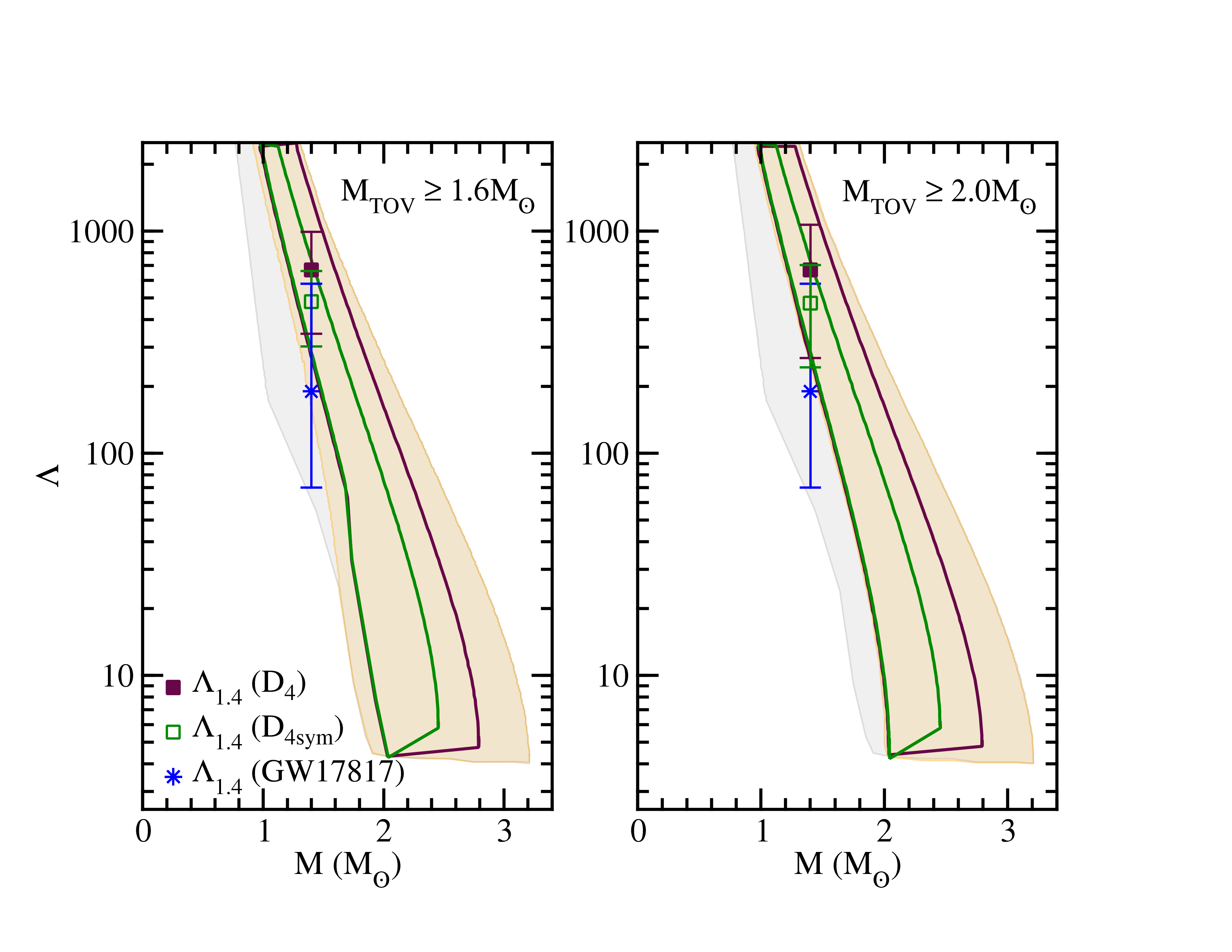}
\caption{$\Lambda$-M correlation for the group D$_0$ group (grey band in all panels) and for the group D$_0$ conditioned by M$_\tov$ (orange band). See Fig.~\ref{fig:inertia} for the description of the curves in the top panels and the contours in the bottom panels. The data for $\Lambda_{1.4}$ shown in the bottom panels is extracted from Ref.~\cite{ligo} by the LIGO/Virgo Collaboration (GW17817 event).} 
\label{fig:lambda}
\end{figure}

Tidal deformability is defined as the quotient of the induced quadrupole moment $Q_{ij}$ to the tidal field $\varepsilon_{ij}$. In terms of the second Love number $k_2$, it is given by $\lambda={2\over{3}} k_{2} R^{5}$. One can also define the dimensionless tidal deformability as $\Lambda = {2\over{3}} k_{2} (R/M)^5 \equiv {2\over{3}} k_{2} C^{-5}$ where $C$ is the compactness. The Love number $k_2$ is defined as,
\begin{eqnarray}
&k_2& =\frac{8C^5}{5}(1-2C)^2[2+2C(y_R-1)-y_R]\nonumber\\
&\times&\Big\{2C [6-3y_R+3C(5y_R-8)] \nonumber\\
&+& 4C^3[13-11y_R+C(3y_R-2) + 2C^2(1+y_R)]\nonumber\\
&+& 3(1-2C)^2[2-y_R+2C(y_R-1)]{\rm ln}(1-2C)\Big\}^{-1},\qquad
\label{k2}
\end{eqnarray}
with $y_R\equiv y(R)$ and $y(r)$ being the solution of the differential equation, 
\begin{align}
r\frac{dy}{dr} + y^2 + yF(r) + r^2Q(r) = 0,     
\label{dydr}
\end{align}
where the functions $F(r)$ and $Q(r)$ are given by
\begin{eqnarray}
F(r) &=& \frac{1 - 4\pi r^2[\epsilon(r) - p(r)]}{f(r)}, 
\\
Q(r)&=&\frac{4\pi}{f(r)}\left[5\epsilon(r) + 9p(r) + 
\frac{\epsilon(r)+p(r)}{c_s^2(r)}- \frac{6}{4\pi r^2}\right]
\nonumber\\ 
&-& 4\left[ \frac{m(r)+4\pi r^3 p(r)}{r^2f(r)} \right]^2,
\label{qr}
\end{eqnarray}
in which $c_s^2(r)=\partial p(r)/\partial\epsilon(r)$ is the square of the sound speed and $f(r)=1-2m(r)/r$~\cite{tanj10,Prakash,hind08,damour,tayl09,had4}. 

The dimensionless tidal deformability $\Lambda$ is shown as a function of M in Fig.~\ref{fig:lambda} for the six sets, as indicated in the legend. We can compare the envelop of the best EoS~(group D$_4$) with the prior from group D$_0$ (grey band). The orange band represents the envelop of the group D$_0$ conditioned by the constraint on M$_\tov$. As discussed previously for the M-R relation, as well as for the I-R one, there is only a small impact  of a better reproduction of the low-energy nuclear data, with most of the uncertainty originating from the unknown density dependence of the EoS. We also indicate the point reported by the LIGO/Virgo collaboration for the dimensionless tidal deformability of a canonical star, namely, $\Lambda_{1.4}=190^{+390}_{-120}$~\cite{ligo}. As we see, the softest interactions are more compatible with this specific restriction. We obtain $\Lambda^{\rm{mean}}_{1.4}=669\pm323$ (482$\pm179$) for the D$_4$ (D$_{4\sym}$) for M$_\tov\geq 1.6$M$_\odot$ and $\Lambda^{\rm{mean}}_{1.4}=760\pm 400$ (474$\pm 231$) for M$_\tov\geq 2.0$M$_\odot$.

\begin{figure}[tb]
\centering
\includegraphics[width=0.55\textwidth,trim=30 140 0 300, clip=true]{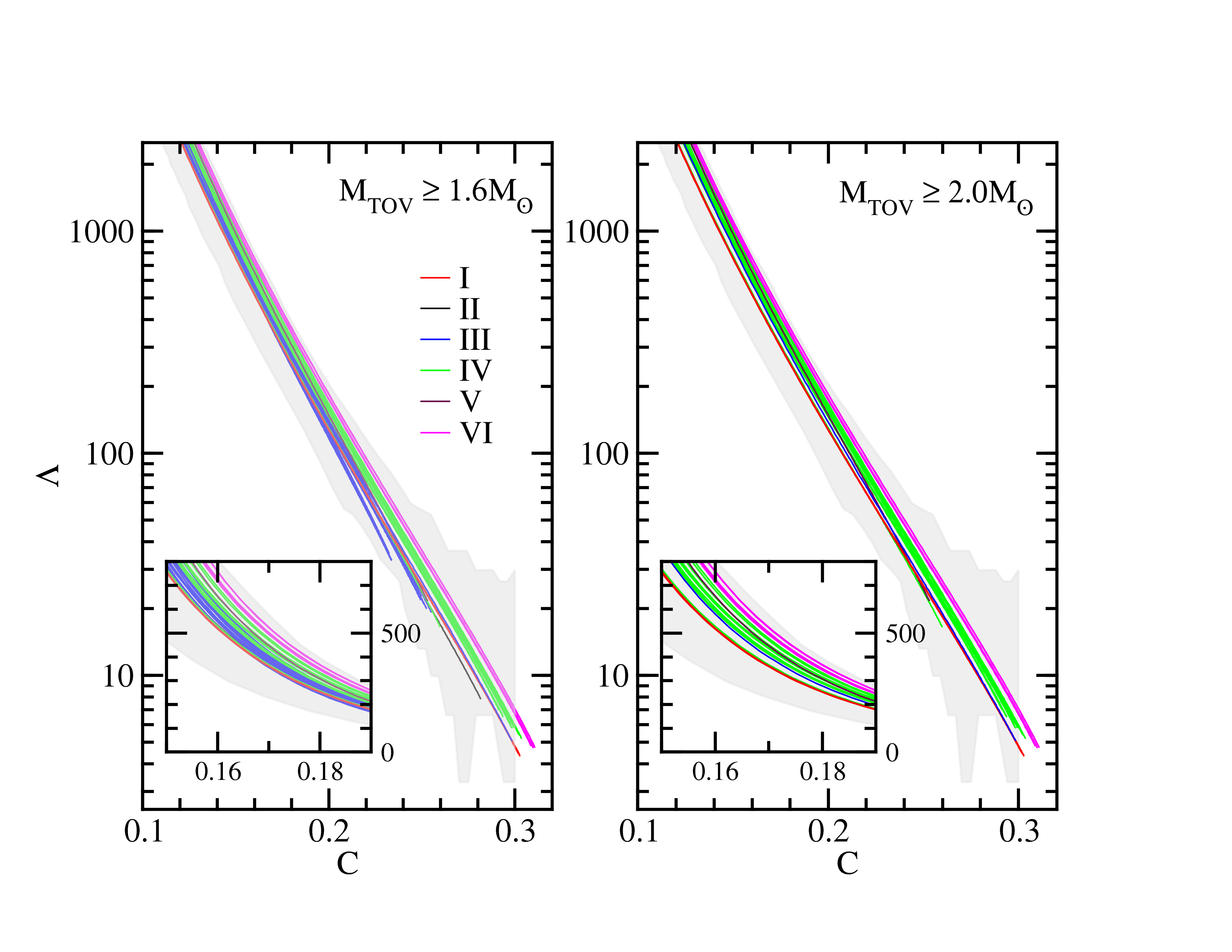}
\caption{Dimensionless tidal deformability $\Lambda$ as a function of the compactness $C=M/R$ for the D$_4$ group. The grey band represents the contour of the D$_0$ group (prior). The insets show a zoom of the curve in linear scale and for a small region of $C$.}
\label{fig:lambda-c}
\end{figure}

We investigate in Fig.~\ref{fig:lambda-c} the $\Lambda$-$C$ universal relation suggested in Ref.~\cite{Yagi2014}. The contributions for the six sets are shown in different colors. We confirm the universal relation, and we show that the dispersion in this relation is mainly given by $L_{sym,2}$. In addition, the dispersion is even larger when all EDFs in the D$_0$ group are considered. The origin of the small dispersion is thus due to the interaction selection of the G$_i$/D$_i$ groups. We have checked that an accurate description of the low-energy nuclear data by the EDFs is also less important here than an improved determination of the density dependence of the EoS.

\begin{figure}[tb]
\centering
\includegraphics[scale=0.36]{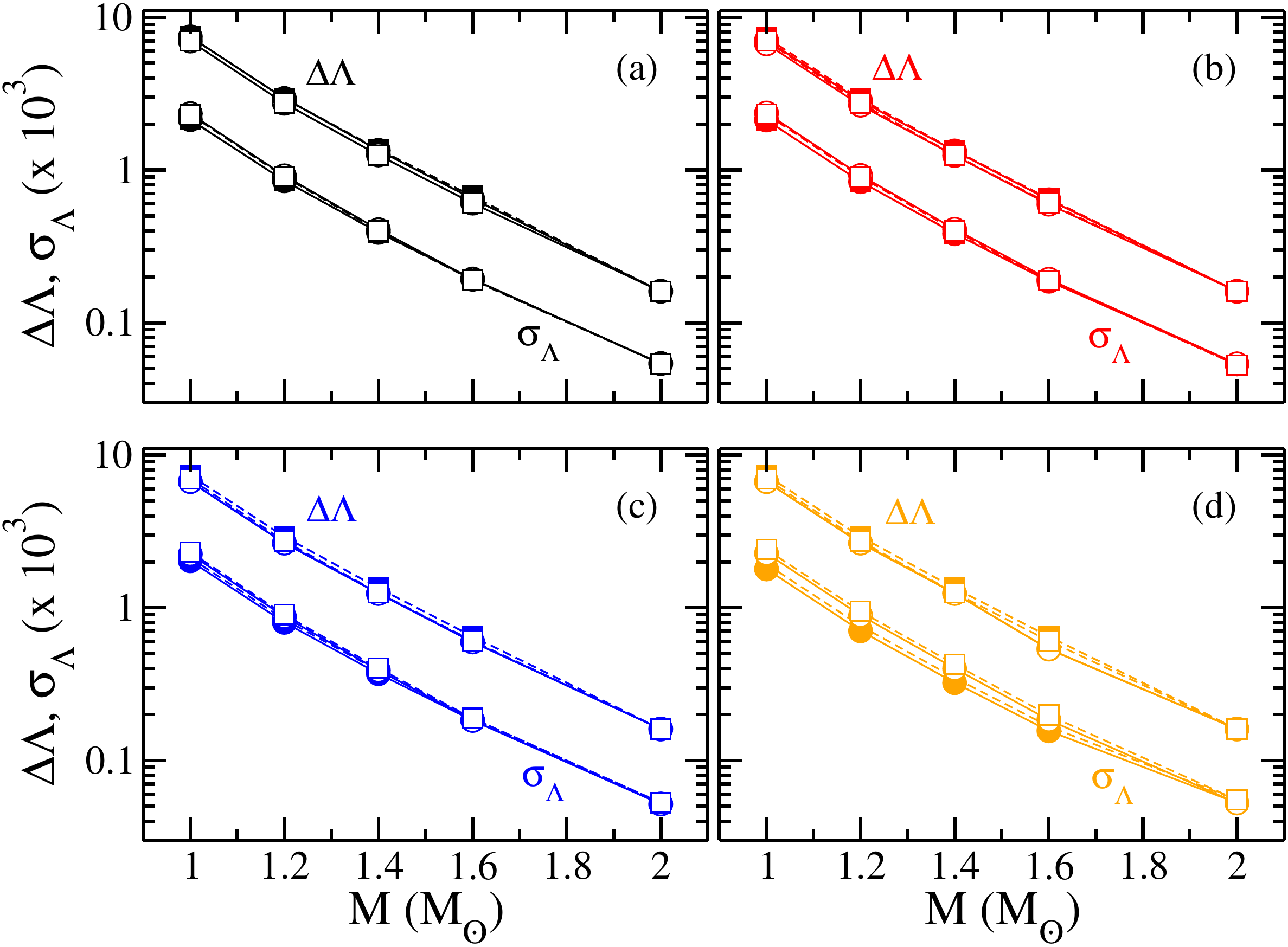}
\caption{Uncertainties in $\Lambda$ represented by the largest $\Lambda$ uncertainty $\Delta\Lambda$ and the standard deviation $\sigma_\Lambda$ for each groups G$_i$ and D$_i$ with $i=1$ (black), 2 (red), 3 (blue), and 4 (orange). The results correspond to those interactions satisfying M$_\tov\geq 1.6$M$_\odot$ (closed symbols) and M$_\tov\geq 2.0$M$_\odot$ (open symbols).}
\label{statL}
\end{figure}

In Fig.~\ref{statL} we present the maximal uncertainty and the standard deviation related to the dimensionless tidal deformability, defined similarly to the moment of inertia and to the radius shown before. We clearly see a reduction of $\Delta\Lambda=\Lambda_{\mbox{\tiny max}}-\Lambda_{\mbox{\tiny min}}$ and $\sigma_\Lambda$ as a function of the neutron star mass, at variance with the analysis of R and I. This is related to the fact that $\Lambda$ is strongly decreasing with M, as shown in Fig.~\ref{fig:lambda}. The correlation of the experimental nuclear data with the tidal deformability is extremely small. Among the quantities we have investigated in this study, the tidal deformability is perhaps the quantity on which the constraints provided by the experimental nuclear data has the smallest impact.

\begin{figure}[tb]
\centering
\includegraphics[scale=0.35]{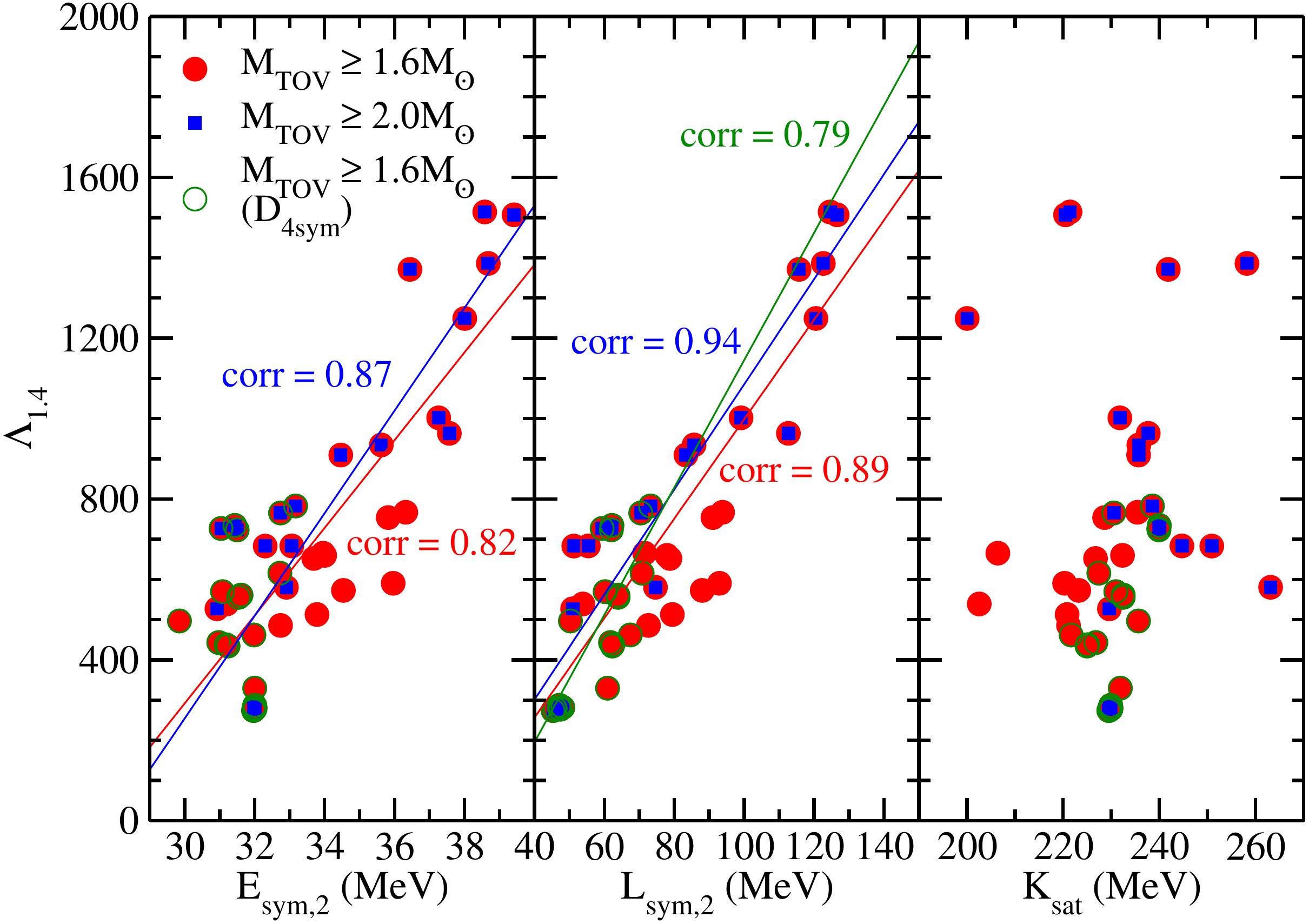}
\caption{Dimensionless tidal deformability $\Lambda_{1.4}$ as a function of $E_{\sym,2}$ (left), $L_{\sym,2}$ (middle), and $K_\sat$ (right) for the D$_4$ group conditioned by M$_\tov$. We also present results for the D$_{4\sym}$ subgroup restricted to M$_\tov\geq 1.6M_\odot$. Full lines: fitting curves with the respective correlation coefficients.}
\label{fig:lambdaxsat}
\end{figure}

Finally, we show in Fig.~\ref{fig:lambdaxsat} the dimensionless tidal deformability of a \mbox{M=$1.4$M$_\odot$} star, namely, $\Lambda_{1.4}$, as a function of the NEPs $E_{\sym,2}$, $L_{\sym,2}$, and $K_\sat$. This figure is similar to Fig.~5 from Ref.~\cite{Wei2020}, which showed no correlation for a reduced set of nuclear interactions. From this figure, the authors of Ref.~\cite{Wei2020} concluded that there is no correlation between global properties of NS and saturation properties of nuclear matter. We find that a correlation does indeed exist, when considering the D$_4$ group conditioned by M$_\tov$. We find a good correlation for $\Lambda_{1.4}$-$E_{\sym,2}$ and $\Lambda_{1.4}$-$L_{\sym,2}$ for the D$_4$ group with both constraints on M$_\tov$. In the case of the D$_{4\sym}$ subgroup, we see that the correlation of $E_{\sym,2}$ with $\Lambda_{1.4}$ n longer exists, but for $L_{\sym,2}$ there is still a correlation, although smaller than for the D4 group. Concerning  $\Lambda_{1.4}$-$K_\sat$, we find the Pearson correlation coefficients to be smaller than $0.6$ in all cases. We thus conclude that correlations do exist between global properties of NSs and saturation properties, especially for the isovector QNEPs, although there is a large dispersion in these correlations, which limits the value of the Pearson correlation coefficient.

\section{Conclusions}

In this study, we have analysed the link between the constraints on mean field EDFs generated by low-energy nuclear experimental data and their corresponding predictions for NSs. To do so, we have investigated 415 mean field interactions, both relativistic and non-relativistic, for which we have calculated several quantities that can be directly compared to the experimental data. These quantities are the masses, radii and GMR energies of a number of doubly magic nuclei (chosen to minimize the impact of uncontrolled approximations such as pairing, deformation, etc). We have defined five groups, from G$_0$ to G$_4$, where G$_0$ is the set of interactions reproducing the experimental nuclear masses with the largest tolerance, G$_2$ with the smaller tolerance, while G$_3$ and G$_4$ add successively the constraint on the charge radius and the giant monopole resonance. In these groups, we have evaluated the reproduction of the experimental data globally. They are contrasted with another set of groups, called D$_0$ to D$_4$, for which a more detailed evaluation is performed by separating the $N=Z$ nuclei from the others: To be well ranked in the groups D$_i$, the interactions must reproduce equally well the $N=Z$ and $N \ne Z$ nuclei. From this first step of our analysis, we find that
\begin{enumerate}
\item The group D$_4$ exhibits a fairly strong correlation between $E_{\sym,2}$ and $L_{\sym,2}$.
\item By combining the low-energy nuclear data and an analysis of the density dependence of the symmetry energy~\cite{Danielewicz2013}, we have isolated a group D$_{4\sym}$ that further reduces the uncertainty in the symmetry energy. We find $E_{\sym,2}=31.8\pm0.7$~MeV and $L_{\sym,2}=58.1\pm 9.0$~MeV.
\end{enumerate}

In a second step, we have confronted the different groups G$_i$ and D$_i$ with global observational quantities related to stable NSs, such as radii, moments of inertia and tidal deformabilities. We have compared the priors, identified as the G$_0$ or D$_0$ groups, which include all viable EoSs, with the best interactions of the groups G$_4$ and D$_4$. 
From this comparison, we find that 
\begin{enumerate}[resume]
\item The selection of interactions according to their adequacy in reproducing the experimental nuclear data has a weak impact on the reduction of uncertainties of global NS properties 
with masses around or above the canonical one. This reveals that the density dependence of the EoS is not constrained by precision measurements of low-energy nuclear data.
\item The selection of the groups D$_i$ has a greater impact on the results than the selection of the groups G$_i$, showing the importance of having control over the isotopic predictions of the interactions. The charge radius plays an important role in this selection.
\item The 1.4M$_\odot$ neutron star (NS) radius lies between 12 and 14~km for the ``better'' nuclear interactions. 
\item To a large degree, the density dependence of the symmetry energy  explains the observed dispersion in NS properties, so that a more detailed knowledge of the symmetry energy should result in a reduction of the uncertainties in NS radii, at least for canonical to low-mass NS, where there is no phase transition to exotic matter.
\end{enumerate}

The fourth point is not surprising, since NS matter is an extrapolation of current nuclear interactions towards large isospin asymmetries. The third point, however, is a bit more surprising. It tells us that the constraints of experimental nuclear data near saturation density are only weakly correlated with the behavior of the EoS at several times saturation density. This confirms the conclusions of Ref.~\cite{Margueron2018b}, where the uncertainties in the extrapolation of the nucleonic EoS are found to be fairly uncontrolled above the densities at which experimental data exist. We therefore emphasize that the reduction of the uncertainties in NS global properties will not originate from better data related to low-energy nuclear physics, since the density or energy region for which constraints are required is outside the reach of standard nuclear physics. 

The experimental data on the symmetry energy is found to be much more constraining, however. We find that the slope of the symmetry energy $L_{\sym,2}$ near saturation density is well correlated with NS radii and masses. We also observe that the experimental data on IAS+$\Delta r_{np}$ (group D$_{4\sym}$) has a large impact on further selection among our best set of interactions D$_4$. 
The D$_{4\sym}$ group furnishes, for the most part, values of $L_{\sym,2}\leq 90$~MeV, depending on the value of $K_{\sym,2}$. It also determines boundaries for a few QNEPs, $L_{\sym,2}$, $K_{\sym,2}$, and $Q_{\sym,2}$, which are tighter than the ones proposed in Ref.~\cite{Margueron2018a}.

In conclusion, we have shown in our analysis that the experimental nuclear masses, radii and GMR energies of a set of doubly magic nuclei show little correlation with the properties of nucleonic matter at several times saturation density. The experimental data related to the symmetry energy, however, are somewhat better correlated with these properties. In the future, we plan to perform a complementary analysis including data from heavy-ion collision exploring densities above $n_\sat$, the saturation density of nuclear matter. This appears to be a necessary condition for making substantial progress on the understanding of the properties of dense nuclear matter.

\begin{acknowledgements}
This work is supported by the project INCT-FNA proc. No. 464898/2014-5, as well as by the Conselho Nacional de Desenvolvimento Cient\'ifico e Tecnol\'ogico (CNPq) under Grants No. 303131/2021-7 (B.V.C.), 312410/2020-4 (O.L.), and 433369/2018-3 and 308528/2021-2 (M.D.). We also acknowledge support from the Funda\c{c}\~ao de Amparo \`a Pesquisa do Estado de S\~ao Paulo (FAPESP) under Thematic Project 2017/05660-0 (O.L., M.D., B.V.C) and Project 2020/05238-9 (O.L., M.D.). 
J.M. is supported by the CNRS/IN2P3 NewMAC project, and benefits from PHAROS COST Action MP16214 and the LABEX Lyon Institute of Origins (ANR-10-LABX-0066) of the \textsl{Universit\'e de Lyon}.
The authors acknowledge support from the CNRS International Research Project ``SUBATOMICS", which enabled a visit by J.M. to ITA, where part of this work was carried out, and we thank LAB-CCAM from ITA for computational support.
\end{acknowledgements}

\appendix

\section{Non-relativistic spin-orbit parameters}
\label{ap:sof}

For a Skyrme-type force the mean-field Hamiltonian for neutrons reads
\begin{equation}
\hat{h} = \hat{p} B \hat{p} + U + W\cdot(\sigma\times\hat{p})\, ,
\end{equation}
where the mean-field $U$, the inverse mass $B$ and the spin-orbit potential $W$ are local functionals of the neutron and proton densities $n_n$ and $n_p$.
The density dependence is linear for the mass and spin-orbit potentials, i.e.
\begin{eqnarray}
B &=& \frac{\hbar^2}{2m} + b_1 n + b_1^\prime n_n \, , \\
W &=& \frac 1 2 W_0 ( \nabla n + \nabla n_n) \, ,
\end{eqnarray}
and for interactions with generalized spin-orbit interactions~\cite{Reinhard1995}
\begin{eqnarray}
W &=& b_4 \nabla n + b_4^\prime \nabla n_n \, .
\end{eqnarray}
For standard Skyrme interactions, the ratio $b_4^\prime/b_4$ is unity, while it is zero for $\sigma\omega$-RMF.
 
We can link the parameters $b_4$ and $b_4^\prime$ to the EDF coefficients $C_t^{\nabla J}$~\cite{Bender2003} as
\begin{equation}
C_0^{\nabla J} = -\frac 3 4 W_0, , \hspace{0.5cm} C_1^{\nabla J} = - \frac 1 4 W_0 \, .
\end{equation}
and for interactions with generalized spin-orbit interaction,
\begin{equation}
C_0^{\nabla J} = -b_4 - \frac 1 2 b_4^\prime\, , \hspace{0.5cm} C_1^{\nabla J} = - \frac 1 2 b_4^\prime \, .
\end{equation}

\section{Spin-orbit matrix elements}
\label{ap:so}

We have
\begin{eqnarray}
\left\langle \vec{\sigma}\cdot\vec{l}\right\rangle _{lj} &= j\left(j+1\right)-l\left(l+1\right)-3/4
\nonumber\\
&= \left\{ \begin{array}{ll} l, & j=l+1/2\\
-\left(l+1\right), & j=l-1/2
\end{array}\right.
\end{eqnarray}
so that 
\begin{align}
\left(2j+1\right)\left\langle \vec{\sigma}\cdot\vec{l}\right\rangle _{lj}=\left\{ \begin{array}{ll}
2l\left(l+1\right) & \qquad j=l+1/2\\
-2l\left(l+1\right) & \qquad j=l-1/2
\end{array}\right.\,.
\end{align}
Since the occupation numbers of the nuclei under consideration are
$v_{nlj\tau}=1$, we see that the only contributions to the spin orbit
sum will come from filled $j=l+1/2$ orbitals for which the higher
energy $j=l-1/2$ orbital is empty. We tabulate these values in Table~\ref{spinorbit}. 

\begin{table}[!htb]
\centering
\setlength{\tabcolsep}{1pt}
\renewcommand{\arraystretch}{1.3}
\caption{Spin orbit quantities.}
\begin{ruledtabular}
\begin{tabular}{rrrcccc}
  $Z$ & $N$ & nucleus & ~proton &  neutron  &
  $\sum_{nlj\tau} l\left(l+1\right) \mu_\tau $ &
  $\langle \Delta R_{ch}^{so}\rangle^2$ \\
& & & & & & (fm$^2$) \\
\hline
8  &  8  & $^{16}$O   & - & - & 0 & ~0.000\\ 
14 & 20  & $^{34}$Si  & 1d5/2 & - & $12\mu_p$ & ~0.026 \\
20 & 20  & $^{40}$Ca  & - & - & 0 & ~0.000 \\
20 & 28  & $^{48}$Ca  & - & 1f7/2 & $24\mu_n$ & -0.025\\
20 & 32  & $^{52}$Ca  & - & 1f7/2,2p3/2 & $28\mu_n$ & -0.030\\
20 & 34  & $^{54}$Ca  & - & 1f7/2 & $24\mu_n$ & -0.025 \\
28 & 20  & $^{48}$Ni  & 1f7/2 & - & $24\mu_p$ & ~0.026 \\
28 & 28  & $^{56}$Ni  & 1f7/2 & 1f7/2 & $24\mu_p +24\mu_n$ & ~0.008 \\
28 & 50  & $^{78}$Ni  & 1f7/2 & 1g9/2 & $24\mu_p +40\mu_n$ & -0.004\\
40 & 50  & $^{90}$Zr  & - & 1g9/2 & $40\mu_n$ & -0.021\\
50 & 50  & $^{100}$Sn & 1g9/2 & 1g9/2 & $40\mu_p+40\mu_n$ & ~0.008 \\
50 & 82  & $^{132}$Sn & 1g9/2 & 1h11/2 & $40\mu_p+60\mu_n$ & -0.001 \\ 
82 & 126 & $^{208}$Pb & 1h11/2 & 1i13/2 & $60\mu_p+84\mu_n$ & ~0.001 \\
\end{tabular}
\end{ruledtabular}
\label{spinorbit}
\end{table}

The smallest radius in the Table~\ref{spinorbit} that we compare to experimental data and for which $\langle \Delta R_{ch}^{so}\rangle^2$ is nonzero  is that of $^{34}$Si. Taking its charge radius to be $R_{ch}(^{34}$Si$)\approx 3.2 $~fm, we find the relative change to be
\[\frac{\Delta R_{ch}}{R_{ch}}\approx \frac 1 2 \frac{\Delta \langle R_{ch}^{so}\rangle^2}{R_{ch}^2} \approx -0.001. \]
The effect is thus quite small and will not make a perceptible difference in our results. Kurasawa and Suzuki have shown that the spin-orbit correction is nonzero in the relativistic case, even for closed shell nuclei~\cite{Kurasawa2019}. However, the size of the correction is still of the same order of magnitude of those shown here.

\section{Dense matter equations of state}
\label{sec:densematter}

We briefly review the equations of state for dense matter and neutron stars at beta equilibrium.

\subsection{Skyrme and relativistic mean field models}

For the Skyrme model, the energy density (including the nucleon rest mass) and pressure are given by, 
\begin{align}
\rho_{\mbox{\tiny sky}} &= \frac{3}{10M_{\mbox{\tiny nuc}}}\left(\frac{3\pi^2}{2}\right)^{2/3}n^{5/3}H_{5/3} \nonumber\\
&+ \frac{t_0}{8}n^2[2(x_0+2)-(2x_0+1)H_2] \nonumber \\
&+ \frac{1}{48}\sum_{i=1}^{2}t_{3i}n^{\sigma_{i}+2} [2(x_{3i}+2)-(2x_{3i}+1)H_2]
\nonumber\\
&+ \frac{3}{40}\left(\frac{3\pi^2}{2}\right)^{2/3}n^{8/3}\left(aH_{5/3}+bH_{8/3}\right) + nM_{\mbox{\tiny nuc}}
\label{desky}
\end{align}
and
\begin{align}
p_{\mbox{\tiny sky}}&= \frac{1}{5M_{\mbox{\tiny nuc}}}\left(\frac{3\pi^2}{2}\right)^{2/3}n^{5/3}H_{5/3} \nonumber\\
&+\frac{t_0}{8}n^2[2(x_0+2)-(2x_0+1)H_2]\nonumber\\
&+\frac{1}{48}\sum_{i=1}^{2}t_{3i}(\sigma_i+1)n^{\sigma_i+2}
[2(x_{3i}+2)-(2x_{3i}+1)H_2]\nonumber\\
&+\frac{1}{8}\left(\frac{3\pi^2}{2}\right)^{2/3}n^{8/3}\left(aH_{5/3}+bH_{8/3}\right)
\label{pressky}
\end{align}
with
\begin{align}
a&=t_1(x_1+2)+t_2(x_2+2),\\
b&=\frac{1}{2}\left[t_2(2x_2+1)-t_1(2x_1+1)\right],
\label{eq:b} \\
H_l(y)&=2^{l-1}[y^l+(1-y)^l].
\end{align}
where $y=Z/A=n_p/n$ is the proton fraction of the system, and $M_{\mbox{\tiny nuc}}$ is the nucleon rest mass. The set of constants $x_0$, $x_1$, $x_2$, $x_{31}$, $x_{32}$, $t_0$, $t_1$, $t_2$, $t_{31}$, and $t_{32}$ defines a particular parametrization of the interaction. The nucleon chemical potential is obtained as follows
\begin{align}
&\mu_q^{\mbox{\tiny sky}}(n,y) = \frac{\partial\mathcal{E}_{\mbox{\tiny sky}}}{\partial n_q} =
\frac{1}{2M_{\rm nuc}}\left(\frac{3\pi^2}{2}\right)^{2/3}n^{2/3}H_{5/3}(y)
\nonumber\\
&+ \frac{1}{5}\left(\frac{3\pi^2}{2}\right)^{2/3}n^{5/3}[aH_{5/3}(y) + bH_{8/3}(y)]
\nonumber\\
&+\frac{t_0}{4}n[2(x_0+2)-(2x_0+1)H_2(y)]
\nonumber\\
&+ 
\frac{1}{48}\sum_{i=1}^{2}t_{3i}(\sigma_i+2)n^{\sigma_i+1}[2(x_{3i}+2)-(2x_{3i}
+1)H_2(y)]
\nonumber\\
&\pm \frac{1}{2}\left[1 \mp 
(2y-1)\right]\left\{\frac{3}{10M_{\rm nuc}}\left(\frac{3\pi^2}{2}\right)^{2/3} 
n^{2/3}H'_{5/3}(y) \right.
\nonumber\\
&\left.-\frac{t_0}{8}n(2x_0+1)H'_2(y) 
-\frac{1}{48}\sum_{i=1}^{2}t_{3i}n^{\sigma_i+1}(2x_{3i} +1)H'_2(y) \right.
\nonumber\\
&+\left. \frac{3}{40} 
\left(\frac{3\pi^2}{2}\right)^{2/3}n^{5/3}[aH'_{5/3}(y) + bH'_{8/3}(y)]\right\},
\label{muqsk}
\end{align}
where $q=p,n$ stands for protons and neutrons, respectively, and $H'_l(y)=dH_l/dy$.

For the relativistic mean field (RMF) model, the Hartree approximation to the $T_{00}$ and $T_{ii}/3$ components of the energy-momentum tensor leads to
\begin{align}
&\rho_{\mbox{\tiny RMF}} = \frac{1}{2}m^2_\sigma\sigma^2 
+ \frac{A}{3}\sigma^3 + \frac{B}{4}\sigma^4 - \frac{1}{2}m^2_\omega\omega_0^2 
- \frac{C}{4}(g_\omega^2\omega_0^2)^2 \nonumber\\
&- \frac{1}{2}m^2_\rho\bar{\rho}_{0(3)}^2
+g_\omega\omega_0n+\frac{g_\rho}{2}\bar{\rho}_{0(3)}n_3 + \rho_{\mbox{\tiny kin}}^p + \rho_{\mbox{\tiny kin}}^n
\nonumber \\
&+ \frac{1}{2}m^2_\delta\delta^2_{(3)} - g_\sigma g_\omega^2\sigma\omega_0^2
\left(\alpha_1+\frac{1}{2}{\alpha'_1}g_\sigma\sigma\right) \nonumber\\
&- g_\sigma g_\rho^2\sigma\bar{\rho}_{0(3)}^2 
\left(\alpha_2+\frac{1}{2}{\alpha'_2} g_\sigma\sigma\right) - \frac{1}{2}{\alpha'_3}g_\omega^2 g_\rho^2\omega_0^2\bar{\rho}_{0(3)}^2,
\label{dermf}
\end{align}
and
\begin{align}
&p_{\mbox{\tiny RMF}} = - \frac{1}{2}m^2_\sigma\sigma^2 - \frac{A}{3}\sigma^3 -
\frac{B}{4}\sigma^4 + \frac{1}{2}m^2_\omega\omega_0^2 + \frac{C}{4}(g_\omega^2\omega_0^2)^2 \nonumber\\
&+ \frac{1}{2}m^2_\rho\bar{\rho}_{0(3)}^2 
+ p_{\mbox{\tiny kin}}^p + p_{\mbox{\tiny kin}}^n
\nonumber\\
&-\frac{1}{2}m^2_\delta\delta^2_{(3)} + g_\sigma g_\omega^2\sigma\omega_0^2
\left(\alpha_1+\frac{1}{2}{\alpha'_1}g_\sigma\sigma\right)
\nonumber\\
&+ g_\sigma g_\rho^2\sigma\bar{\rho}_{0(3)}^2 
\left(\alpha_2+\frac{1}{2}{\alpha'_2} g_\sigma\sigma\right)
+ \frac{1}{2}{\alpha'_3}g_\omega^2g_\rho^2\omega_0^2\bar{\rho}_{0(3)}^2,
\label{presrmf}
\end{align}
with the following kinetic energy terms,
\begin{align}
\rho_{\mbox{\tiny kin}}^{p,n}&=\frac{1}{\pi^2}\int_0^{{k_F}_{p,n}}
\hspace{-0.2cm}k^2(k^2+M^{*2}_{p,n})^{1/2}dk,
\\
p_{\mbox{\tiny kin}}^{p,n} &=
\frac{1}{3\pi^2}\int_0^{{k_F}_{p,n}}
\hspace{-0.5cm}\frac{k^4dk}{(k^2+M^{*2}_{p,n})^{1/2}},
\end{align}
in which the Fermi momentum of the nucleon is ${k_F}_{p,n}$, and $n_3=n_p-n_n$. From the energy density, we obtain the chemical potentials for protons and neutrons as
\begin{align}
\mu_{p,n}^{\mbox{\tiny RMF}} &= (k_{Fp,n}^2+{M_{p,n}^*}^2)^{1/2} + g_\omega\omega_0 \pm \frac{g_\rho}{2}\bar{\rho}_{0(3)},
\end{align}
with $+$ ($-$) for protons (neutrons). The meson masses are $m_\sigma$, $m_\omega$, $m_\rho$, and $m_\delta$. In parametrizations of the interaction that include the $\delta$ meson, the effective masses of protons and neutrons are different,
\begin{align}
M_{p,n}^*=M_{\mbox{\tiny nuc}}-g_\sigma\sigma\mp g_\delta\delta_{(3)},
\label{emrmf}
\end{align}
with $-$ ($+$) for protons (neutrons). In all of the above expressions, $\sigma$, $\omega_0$, $\bar{\rho}_{0(3)}$ and $\delta_{(3)}$ are the ``classical'' fields of the model. Furthermore, $g_\sigma$, $g_\omega$, $g_\rho$, $g_\delta$, $A$, $B$, $C$, $\alpha_1$, $\alpha'_1$, $\alpha_2$, $\alpha'_2$ and $\alpha'_3$ are the coupling constants of the interaction. For the case of the density-dependent version of the model, similar expressions are found by taking $g_\sigma\to\Gamma_\sigma(n)$, $g_\omega\to\Gamma_\omega(n)$, $g_\rho\to\Gamma_\rho(n)$, $g_\delta\to\Gamma_\delta(n)$, and $A=B=C=\alpha_1=\alpha'_1=\alpha_2=\alpha'_2=\alpha'_3=0$ in Eqs.~(\ref{dermf})-(\ref{emrmf}). Additionally, in this case, the pressure also contains a term given by $n\Sigma_R(n)$, with $\Sigma_R(n)$ (the rearrangement term) defined in terms of the field derivatives with respect to the density. Likewise, $\Sigma_R(n)$ is also added to the proton/neutron chemical potential in this case. We refer the reader to Ref.~\cite{Dutra2014} for more details regarding the thermodynamics of the RMF model (in infinite matter), as well as for the explicit expressions of the field equations.

\subsection{Equation of state for neutron star matter}

Cold catalyzed NSs are at $\beta$ equilibrium and are locally charge neutral. These conditions impose the following equilibrium equations between nucleons ($n$ and $p$) and leptons ($e$ and $\mu$): $\mu_n-\mu_p=\mu_e=\mu_\mu$ and $n_p=n_e+n_\mu$. 

The total energy density and pressure are then given by 
\begin{align}
\rho_\tot = \rho_\nuc + \frac{\mu_e^4}{4\pi^2} 
+ \frac{1}{\pi^2}\int_0^{\sqrt{\mu_\mu^2-m^2_\mu}}\hspace{-0.6cm}dk\,k^2(k^2+m_\mu^2)^{1/2},
\label{eq:totaled}
\end{align}
and
\begin{align} 
p_\tot = p_\nuc + \frac{\mu_e^4}{12\pi^2} +\frac{1}{3\pi^2}\int_0^{\sqrt{\mu_\mu^2-m^2_\mu}}\hspace{-0.5cm}\frac{dk\,k^4}{(k^2+m_\mu^2)^{1/2}},
\label{eq:totalp}
\end{align}
where $\rho_\nuc$ and $p_\nuc$ represent the nucleonic contribution to the energy density and to the pressure. They are described by the non-relativistic Skyrme model (sky) or the relativistic mean field (RMF) previously described in Section~\ref{sec:densematter}.
Furthermore, in the ground state the electron density $n_e$ is related to $\mu_e$ through $n_e=\mu_e^3/(3\pi^2)$ and if $\mu_\mu\geq m_\mu$ ($m_\mu=105.7$~MeV) the muon density reads $n_\mu=[(\mu_\mu^2 - m_\mu^2)^{3/2}]/(3\pi^2)$.


\begin{thebibliography}{99}

\bibitem{Lattimer2016} J. M. Lattimer and M. Prakash, Physics Reports 621 (2016) 127–164

\bibitem{Haensel2007} P. Haensel, A. Y. Potekhin, and D. G. Yakovlev, Neutron Stars I (Springer, Berlin, 2007).

\bibitem{Steiner2005} A. W. Steiner, M. Prakash, J. M. Lattimer, and P. J. Ellis, Phys. Rep. 411, 325 (2005).

\bibitem{ligo} B. P. Abbott {\it et al}. (The LIGO Scientific Collaboration and the Virgo
Collaboration). Phys. Rev. Lett. 121, 161101 (2018).

\bibitem{nicer1a} M. C. Miller {\it et al}., Astrophys. J. Lett. 887, L24 (2019).

\bibitem{nicer1b} T. E. Riley {\it et al}., Astrophys. J. Lett. 887, L21 (2019).

\bibitem{nicer2a} M. C. Miller {\it et al}., Astrophys. J. Lett. 918, L28 (2021).
 
\bibitem{nicer2b} T. E. Riley {\it et al}., Astrophys. J. Lett. 918, L27 (2021).

\bibitem{Wei2020} Jin-Biao Wei, Jia-Jing Lu, G. F. Burgio, Zeng-Hua Li, and H.-J. Schulze, Eur. Phys. J. A 56, 63 (2020).

\bibitem{Prakash2001} J. M. Lattimer, M. Prakash, Astrophys. J. 550, 426 (2001).

\bibitem{Souza2020} L. A. Souza, M. Dutra, C. H. Lenzi and O. Louren\c{c}o, Phys. Rev. C 101, 065202 (2020).

\bibitem{Bender2003} M. Bender, P. H. Heenen, and P. -G. Reinhard, Rev. Mod. Phys. 75, 121 (2003).

\bibitem{Reinhard1989} P. -G. Reinhard, Rep. Prog. Phys. 52, 439 (1989).

\bibitem{Typel2018} S. Typel, Particles 1, 3 (2018). 

\bibitem{Stone2007} J. Stone and P.G. Reinhard, Prog. Part. Nucl. Phys. 58(2), 587–657 (2007). 

\bibitem{Dutra2012} M. Dutra, O. Lourenço, J. S. Sá Martins, A. Delfino, J. R. Stone, P. D. Stevenson, Phys. Rev. C 85, 035201 (2012). 

\bibitem{Dutra2014} M. Dutra, O. Lourenço, S. S. Avancini, B. V. Carlson, A. Delfino, D. P. Menezes, C. Providência, S. Typel, J. R. Stone, Phys. Rev. C 90, 055203 (2014).

\bibitem{Annala2020} E. Annala, T. Gorda, A. Kurkela, J. N\"attil\"a, and A. Vuorinen, Nature Phys. 16, 907 (2020)

\bibitem{Somasundaram2022b} R. Somasundaram, I. Tews, J. Margueron, arXiv:2112.08157 [nucl-th].

\bibitem{Li2021} B.-A. Li, B.-J. Cai, W.-J. Xie and N.-B. Zhang, Universe 7, 182 (2021).

\bibitem{skyrmeligo} O. Louren\c{c}o, M. Dutra, C. H. Lenzi, S. K. Biswal, M. Bhuyan, and D. P. Menezes, Eur. Phys. J. A {\bf 56}, 32 (2020).

\bibitem{AMDC2016} G. Audi, F.G. Kondev, M. Wang, W.J. Huang, S. Naimin Chin. Phys. C 41, 030001 (2017).

\bibitem{unedf} \url{https://www.phy.ornl.gov/workshops/lacm08/UNEDF/database.html}

\bibitem{Angeli2013} I. Angeli, and K.P. Marinova, Atomic Data and Nuclear Data Tables 99, 69 (2013).

\bibitem{Chabanat1997} E. Chabanat, P. Bonche, P. Haensel, J. Meyer, and R. Schaeffer, Nucl. Phys. A 627, 710 (1997).

\bibitem{Fricke1995} G. Fricke, C. Bernhardt, and K. Heilig, Atomic Data and Nuclear Data Tables 60, 177 (1995).

\bibitem{Chabanat1998} E. Chabanat, P. Bonche, P. Haensel, J. Meyer, and R. Schaeffer, Nucl. Phys. A 635, 231 (1998).

\bibitem{BSK18} N. Chamel, S. Goriely, and J. M. Pearson, Phys. Rev. C 80, 065804 (2009).

\bibitem{unedf0} M. Kortelainen, T. Lesinski, J. Mor\'e, W. Nazarewicz, J. Sarich, N. Schunck, M. V. Stoitsov, and S. Wild, Phys. Rev. C 82, 024313 (2010).

\bibitem{ddme2} G. A. Lalazissis, T. Niksic, D. Vretenar, and P. Ring, Phys. Rev. C 71, 024312 (2005).

\bibitem{nl3s}  G. A. Lalazissis {\it et al}., Phys. Lett. B 671, 36 (2009).

\bibitem{NLRA1} M. Rashdan, Phys. Rev. C 63, 044303 (2001).

\bibitem{AME2003} G. Audi, A. H. Wapstra, and C. Thibault, Nucl. Phys. A 729, 337 (2003).

\bibitem{JYFLTRAP}\url{http://research.jyu.fi/igisol/JYFLTRAP$\_$masses/}

\bibitem{Dobaczewski2015} J. Dobaczewski, W. Nazarewicz and P.-G. Reinhard, J. Phys. G: Nucl. Part. Phys. 41, 074001 (2015).

\bibitem{Bertozzi1972} W. Bertozzi, J. Friar, J. Heisenberg, and J. W. Negele, Phys. Lett. 41B, 408 (1972).

\bibitem{Friar1975} J. L. Friar and J. W. Negele, Adv. in Nucl. Phys. 8, 219 (1975).

\bibitem{Friar1997} J. L. Friar, J. Martorell, and D. W. L. Sprung, Phys. Rev. A 56, 4579 (1997).

\bibitem{Jentschura2011} U.D. Jentschura, Eur. Phys. J. D 61, 7 (2011).

\bibitem{Nishizaki1988} S. Nishizaki, H. Kurasawa, T. Suzuki, Phys. Lett. B 209, 6 (1988).

\bibitem{pdg} P.A. Zylaet al.(Particle Data Group), Prog. Theor. Exp. Phys. 083C01 (2020) and 2021 update. \url{http://pdg.lbl.gov}

\bibitem{Karr2020} J.-P. Karr, D. Marchand, and E. Voutier, Nature Reviews Physics 2, 601 (2020).

\bibitem{Atac2021} H. Atac, M. Constantinou, Z.-E. Meziani, M. Paolone, and N. Sparveris, Eur. Phys. J. A 57, 65 (2021).

\bibitem{RocaMaza2008} X. Roca-Maza, M. Centelles, F. Salvat, and Xavier Vi\~nas, Phys. Rev. C 78, 044332 (2008).

\bibitem{Blaizot1980} J.-P. Blaizot, Phys. Rep. 64, 171 (1980). 

\bibitem{Blaizot1995} J.-P. Blaizot, J.-F. Berger, J. Dechargé, and M. Girod, Nucl. Phys. A 591, 435 (1995).

\bibitem{Stone2014} J.R. Stone, N.J. Stone, and S.A. Moszkowski, Phys. Rev. C 89, 044316 (2014).

\bibitem{Garg2018} U. Garg and G. Col\`o, Prog. Part. Nucl. Phys. 101, 55 (2018).

\bibitem{Bohigas1979} O. Bohigas, A. M. Lane, and J. Martorell, Phys. Rep. 51, 267 (1979).

\bibitem{Thouless1961} D. J. Thouless, Nucl. Phys. 22, 78 (1961).

\bibitem{Colo2004} G. Colo, N. V. Giai, J. Meyer, K. Bennaceur, and P. Bonche, Phys. Rev. C 70, 024307 (2004).

\bibitem{Avogadro2013} P. Avogadro and C.A. Bertulani, Phys. Rev. C 88, 044319 (2013).

\bibitem{ratp} M. Rayet, M. Arnould, F. Tondeur, and G. Paulus, Astron. Astrophys. 116, 183 (1992).

\bibitem{sgii} N. V. Giai and H. Sagawa, Phys. Lett. B, 106, 379 (1981).

\bibitem{siii} M. Beiner, H. Flocard, N. V. Giai, P. Quentin, Nucl. Phys. A 238, 29 (1975).

\bibitem{Khan2012} E. Khan, J. Margueron, and I. Vida\~na, Phys. Rev. Lett. 109, 092501 (2012); E. Khan and J. Margueron, Phys. Rev. C 88, 034319 (2013).

\bibitem{Margueron2018a} J. Margueron, R. H. Casali, and F. Gulminelli, Phys. Rev. C 97, 025805 (2018).

\bibitem{Somasundaram2021} R. Somasundaram, C. Drischler, I. Tews, and J. Margueron, Phys. Rev. C 103, 045803 (2021).

\bibitem{Lattimer2013} J. Lattimer and Y. Lim, Astroph J. 771, 51 (2013).

\bibitem{Tsang2009} M.B. Tsang, Y. Zhang, P. Danielewicz, M. Famiano, Z. Li, W.G. Lynch, A.W. Steiner, Phys. Rev. Lett. 102, 122701 (2009).

\bibitem{RocaMaza2015} X. Roca-Maza, X. Viñas, M. Centelles, B.K. Agrawal, G. Colò, N. Paar, J. Piekarewicz, D. Vretenar, Phys. Rev. C 92, 064304 (2015) 

\bibitem{Chen2010} L.W. Chen, C.M. Ko, B.A. Li, J. Xu, Phys. Rev. C 82, 024321 (2010) 

\bibitem{FRDM} P. M\"oller, W.D. Myers, H. Sagawa, S. Yoshida, Phys. Rev. Lett. 108, 052501 (2012) 
\bibitem{Danielewicz2013} P. Danielewicz and J. Lee, Nucl. Phys. A 922, 1 (2014)

\bibitem{Steiner2013} A.W. Steiner, J.M. Lattimer, E.F. Brown, Astrophys. J. Lett. 765, L5 (2013).

\bibitem{Tews2017} I. Tews, J. M. Lattimer, A. Ohnishi, E. E. Kolomeitsev, Astrophys. J. 848, 105 (2017).

\bibitem{Adhikari2021} D. Adhikari, H. Albataineh, D. Androic, K. Aniol, D. S. Armstrong, T. Averett, C. Ayerbe Gayoso, {\it et al}. Phys. Rev. Lett. 126, 172502 (2021).

\bibitem{Adhikari2022} D. Adhikari, H. Albataineh, D. Androic, K. A. Aniol, D. S. Armstrong, T. Averett, C. Ayerbe Gayoso, et al., arXiv, 16 juin 2022. http://arxiv.org/abs/2205.11593.

\bibitem{Reed2021} B. T. Reed, F. J. Fattoyev, C. J. Horowitz, and J. Piekarewicz, Phys. Rev. Lett. 126, 172503 (2021).

\bibitem{Reinhard2021} P.-G. Reinhard, X. Roca-Maza, and W. Nazarewicz, Phys. Rev. Lett. 127, 232501 (2021).

\bibitem{Zhang2022} Z. Zhang and Lie-Wen Chen. arXiv, 7 juillet 2022. http://arxiv.org/abs/2207.03328.

\bibitem{Reinhard2022} P.-G. Reinhard, X. Roca-Maza, and W. Nazarewicz, arXiv, 7 juin 2022. http://arxiv.org/abs/2206.03134.

\bibitem{Yuksel2022} E. Y\"uksel, and N. Paar, arXiv2206.06527, http://arxiv.org/abs/2206.06527.

\bibitem{tov39} R. C. Tolman, Phys. Rev. 55, 364 (1939).

\bibitem{tov39a} J. R. Oppenheimer and G. M. Volkoff, Phys. Rev. 55, 374 (1939).

\bibitem{glen} N. K. Glendenning, {\it Compact Stars}, 2nd ed. (Springer, New York, 2000).

\bibitem{DH2001} F. Douchin and P. Haensel, Astrop. \& Astron. 380, 151 (2001).

\bibitem{fortin} M. Fortin, C. Provid\^encia, A. R. Raduta, F. Gulminelli, J. L. Zdunik, P. Haensel, and M. Bejger, Phys. Rev. C 94, 035804 (2016).


\bibitem{Margueron2018b} J. Margueron, R. H. Casali, and F. Gulminelli, Phys. Rev. C 97, 025806 (2018).

\bibitem{Tews2018} I. Tews, J. Margueron, and S. Reddy, Phys. Rev. C 98, 045804 (2018).

\bibitem{Wolf2013} R. N. Wolf, D. Beck, K. Blaum, Ch. B\"ohm, Ch. Borgmann, M. Breitenfeldt, N. Chamel, et al., Phys. Rev. Lett. 110, 041101 (2013).

\bibitem{Fortin2016} M. Fortin, C. Provid\^encia, Ad. R. Raduta, F. Gulminelli, J. L. Zdunik, P. Haensel, and M. Bejger, Phys. Rev. C 94, 035804 (2016).

\bibitem{Antic2019} S. Antić, D. Chatterjee, T. Carreau, and F. Gulminelli. J. Phys. G 46, 065109 (2019).

\bibitem{Grams2022a} G. Grams, J. Margueron, R. Somasundaram, S. Reddy, Eur. Phys. J. A 58, 56 (2022).

\bibitem{Grams2022b} G. Grams, R. Somasundaram, J. Margueron, S. Reddy, Phys. Rev. C 105, 035806 (2022).

\bibitem{Xie2021} Wen-Jie Xie and Bao-An Li, Phys. Rev. C 103, 035802 (2021).

\bibitem{Baillot2019} N. Baillot d’Etivaux, S. Guillot, J. Margueron, N. Webb, M. Catelan, and A. Reisenegger, Astrop. J. 887, 48 (2019).

\bibitem{Sagawa2019} H. Sagawa, S. Yoshida, and Li-Gang Cao, AIP Conference Proceedings 2127, 020002 (2019).

\bibitem{Carson2019} Z. Carson, A. W. Steiner, and K. Yagi, Phys. Rev. D 99, 043010 (2019).

\bibitem{Guven2020} H. G\"uven, B. Kutsal, E. Khan, J. Margueron , Phys. Rev. C 102, 015805 (2020)

\bibitem{Abbott2019} B. P. Abbott, R. Abbott, T. D. Abbott, F. Acernese, K. Ackley, C. Adams, T. Adams, P. Addesso, R. X. Adhikari, V. B. Adya et al. (LIGO Scientific Collaboration and Virgo Collaboration), Phys. Rev. X 9, 011001 (2019).

\bibitem{De2018} S. De, D. Finstad, J. M. Lattimer, D. A. Brown, E. Berger, and C. M. Biwer, Phys.Rev.Lett.121, 091102 (2018).

\bibitem{Coughlin2019} M. W. Coughlin, T. Dietrich, B. Margalit, and B. D. Metzger, Mon. Not. R. Astron. Soc. 489, L91 (2019).

\bibitem{Chabanat-thesis} E. Chabanat, Interactions effectives
pour desconditions extrêmes d'isospin, Doctoral dissertation, Université Claude Bernard Lyon-1, 1995.

\bibitem{sd1} T. Lesinski, K. Bennaceur, C. Simenel, 2006, private communication. See also supplemental material.

\bibitem{iufsus} B.K.Agrawal, A.Sulaksono, P.-G.Reinhard, Nucl. Phys. A 882, 1 (2012).

\bibitem{sinpab} C. Mondal, B. K. Agrawal, J. N. De, and S. K. Samaddar
Phys. Rev. C 93, 044328 (2016).

\bibitem{BSR}S. K. Dhiman, R. Kumar, and B. K. Agrawal, Phys. Rev. C 76, 045801 (2007).

\bibitem{fsugz} R. Kumar, B. K. Agrawal, and S. K. Dhiman, Phys. Rev. C 74, 034323 (2006).

\bibitem{ska} H. S. K\"ohler, Nucl. Phys. A 258, 301 (1976).

\bibitem{fsugarnet}  Wei-Chia Chen and J. Piekarewicz, Phys. Letts. B 748, 284 (2015).

\bibitem{ddme1} T. Niksic, D. Vretenar, P. Finelli, and P. Ring, Phys. Rev. C 66, 024306 (2002).

\bibitem{fsugold2} Wei-Chia Chen and J. Piekarewicz, Phys. Rev. C 90, 044305 (2014).

\bibitem{q1} R. J. Furnstahl, B. D. Serot, and H. B. Tang, Nucl. Phys. A 615, 441 (1997).

\bibitem{fama1} J. Piekarewicz, Phys. Rev. C 66, 034305 (2002).

\bibitem{eer} M. Rufa, P.-G. Reinhard, J. A. Maruhn, W. Greiner, and M. R. Strayer, Phys. Rev. C 38, 390 (1988).

\bibitem{inertia1} J. B. Hartle and D. H. Sharp, Astrophys. J. 147, 317 (1967).

\bibitem{inertia2} I. A. Morrison, T. W. Baumgarte, S. L. Shapiro, and V. R. Pandharipande, Astrophys. J. 617, L135 (2004).

\bibitem{Prakash} S. Postnikov, M. Prakash and J. M. Lattimer, Phys. Rev. D 82, 024016 (2010). 

\bibitem{tanj10} T. Hinderer, B. D. Lackey, Ryan N. Lang, J. S. Read, Phys. Rev. D 81, 123016 (2010).

\bibitem{hind08} T. Hinderer, Astrophys. J. 677, 1216 (2008).

\bibitem{damour}  T. Damour, A. Nagar, Phys. Rev. D 81, 084016 (2010).

\bibitem{tayl09} T. Binnington, E. Poisson, Phys. Rev. D 80, 084018 (2009).

\bibitem{had4} O. Louren\c{c}o, M. Dutra, C. H. Lenzi, S. K. Biswal, M. Bhuyan, and D. P. Menezes, Eur. Phys. J. A 56, 32 (2020).

\bibitem{Yagi2014} K. Yagi and N. Yunes, Science 341, 365 (2014).

\bibitem{Reinhard1995} P.-G. Reinhard and H. Flocard, Nucl. Phys. A 584, 467 (1995).

\bibitem{Kurasawa2019} H. Kurasawa, T. Suzuki, Prog. Theor. Exp. Phys. 113D01 (2019).

\end{thebibliography}
\end{document}